\documentclass[prb,reprint,twocolumn,aps,amssymb,graphicx,floatfix,amsmath,superscriptaddress,showpacs,longbibliography]{revtex4-1}
\usepackage{bm}
\usepackage{float}
\usepackage{graphicx}
\usepackage{color}
\usepackage{amsfonts}
\usepackage{amsmath}
\usepackage{textcomp}
\usepackage{subcaption}
\usepackage{hyperref}
\usepackage{comment}
\hypersetup{colorlinks=true,linkcolor=blue,anchorcolor=blue,citecolor=blue,filecolor=blue,urlcolor=blue,bookmarksnumbered=true,pdfview=FitB}
\newcommand{\w}{\omega}
\newcommand{\e}{{c(s)}}

\newcommand{\beq}{\begin{equation}}
\newcommand{\eeq}{\end{equation}}
\newcommand{\bea}{\begin{eqnarray}}
\newcommand{\eea}{\end{eqnarray}}
\newcommand{\dd}{\tilde{\boldsymbol{\Delta}}^\e}
\newcommand{\nn}{\nonumber}
\begin{document}

\title{Collective modes near a Pomeranchuk instability}
\author{Avraham Klein}
\affiliation{School of Physics and Astronomy, University of Minnesota, Minneapolis, MN 55455}
\author{Dmitrii L. Maslov}
\affiliation{Department of Physics, University of Florida, P.O. Box 118440, Gainesville, Florida 32611-8440, USA}
\author{Lev P. Pitaevskii}
\affiliation{INO-CNR BEC Center and Dipartimento di Fisica, Universita di Trento}
\affiliation{Kapitza Institute for Physical Problems, Russian Academy of Sciences, Moscow, Russia}
\author{Andrey V. Chubukov}
\affiliation{School of Physics and Astronomy, University of Minnesota, Minneapolis, MN 55455}

\begin{abstract}
  We consider collective excitations of a Fermi liquid.
    For each value of the angular momentum $l$, we study the evolution of longitudinal and transverse collective modes in the charge ($c$) and spin ($s$) channels with the Landau parameter $F^\e_l$, starting from positive  $F^\e_l$ and all the way to the Pomeranchuk transition at $F^\e_l=-1$. In each case, we identify a critical zero-sound mode, whose velocity vanishes at the Pomeranchuk instability. For $F^\e_l <-1$, this mode is located in the upper frequency half-plane, which signals an instability of the ground state. In a clean Fermi liquid the critical mode may be either purely relaxational or almost propagating, depending on the parity of $l$ and on whether the response function is longitudinal or transverse.
  These differences
  lead
  to qualitatively different types of time evolution of the order parameter following an initial perturbation.
  A special situation occurs for
  the $l=1$ order parameter that coincides with the spin or charge current.
  In this case the residue of the critical mode vanishes at the Pomeranchuk transition.
  However,
the critical mode
  can be identified at any distance from the transition, and
  is still located in the upper frequency half-plane for $F^\e_1 <-1$.
  The only peculiarity of
the charge/spin current order parameter is that
  its time evolution occurs on longer scales than for other order parameters.
  We also analyze collective modes away from the critical point, and find that the modes
 evolve
with $F^\e_l$
on a multi-sheet Riemann surface.
For certain intervals of $F^\e_l$, the modes
  either move to an unphysical Riemann sheet or stay on
the  physical
 sheet but away from the real frequency axis. In that case,
the modes
  do not give rise to peaks in the imaginary parts of the corresponding susceptiblities.
\end{abstract}

\date{\today}

\maketitle
\tableofcontents
\section{Introduction}

A Pomeranchuk transition is an instability of a Fermi liquid (FL) towards a spontaneous order which breaks rotational symmetry but leaves translational symmetry intact~\cite{Pomeranchuk1959}. Examples include ferromagnetism \cite{Dzyaloshinskii1976,Kondratenko1965,Kondratenko1964} and various forms of nematic order in quantum Hall systems, Sr$_3$Ru$_2$O$_7$, and cuprate and Fe-based superconductors \cite{Fernandes2014,Fradkin2010}. For a rotationally-invariant system in two dimensions (2D),
  deformations of the Fermi surface (FS) can be classified by the value of the angular momentum $l$.
In general, a deformation with only one particular  $l$ develops at a Pomeranchuk transition.
 A Pomeranchuk order parameter
$\Delta^{c(s)}_l ({\bf q})  =
\sum_{\bf k}
 f^{c(s)}_l ({\bf k})
\langle a^\dagger_{{\bf k}
+{\bf q}/2,
\alpha} t^{c(s)}_{\alpha, \alpha'}  a_{{\bf k}
-{\bf q}/2,
\alpha'}^{\phantom{\dagger}}
\rangle
$ is bilinear in fermions and has the spin structure $t^{c}_{\alpha, \alpha'} = \delta_{\alpha,\alpha'}$ or $t^{s}_{\alpha, \alpha'} = \sigma^z_{\alpha,\alpha'}$ in the charge ($c$) and spin ($s$) channels, correspondingly ($\sigma^z$ is the Pauli matrix).
The order parameter is assumed to vary slowly, i.e.,  $q\ll \min\{a^{-1}_0,k_F\}$, where $a_0$ is the lattice constant and $k_F$ is the Fermi momentum.
Under rotations, the form-factors
  $f^{c(s)}_l ({\bf k})$ transform as basis functions of the angular momentum and, in general, also depend on the magnitude of  $ |{\bf k}|  \equiv k$. For example, $f^{c(s)}_{1} ({\bf k}) = \cos \theta f^{c(s)}(k)$ or $f^{c(s)}_{1} (k) = \sin \theta f^{c(s)}(k)$,
 where $\theta$ is the azimuthal angle of ${\bf k}$.
 According to the FL theory \cite{Landau1980,Abrikosov1975}, a Pomeranchuk order with angular momentum $l$ emerges when the corresponding Landau parameter $F_l^{c(s)}$ approaches the critical value of $-1$
 from above.

In this paper we focus on dynamical aspects of a Pomeranchuk instability. We consider primarily the 2D case, because examples of Pomeranchuk transitions have been discussed mostly for 2D systems \cite{Oganesyan2001a, Wu2004, DellAnna2006, Wu2007, Chubukov2009, Maslov2010,Lederer2017,Chubukov2015,Wang2015,Hartnoll2014,Chu2012,Chu2010,Fernandes2012,Metlitski2010a,Klein2018c,Klein2018,Klein2018a,Wu2018}. We consider an isotropic FL but do not specifically
assume
 Galilean invariance, i.e.,
the single-particle dispersion in our model is not necessarily quadratic in $|{\bf k}|$.
 The main object of our study is the dynamical susceptibility, $\chi_l (q,\omega)$, which corresponds to a particular order parameter $\Delta^\e_l$. In 2D, there are two types of susceptibilities for any given $l$: a longitudinal one, with the form-factor proportional to $\cos{l \theta}$, and a transverse one, with the form-factor proportional to $\sin {l \theta}$ (for $l=0$, there is only one type, with an isotropic form-factor).
 In the low energy limit ($q \to 0$ and $\omega \to 0$)
 the dynamical susceptibility is a function of the ratio $s = \omega / v_F^* q$, where $v_F^*$ is the renormalized Fermi velocity.
  As a function of complex variable, $\chi(s)$ has both poles and branch cuts in the complex plane. We focus on the retarded susceptibility, which is analytic in the upper half plane $\mbox{Im} s > 0$, except for the case when the system is below the Pomeranchuk instability,
   and,
   in general, has poles and branch cuts in the lower half plane $\mbox{Im}s < 0$. The poles of
 $\chi(s)$ correspond to zero-sound collective modes whose frequency and momentum are related by
 $\omega = s v_F^*q$.
If the Landau parameter is positive and non-zero for only one value of $l$, there is one longitudinal and at most one transverse zero-sound mode for any $l\neq 0$. These are conventional propagating modes with $\text{Re}s > 1$ and infinitesimally small $\text{Im}s$ in the clean limit, when the fermionic lifetime is infinite.
The branch cuts are a consequence of the non-analyticity of the free-fermion bubble (the Lindhard function). Their significance is that the zero-sound poles are defined on a multi-sheet Riemann surface.
In 2D, this non-analyticity is particularly simple, being just a square-root (see Sec.~\ref{sec:qpsusc}). As a result, the Riemann surface is a genus 0, two-sheet surface. We shall refer to the first sheet, which includes the analytic half plane, as the ``physical sheet''. In our work we mostly discuss the properties of the physical sheet.

We obtain explicit results for the frequencies of collective modes in the whole range of $  -1<F^\e_l<  \infty$. First, we consider a clean FL (Sec.~\ref{sec:qp}). We present explicit results for the zero-sound modes with $l=0,1,2$
 (Secs.~\ref{sec:l=0}, \ref{sec:l=1}, and \ref{sec:l=2}, correspondingly) and analyze the structure of the zero-sound modes for arbitrary $l$ (Sec.~\ref{sec:analysis-general-l}). We show that for $l=0$ and in the transverse channel with $l\neq 0$, all zero-sound modes acquire a finite decay rate already for an arbitrary small negative $F^\e_l$,
 i.e., $s = \pm a -ib$, where  $a$, $b$ are real and $b >0$. A positive $b$ implies that a perturbation of the order parameter decays exponentially with time.  The frequency of one of the modes vanishes at the Pomeranchuk transition. We call this mode the critical one.
 In the longitudinal channel,
  the decay rate of one of the modes remains infinitesimally small until the corresponding $|F^\e_l|$ exceeds a threshold value. Immediately below the threshold, this mode is located at $s = \pm a -ib$  with $a >1$ and $b\ll 1$. Even though the mode frequency is almost real, the corresponding pole is located below the branch cut and thus cannot be reached from the real axis of the physical sheet. Accordingly, the imaginary part of the susceptibility does not have a peak above the particle-hole continuum for real $s$,
 i.e., the mode is ``silent''.
Below the transition, i.e., for $F^\e_l < -1$,  the pole is located in the upper frequency half-plane, and a perturbation of the order parameter grows exponentially with time, which indicates that a FL becomes unstable with respect to a Pomeranchuk order. We obtain explicit results for the frequencies of collective modes in the whole range of $  -1<F^\e_l<  \infty$.

In Sec.~\ref{sec:gamma}, we analyze how the dispersion of collective modes is modified in the presence of impurity scattering. In the dirty limit, the critical modes in both longitudinal and transverse channels become overdamped for all $l$. Impurity scattering also smears the threshold, described in the previous paragraph, i.e., the longitudinal collective modes have non-zero damping rates for any  $F^\e_l <0$.

In Sec.~\ref{sec:time}, we analyze the susceptibility in the time domain,
\begin{equation}
  \chi^\e_l(q,t)=\int \frac{d\omega}{2\pi} \chi^\e_l(q,\omega) e^{i\omega t},
\end{equation}
which determines the time evolution of the order parameter following an initial perturbation.  We obtain explicit forms of
$\chi^\e_l(q,t)$ for $l=0$ and $l=1$.  Above the Pomeranchuk transition,
the time dependence of $\chi^\e_l(q,t)$ in the clean limit is a combination of an exponentially decaying part, which comes from the poles of $\chi^\e_l(q,\omega)$, and of an oscillatory (and algebraically decaying) part, which comes from its branch cuts. At the transition, $\chi^\e_0(q,t)$ reaches a time-independent limit at $t\to\infty$, while $\chi^\e_1(q,t)$ grows linearly with time in the clean case and saturates at a finite value in the presence of disorder. Below the transition, the poles of $\chi^\e_{0,1}(q,\omega)$
are located in the upper half-plane of $\omega$.  Consequently, both $\chi^\e_0(q,t)$ and $\chi^\e_1(q,t)$ increase exponentially with time. This means that any small fluctuation of the corresponding order parameter is amplified, and thus the ground state with no Pomeranchuk order is unstable.
In the case of finite disorder, the branch-cut contribution also begins to decay exponentially, on top of its algebraic and oscillatory behavior.

In Sec.~\ref{sec:current}, we consider the special case of an order parameter that coincides with either the charge or spin current. Previous studies\cite{Kiselev2017,Wu2018,Chubukov2018} found that the corresponding static susceptibility, $\chi^\e_{1}(q,0)$,  does not diverge at the tentative Pomeranchuk instability at $F^\e_1=-1$ because of the Ward identities that follow from conservation of total charge and spin.
We analyze the dynamical susceptibility for such an order parameter. We show, using both general reasoning and direct perturbation theory for the Hubbard model, that while the static susceptibility indeed remains finite at $F^\e_1=-1$, the {\it dynamical} one still has a pole, which moves to the upper frequency half-plane below the transition. The residue of this pole vanishes as $(1 + F^\e_1)^2$ at $F^\e_1 = -1$, but is finite both for $ F^\e_1>-1$ and  $F^\e_1<-1$. We argue that the presence of the pole in the upper frequency half-plane for $F^\e_1 <-1$  indicates that the state with no Pomeranchuk order becomes unstable, like for any other type of the order parameter.
We derive a Landau functional for the charge/spin current order parameter and show
that it has a conventional form,
except that
the coupling between the order parameter and an external perturbation has an additional
factor of
$1+ F^\e_1$.
We argue that the charge/spin current order does develop at $1 + F^\e_1 <0$,
 just as
for a generic  $l=1$ order parameter, but
it takes longer
to reach
equilibrium
after an instantaneous perturbation. This result differs from earlier claims that there is no Pomeranchuk transition to a state with the charge/spin current order parameter~\cite{Kiselev2017,Chubukov2018,Wu2018}.

Before we move on, a comment is in order. It is well known that the range of FL behavior shrinks as the system approaches a Pomeranchuk instability and disappears at the transition point, where the system displays non-Fermi-liquid behavior down to the lowest energies. In our analysis, we will be studying the collective modes at finite $s = \omega/(v^*_F q)$ and assume that $\omega$ and $ q$ are both small enough so that at any given distance to the critical point the system remains a Fermi liquid.

\section{Dynamical quasiparticle susceptibility near a Pomeranchuk transition in a clean Fermi liquid} \label{sec:qp}

\subsection{Quasiparticle susceptibility}
\label{sec:qpsusc}
According to the Kubo formula, the correlation function of an order parameter $\Delta^\e_l$ is related to the susceptibility $\chi^\e_l (q,\omega)$ with respect to the conjugated ``field''. In its turn, $\chi^\e_l (q,\omega)$ is given by a fully renormalized particle-hole bubble with external momentum $q$ and external frequency $\omega$. For free fermions, the particle-hole bubble is just a convolution of two fermionic Green's functions, whose momenta (frequencies) differ by $
{\bf q}$ ($\omega$).
In the time-ordered representation, we define the normalized susceptibility as
 \begin{widetext}
   \begin{eqnarray}
     \chi^\e_{\text{free},l} (q, \omega)=-\frac{2i}
     {     N_F} \int \frac{d^ D k}{(2\pi)^D} \int \frac{d
     \varepsilon
     }{2\pi} \bigg\vert f^\e_l({\bf k})\bigg\vert ^2G_{\text{free}} \left( {\bf k}+\frac 12 {\bf q},
     \varepsilon+\frac 12\omega \right) G_{\text{free}} \left( {\bf k}-\frac 12 {\bf q},
     \varepsilon-\frac 12\omega \right)
   \end{eqnarray}
 \end{widetext}
where $N_F$ is the density of states at the Fermi energy $E_F$,
$G_ {\text{free}} ( {\bf k},\varepsilon) = 1/ \left[\varepsilon -\epsilon_k+E_F+i\delta \text{sgn} \varepsilon \right]$
is the (time-ordered) Green's function, $\epsilon_k$ is the single-particle dispersion,
 $D$ is the spatial dimensionality, and a factor of two comes from summing over spins. We will be interested only in the case of small $q$ and $\omega$, i.e., $q\ll k_F$ and $\omega\ll E_F$. In this case,  integration over the internal fermionic momentum and frequency is confined to the regions of small $
\varepsilon$ and $ \epsilon_k-E_F$, i.e., the susceptibility comes from the states near the FS or, for brevity, from ``low-energy fermions''.

For interacting fermions, the particle-hole bubble is modified in several ways \cite{Landau1980,Abrikosov1975,Leggett1965}. First, the self-energy corrections transform a free-fermion Green's function near the FS into a quasiparticle Green's function, in which the bare velocity $v_F$ is replaced by the renormalized velocity $v^*_F  = v_F (m/m^*)$, where $m^*$ is the renormalized mass, and the Green's function is multiplied by the quasiparticle residue $Z<1$. Second, interactions between low-energy fermions generate multi-bubble contributions to the susceptibility. These renormalizations transform the free-fermion susceptibility $\chi^\e _{\text{free},l} (q,\omega)$ into the quasiparticle susceptibility $\chi^\e_{\text{qp},l} (q,\omega)$. (The effect of damping due to the residual interaction between quasiparticles is a subleading effect in the range of $q$ and $\omega$ of interest to us,
and will not be considered here.) Third, fermions far away from the FS (``high-energy fermions'') also contribute to the full susceptibility $\chi^\e_l (q, \omega)$.

The general expression for the dynamic susceptibility is~\cite{Leggett1965}
\begin{equation}
\chi^\e_l (q, \omega) =  (\Lambda^\e_l)^2 \chi^\e_{\text{qp},l} (q,\omega) + \chi^\e_{\text{inc},l}. \label{new_1}
\end{equation}
Here $\Lambda^\e_l$ is the side vertex, renormalized by high-energy fermions, and the stand-alone term  $\chi^\e_{\text{inc},l}$ represents the contribution solely from high-energy fermions. This last term does not have a singular dependence on $q$ and $\omega$ and will not play any crucial role in our analysis.

The quasiparticle contribution to the susceptibility depends on the fully renormalized (and antisymmetrized) interaction between low-energy fermions.
 usually denoted by
 $ \Gamma_{\alpha \beta, \gamma \delta} ({\bf k} -{\bf p})$. This interaction includes renormalizations by high-energy fermions but not by low-energy fermions. For a rotationally- and $SU(2)$-invariant system, which we consider here, $\Gamma _{\alpha \beta, \gamma \delta} ({\bf k}-{\bf p})$ can be expanded over harmonics characterized by orbital momenta $l$, and the
  properly normalized coefficients of this expansions are known as Landau parameters $F^\e_l$:
\begin{eqnarray}
 && \Gamma_{\alpha \beta, \gamma \delta} = \Gamma^c ({\bf k} - {\bf p}) \delta_{\alpha \gamma} \delta_{\beta \delta} + \Gamma^s ({\bf k} - {\bf p}) \boldsymbol{\sigma}_{\alpha \gamma} \cdot \boldsymbol{ \sigma}_{\beta \delta}, \nonumber \\
  && \Gamma^\e ({\bf k} - {\bf p}) = \Gamma^\e_0 + 2 \sum_{l=1}^\infty \Gamma^\e_l \cos{l(\theta _{{\bf k}} -\theta_{{\bf p}})} \nonumber \\
  &&  F^\e_l = \nu_F \Gamma^\e_l, \label{ex_a}
\end{eqnarray}
where
\begin{eqnarray}
\nu_F=2 N_F Z^2 \frac{m^*}{m}.
\end{eqnarray}
and
 $\theta_{\bf k},\theta_{\bf p}$
are the azimuthal angles of ${\bf k}$, ${\bf p}$. The static quasiparticle susceptibility, $\chi^\e_{\text{qp},l} (q,0)$, is expressed in terms of just a single $F^\e_l$:
\begin{equation}
  \chi^\e_{\text{qp},l} (q, 0) = \frac{\nu_F}{1 +  F^\e_l}. \label{new_2_1}
\end{equation}
 The dynamical quasiparticle susceptibility cannot, in general, be expressed in terms of a single Landau parameter, unless all Landau parameters except for a single $F^\e_l$  are small. In this special case,
\begin{equation}
  \chi^\e_{\text{qp},l} (q,\omega) = \nu_F \frac{\chi_{\text{free},l} (q^*,\omega)}{1 +  F^\e_l \chi_{\text{free},l} (q^*,\omega)}, \label{new_2}
\end{equation}
where $q^* = (m/m^*)q$  and $\chi_{\text{free},l} (q^*,\omega)$ is normalized to $\chi_{\text{free},l} (q^*,0) =1$ (we recall that we consider small $q^* \ll k_F$). For all order parameters, except for the charge or spin current, the vertex $\Lambda^\e_l$ in Eq.~(\ref{new_1}) is expected to remain finite at the Pomeranchuk transition.  The behavior of the full susceptibility is then determined entirely by the quasiparticle $\chi^\e_{\text{qp},l} (q,\omega)$.

Although the calculations are straightforward and some of the results have appeared before, \cite{Pethick1988,Beal-Monod1994,Anderson2011,Li2012,Oganesyan2001,Wu2018,Chubukov2018,Zyuzin2018,Sodemann2018,
Lucas2018,Torre2019,Dupuis,Tremblay} we include below the details of the derivation of $\chi^\e_{\text{qp},l}(q,\omega)$ in 2D, as we will be interested in the pole structure of the susceptibility not only near a Pomeranchuk transition but also away from it. In what follows we consider separately the cases of $l=0,1,2$, and then analyze the case of arbitrary $l$. In these calculations we assume that a single Landau parameter $F^{\e}_l$ is much larger than the rest and compute $\chi^\e_{\text{qp},l} (q,\omega)$ using Eq.~(\ref{new_2}).
 In Sec.~\ref{sec:F0F1}, we consider the case when $F^\e_0$ and $F^\e_1$ are comparable, while all $F^{c(s)}_{l>1}$ can be neglected.

For definiteness, in this and the next two sections we approximate the form-factors by their values on the FS, as $f^\e_l ({\bf k}_F) = \sqrt{2} \cos(l \theta)$ in the longitudinal channel and $f^\e_l ({\bf k}_F) = \sqrt{2} \sin(l \theta)$ in the transverse channel, where ${\bf k}_F=k_F {\bf k}/k$ and $\theta$ is the angle between the direction of ${\bf k}_F$ and the $x$ axis.

 \subsection{$l=0$}
\label{sec:l=0}

In this case the form-factor $f^\e_{0} ({\bf k}_F)$ is just a constant. The form of the retarded free-fermion susceptibility along the real frequency axis is well known
\begin{eqnarray}
\chi_{\text{free},0} (q^*,\omega)           & = & 1+\frac{i\omega}{\sqrt{(v_F q^*)^2-(\omega+i\delta)^2}}\nn                                                                                                                                                                \\
                                            &   & =1+\frac{is}{\sqrt{(1-(s+i\delta)^2}}.
\label{ret}
\end{eqnarray}
where we used that $v_F q^* = v^*_F q$ and defined $s=\omega/v^*_Fq$. Viewed as a function of complex $s$, $\chi_{\text{free},0} (s)$ has branch cuts, which start at $ s= -i \delta$ below the real axis and run along the segments $(-\infty,-1)$ and $(1,\infty)$ along the real axis.

Traditionally, $\delta$ in Eq.~(\ref{ret}) is interpreted as an infinitesimally small damping rate whose physical origin does need not to be specified and whose sole purpose is to shift the branch cut into the lower half-plane of complex $s$. We will see, however, that such approach is not sufficient for our purposes, because it would not allow us to resolve the relative positions of the zero-sound poles and branch cuts of the susceptibility in the complex plane of $s$. For this reason, we will consider a specific damping mechanism, namely, scattering by short-range impurities, and treat $\delta$  as a finite albeit small number.

The order parameter in the $l=0$ channel (charge or spin) is conserved, i.e., the susceptibility must satisfy $\chi^\e_0(q=0,\omega)=0$
 (see, e.g., Refs. ~\onlinecite{Nozieres1999,Halperin1977}). Once $\delta$ is finite, Eq.~(\ref{ret}) does not satisfy this condition because it was obtained either by adding  $i \delta$ self-energy corrections to the Green's functions or, which is equivalent, by solving the kinetic equation in the relaxation time approximation. To ensure that charge and spin are conserved, one also has to include vertex corrections to the particle-hole bubble or go beyond the relaxation time approximation
\footnote{The collision integral in the relaxation-time approximation, $I_{\text{coll}}=-\delta\times (f-f_0)$ with ($f$) $f_0$ being (non)equlibrium distribution function, is not an appropriate form for impurity scattering, which is elastic and therefore must conserve a number of particles with given energy. Consequently, $I_{\text{coll}}$ must vanish upon averaging over the directions of the momentum, which is not the case for  $I_{\text{coll}}$. The correct form of the collision integral for impurity scattering is $-\delta\times (f- \bar f)$, where $\bar f$ is the angular average of $f$.  The kinetic equation with this collision integral reproduces Eq.~(\ref{retd}).}.
The corresponding free-fermion susceptibility is given by~\cite{Zala2001}
\begin{equation}
\chi_{\text{free},0} (s) = 1+\frac{is}{\sqrt{(1-(s+i\delta)^2} - \delta},
\label{retd}
\end{equation}
where $\delta$ now stands for the dimensionless impurity scattering rate. The $-\delta$ term next to $\sqrt{1- (s+i\delta)^2}$ in (\ref{retd}) comes from vertex corrections. Until Sec.~\ref{sec:gamma}, we will be assuming that impurity scattering is weak, i.e., $\delta\ll \min\{\text{Re}s,\text{Im}s\}$.

For $F^\e_0>0$ we expect to have well-defined collective modes with $|s|>1$. In this case, one can safely neglect $\delta$ in Eq.~(\ref{retd}) and replace $ \sqrt{1-(s+i\delta)^2}$ by $-i \text{sgn} s \sqrt{s^2 - 1}$. Equation~(\ref{retd}) is then reduced to
\begin{equation}
\chi_{\text{free},0}(s) = 1 - \frac{|s|}{\sqrt{s^2 - 1}}.
\label{zzz_1}
\end{equation}
Substituting this form into Eq.~(\ref{new_2}), we obtain
\begin{equation}
\chi^\e_{\text{qp},0}(s)
=
\nu_F \frac{1-\frac{|s|}{\sqrt{s^2 - 1}}}{1 + F^\e_0 \left(1-\frac{|s|}{\sqrt{s^2 - 1}}\right)}.
 \label{2}
\end{equation}
The locations of the poles are determined from the equation
\begin{equation}
1 + F^\e_0 - F^\e_0 \frac{|s|}{\sqrt{s^2 -1}} =0. \label{yyya}
\end{equation}
One can check that the solution $s_{1,2} = \pm s_p$ with $s_p>1$
indeed exists only for $F^\e_0 >0$:
\begin{equation}
s_p = \frac{1 + F^\e_0}{\sqrt{1 + 2 F^\e_0}}.
\label{zzz}
\end{equation}

We now widen the scope of our analysis and search for solutions with complex $s$. To this end, we need to keep $\delta$ terms in $\chi_{\text{free}, 0} (s)$. The quasiparticle susceptibility for $s$ in the lower half-plane is obtained by substituting (\ref{retd}) into (\ref{new_2}).  This yields
 \begin{equation}
\chi^\e_{\text{qp},0}(s)
=
\nu_F \frac{1 + i\frac{s}{\sqrt{1 - (s + i \delta)^2} - \delta}}{1 + F^\e_0 \left(1 + i\frac{s}{\sqrt{1 - (s + i \delta)^2} - \delta}\right)}.
 \label{2_1}
\end{equation}
 Using Eq.~(\ref{2_1}), we can study the poles of $\chi^\e_{\text{qp},0}(s)$ everywhere in the lower half-plane of complex $s$ and in the whole range of $F_0^\e$.

The positions of the poles in the lower half-plane of $s$ are determined by
\begin{equation}
  \frac{1 + F^\e_0}{F^\e_0} = - \frac{is}{\sqrt{1 - (s+i\delta)^2} - \delta}. \label{yyyb}
\end{equation}

\begin{figure*}
 \centering
  \begin{subfigure}{0.45\hsize}
      \centering
      \includegraphics[width=\hsize,clip,trim=0 -30 0 -30]{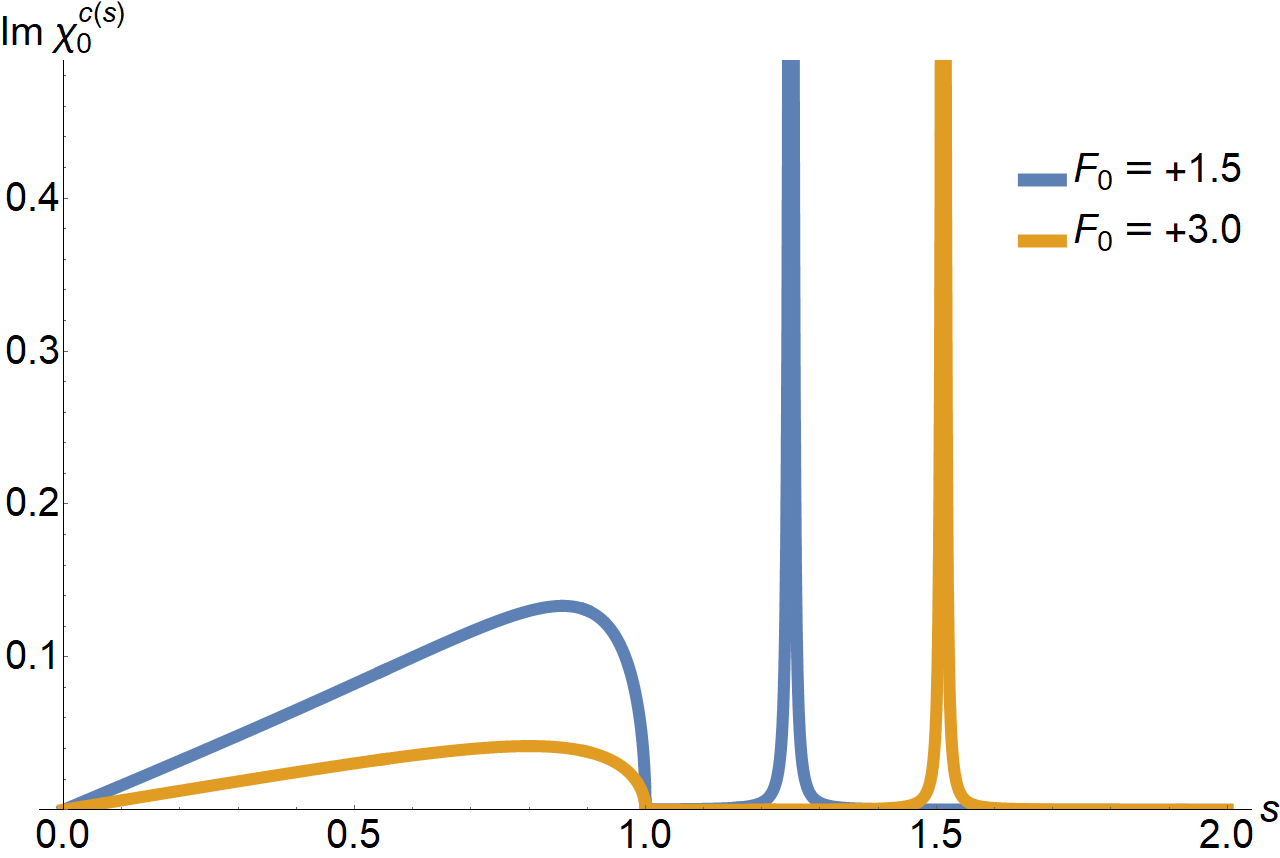}
      \caption{\label{fig:im-chi0-nogamma-pos}}
    \end{subfigure}\hfill
  \begin{subfigure}{0.45\hsize}
      \includegraphics[width=\hsize]{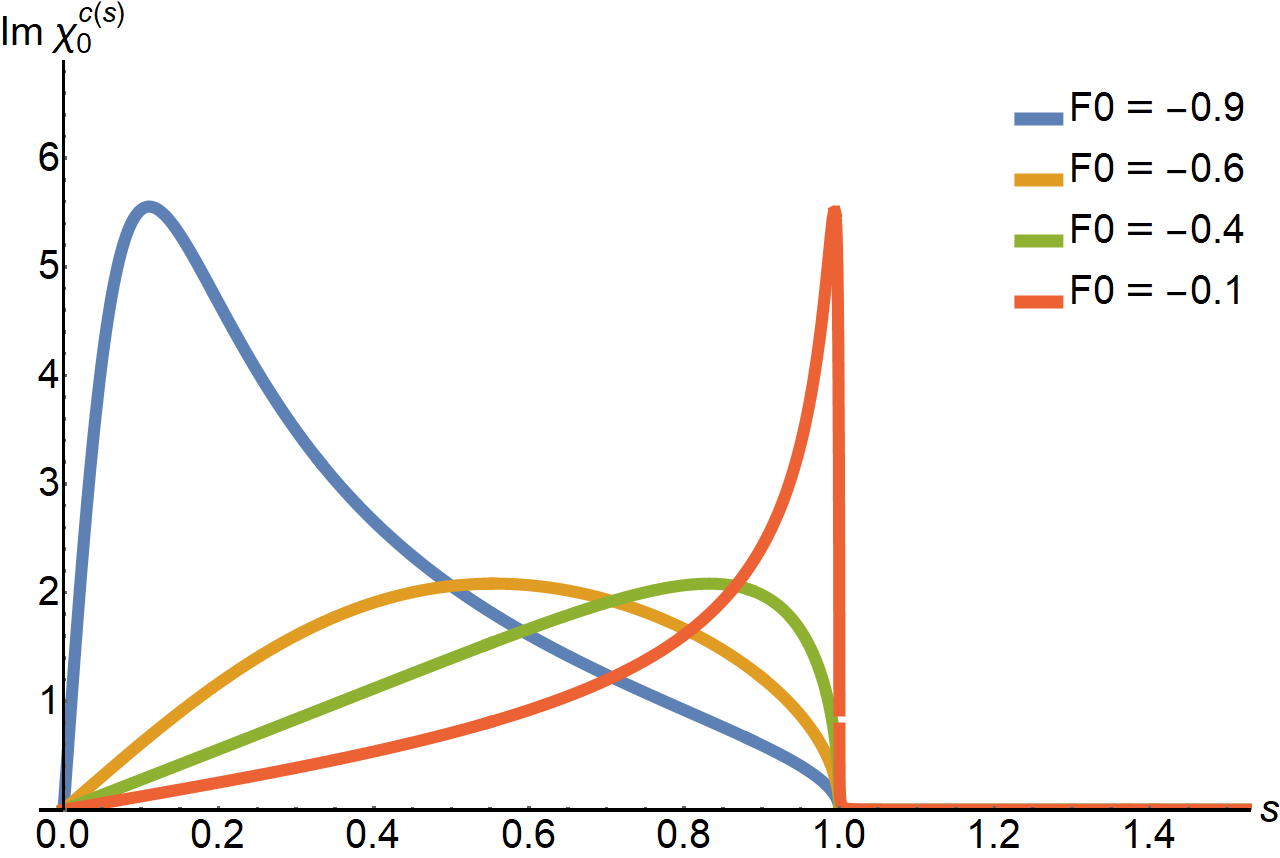}
    \caption{\label{fig:im-chi0-nogamma-neg}}
  \end{subfigure}
  \caption{(color online) The imaginary part of the susceptibility in the $l=0$ channel, $\chi_{\text{qp},0}^\e(s)$, where $s = \omega/v_F^* q$.  (a) $\mbox{Im}\chi_{\text{qp},0}^{c(s)}(s)$ for different $F_0^{\e} > 0$ in a clean system (a small impurity scattering rate  $\delta= 10^{-3}$ was added to make the poles visible). (b) $\mbox{Im}\chi_{\text{qp},0}^{\e}(s)$ for different $F_0^{\e} < 0$ in a clean system. Note that $\mbox{Im}\chi_{\text{qp},0}^\e (s)$ is nonzero only for $|s| < 1$ in this case.
    In this and all other figures, we omit the
``qp'' subscript and
``c(s)''
 superscript of the susceptibiliies,
    and set $\nu_F = 1$.
}
\end{figure*}

For $F^\e_0>0$, the solutions of (\ref{yyyb}) are
\begin{equation}
  s_{1,2} =  \pm s_{p,0} - i{\tilde \delta}_0,
\end{equation} where
\begin{equation}
  s_{p,0} = \frac{1 + F^\e_0}{\sqrt{1 + 2 F^\e_0}}
  \;\text{and}~ {\tilde \delta}_0 =
  \delta \frac{1 + F^\e_0}{1 + 2 F^\e_0}.\label{eq:pole-positive-F0}
\end{equation}
Up to the $-i\tilde\delta$ term, this result coincides with Eq.~(\ref{zzz}), as it should. We see that $\tilde{\delta}$ is positive but {\it smaller} than $\delta$ in (\ref{yyyb}). This implies that the poles are located  {\it above} the branch cuts. The purpose of starting with Eq.~(\ref{retd}) with small but finite $\delta$ was to resolve the difference between $\delta$, which determines the locations of the branch cuts, and ${\tilde\delta}_0$, which determines the distance between the poles and the real axis. Near the poles, the susceptibility reduces to
\begin{flalign}
\chi^\e_{\text{qp},0}(s) \propto \frac{F^\e_0}{(1+2F^\e_0)^{3/2}} \left(\frac{1}{s + s_{p,0} +i{\tilde \delta}_0} -\right.\nonumber\qquad\\ \left.\frac{1}{s-s_{p,0} + i {\tilde \delta}_0}\right).
\label{naa_1}
\end{flalign}
This expression is valid for complex $s$ above the branch cut at $\text{Im}s=-i \delta$. This includes the real axis. For real $s$ and vanishingly small ${\tilde \delta}_0$,  $\text{Im}\chi^\e_0 (s)$ has $\delta-$functional peaks at  $s = \pm s_{p,0}$ with $ s_{p,0}>1$, i.e.,  outside the particle-hole continuum (see Fig.~\ref{fig:im-chi0-nogamma-pos}).

For negative $F^\e_0$ we search for complex solutions of Eq.~(\ref{yyya})
in the form  $s = \pm a -ib$.
For $F^\e_0 < -1/2$, there exists a purely imaginary solution $s = - i s_{i,0}$, where
\begin{equation}
  s_{i,0} =
  \frac{1 -|F^\e_0|}{\sqrt{2 |F^\e_0|-1}}.
  \label{ac1}
\end{equation}
The pole at $s = -i s_{i,0}$   describes a purely relaxational zero-sound mode. As long as $1 + F^\e_0 >0$, the pole in $\chi^\e_{\text{qp},0} (q, \omega)$ is in the lower half-plane of $s$, i.e., excitations decay exponentially with time. Once $1+F^\e_0$ becomes negative, the pole moves into the upper half-plane. Then excitations grow exponentially with time, i.e.,
the system becomes unstable (see Sec. \ref{sec:time} for more detail). This is corroborated by the fact that the static susceptibility diverges as the system approaches a Pomeranchuk instability:
\begin{equation}
  \chi^\e_{\text{qp},0} (q, 0) =
  \frac{ \nu_F }{1 +F^\e_0}.
\end{equation}
As $ F^\e_0$ increases above $-1$, i.e., $|F^\e_0|$ gets smaller, the frequency of the relaxational mode in (\ref{ac1}) increases in magnitude. It reaches
 $s_{i,0} = \infty$
at  $F^\e_0 =-1/2$. At this value of $F^\e_0$, the mode bifurcates into two ($s_{i,0} \to \pm {\bar s}_{p,0}$), and each new mode moves from imaginary to almost real $s$ along infinite quarter-circles in the complex $s$ plane.

\begin{figure*}
  \centering
  \begin{subfigure}{0.56\hsize}
    \fbox{\includegraphics[width=\hsize]{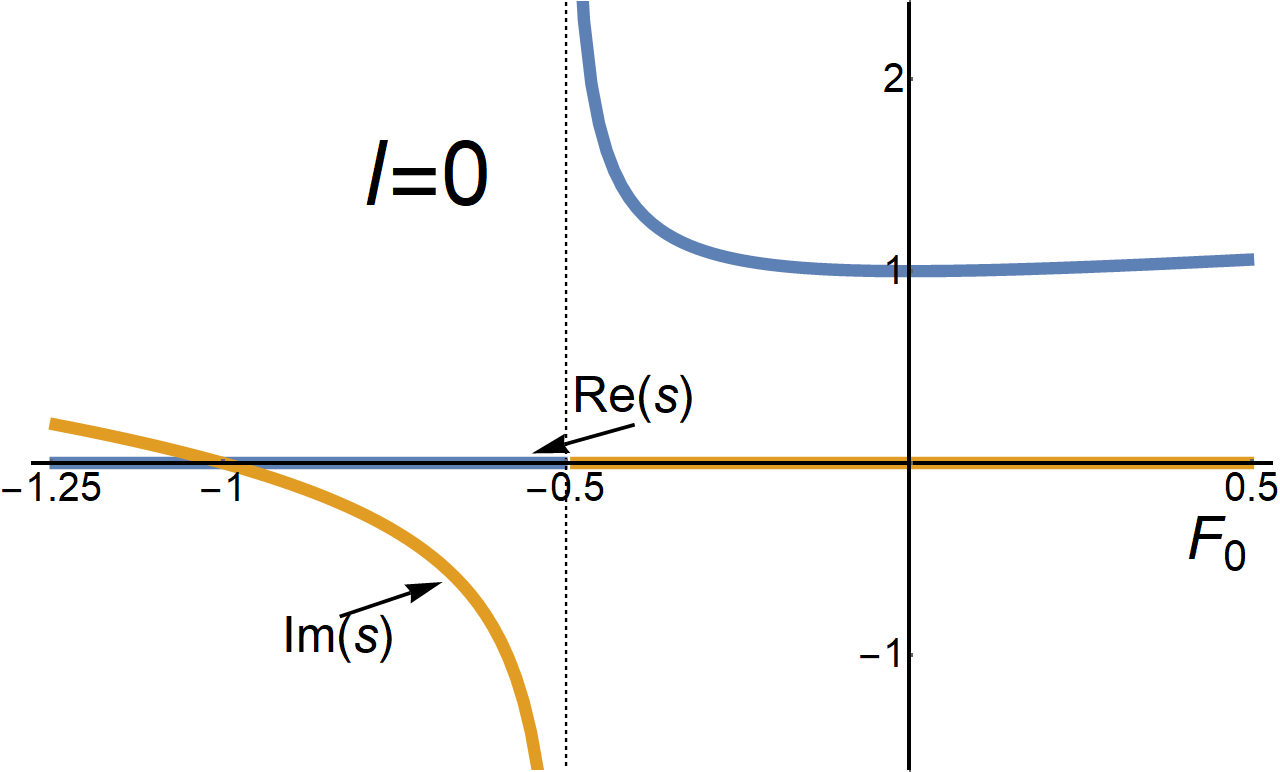}}
    \caption{\label{fig:l0-plot}}
  \end{subfigure}\hfill
  \begin{subfigure}{0.33\hsize}
    \fbox{\includegraphics[width=\hsize]{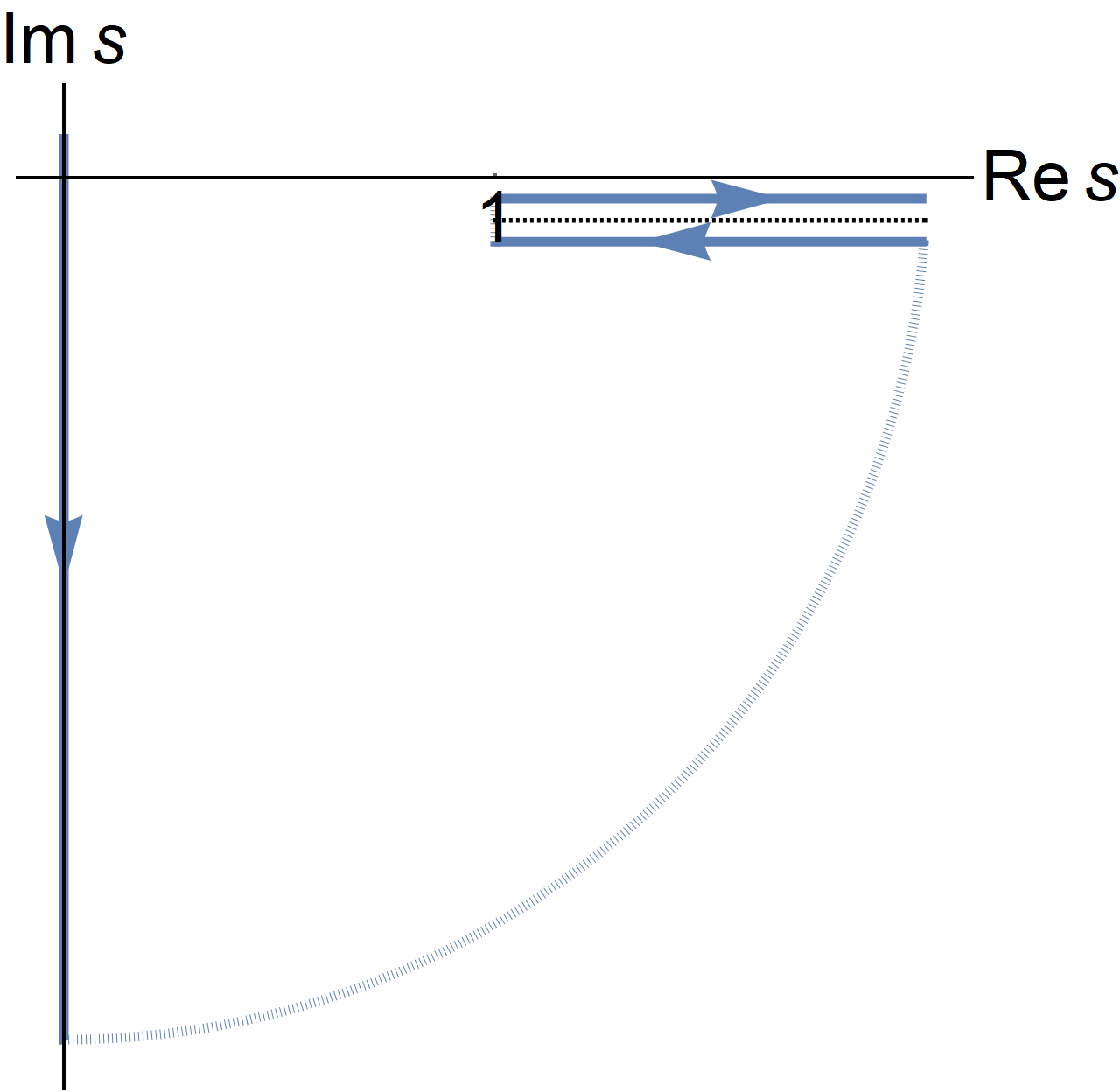}}
    \caption{\label{fig:l0-path}}
  \end{subfigure}
  \caption{
  (color online) Evolution of the poles of the dynamical susceptibility in the $l=0$ channel. (a) The real (blue) and imaginary (yellow) parts of the pole of the quasiparticle susceptibility $\chi^\e_{
  \text{qp},0}(s)$ as a function of the Landau parameter $F_0^\e$. For clarity, we show only one pole (for $F_0^\e > -1/2$ there are two poles with real parts of opposite signs) (b)
 The path followed by the pole in the complex plane with increasing $F_0^\e$.
 For $F_0^\e<-1$ the pole is purely imaginary and above the real axis, which indicates that the FL state is unstable.
 With increasing $F_0^\e$, the pole moves down along the imaginary axis,
 which corresponds to an overdamped zero-sound mode, and reaches $-i\infty$ at $F_0^\e = -1/2$.
 It then  ``jumps'' to the \emph{lower} edge of the branch cut.
 A pole located at the lower edge of the branch cut corresponds to a `silent'' zero-sound mode,
 which cannot be detected in measurements of $\chi^\e_0 (s)$ for real $s$ (i.e., real frequencies).
    At $F_0^\e = 0$ the pole moves to the upper edge of the branch cut,
  where it becomes a well-defined zero-sound mode,
   detectable by spectroscopic methods. } \label{fig:l0-plots}
\end{figure*}

For $ -1/2<F^\e_0<0$ the mode frequency is given by $s = \pm {\bar s}_{p,0} - i {\bar \delta}_{0}$, where
\begin{equation}
  {\bar s}_{p,0} = \frac{1-|F^\e_0|}{\sqrt{1 - 2 |F^\e_0|}},~~{\bar \delta}_0 = \delta \frac{1-|F^\e_0|}{1 - 2 |F^\e_0|}
\label{naa_3}
\end{equation}
The real part varies from ${\bar s}_{p,0} =\infty$  at  $F^\e_0
=-1/2 +0$ to ${\bar s}_{p,0} = 1$ at $F^\e_0 =0$. The pole positions are similar to those for positive $F^\e_0$ (see Eq. (\ref{eq:pole-positive-F0})); however, now ${\bar \delta}_0 \geq \delta$, i.e., the poles are located {\em below} the branch cut at $\text{Im}s = -
\delta$. At vanishingly small $\delta$, which we consider here, the poles are glued to the lower edge of the branch cut immediately below the real axis. The evolution of the real and imaginary parts of the poles with $F^\e_0$  is shown in Fig. \ref{fig:l0-plots}.

The existence of the poles glued to the lower edge of the branch cut is a tricky phenomenon. At first glance, they describe undamped collective excitations with velocity larger than the Fermi velocity ( note that ${\bar s}_{p,0}>1$ in Eq.~(\ref{naa_3})). Indeed, the susceptibility near the poles is
\begin{flalign}
  \chi^\e_{\text{qp},0} (s) \propto \frac{|F^\e_0|}{(1-2|F^\e_0|)^{3/2}} \left(\frac{1}{s + {\bar s}_{p,0} +i{\bar \delta}_0}\nonumber\qquad \right.\\
\left.    - \frac{1}{s-{\bar s}_{p,0} + i {\bar\delta}_0}\right).
 \label{naa_2}
\end{flalign}
This form is very similar to that in Eq. (\ref{naa_1})  for positive $F^\e_0$. However, Eq.~(\ref{naa_2})  is valid only for complex $s$ below the lower edge of the branch cut at $|s|>1$, and {\it cannot} be extended to real $s$.
More precisely, Eq.~\eqref{naa_2} cannot be extended to the real axis on the physical sheet of the Riemann surface, which we recall is the sheet for which $\chi_{\mbox{qp},0}^\e(s)$ is analytic in the upper half-plane. Instead, it can be extended to the real axis of the unphysical sheet, the one for which $\sqrt{1 - (s + i \delta)^2}  = i \sqrt{(s + i \delta)^2-1}$.
This means that the pole below the branch cut has no effect on the behavior of $\text{Im}\chi^\e_{\text{qp},0} (s)$ on the real axis, the imaginary part of the susceptibility for real $s$,
\begin{eqnarray}
\text{Im}\chi^\e_
{\text{qp},0}(s)=\frac{\nu_F\chi''_{\text{free},0}(s)}{\left(1+F^\e_0 \chi'_{\text{free},0}(s)\right)^2+\left(F^\e_0 \chi''_{\text{free},0}(s)\right)^2}\nn                                                                                                                 \\\label{imchi}
\end{eqnarray}
with $\chi'(s)\equiv \text{Re}\chi^\e_{\text{free},0}(s)$ and $\chi''(s)\equiv \text{Im}\chi^\e_{\text{free},0}(s)$, has no peak above the continuum.
Therefore, the modes for $-1/2<F^\e_0<0$  are ``silent'', in a sense that they cannot be detected by a spectroscopic measurement which probes $\text{Im}\chi^\e_{\text{qp},0}(s)$.

\subsection{$l=1$}
\label{sec:l=1}

For $l\geq 1$ we have to distinguish between the longitudinal susceptibility with the form-factor $\sqrt{2} \cos \theta$ and the transverse susceptibility with the form-factor $\sqrt{2} \sin \theta$. We consider the two cases separately. Here and in what follows, we will suppress the $\e$ superscript in the longitudinal and transverse susceptibilities for brevity, i.e., we will relabel $\chi^{\e,\text{long}}_l
  \to
   \chi^{\text{long}}_l$ and $\chi^{\e,\text{tr}}_l \to \chi^{\text{tr}}_l$.

\subsubsection{$l=1$, longitudinal channel}
\label{sec:l=1_a}

The computation of the free-fermion susceptibility with $\sqrt{2} \cos{\theta}$ formfactors at the vertices is quite straightforward. In notations of the previous section, the retarded susceptibility is given by
\begin{eqnarray}
 \chi^{\text{long}}_{\text{free},1} (
 s)
 =1 +  2 s^2 \left(1+\frac{is}{\sqrt{1-(s+i\delta)^2}}\right).\label{ret1}
\end{eqnarray}
For real  $|s|>1$ Eq.~(\ref{ret1})
reduces to
\begin{equation}
 \chi^{\text{long}}_{\text{free},1}(s) =  1 +  2s^2\left(1- \frac{|s|}{\sqrt{s^2 - 1}}\right).
\end{equation}
Substituting this form into Eq. (\ref{new_2}), we obtain an equation for the poles:
\begin{equation}
  \frac{1 + F^\e_1}{F^\e_1} =-2 s^2 + 2 s^2 \frac{|s|}
  {\sqrt{s^2-1}}.
  \label{n_1}
\end{equation}
A solution of Eq.~(\ref{n_1})
in the form
$s_{1,2} = \pm s_{p,1}$ with $ s_{p,1} >1$, i.e., outside the continuum, exists only for $F^\e_1 >0$. For small $ F^\e_1$,  $s_{p,1} =
1 + 2 (F^\e_1)^2$.  As $F^\e_1$ increases, the magnitude of $s_{p,1}$ also increases, and at large $F^\e_1$ becomes $s_{p,1} \approx (3 F^\e_1/4)^{1/2}$. Correspondingly, $\text{Im}\chi_{\text{qp},1}^{\text{long}}(s)$ has peaks on the real axis at $s=\pm s_{p,1}$.

To find the actual position of the poles in the complex plane, we will again need to treat $\delta$ as a finite albeit small quantity. As for the $l=0$ case, we associate $\delta$ with weak impurity scattering. Because a generic $l=1$ order parameter is not a conserved quantity, vertex corrections are not crucial\footnote{In the clean case, $\chi_l (q=0, \omega)$ for an order parameter with $l\neq0$ is finite because the self-energy and vertex corrections due to electron-electron interaction do not cancel each other~\cite{Klein2018}.  For non-interacting fermions, but in the presence of disorder,
  $\chi_{l\neq 0} (q=0, \omega)$ is finite because the self-energy and vertex corrections due to impurity scattering do not cancel each other.}.
Nevertheless, they are necessary to correctly determine the location of the poles.

The expression for $\chi^{\text{long}}_{\text{free},1} (s)$ in the presence of impurity scattering will be derived in Sec.~\ref{sec:gamma}. Here, we just borrow the result:
\begin{equation}
  \chi^{\text{long}}_{\text{free},1} (s) =1 +  2
  s^2 \frac{1  + i \frac{(s+ i \delta)}{\sqrt{1-(s+i\delta)^2}}}{1-\frac{\delta}{\sqrt{1-(s+i\delta)^2}}}.\label{ret1a}
\end{equation}
The equation for the poles becomes
\begin{equation}
  -\frac{1 + F^\e_1}{F^\e_1} =
  2s^2 \frac{\sqrt{1- (s+i\delta)^2} + i(s+ i \delta)}{
    \sqrt{1- (s+i\delta)^2} -\delta}.
  \label{n_2}
\end{equation}
If we assume that $s$ is in the lower half-plane above the branch cut, i.e., $ -\delta \leq \text{Im} s <0$ and $\sqrt{1 - (s+i \delta)^2} = -i \text{sgn} s \sqrt{(s+ i \delta)^2 -1}$, we find that the solution actually exists only for $0
\leq F_1
 ^\e
  \leq
  3/5$.  For these $F^\e_1$, the poles are located at $s_{1,2} = \pm s_{p,1}  - i {\tilde \delta}_1$, where $s_{p,1}$ is the solution of Eq. (\ref{n_1})  (which exists for all $F^\e_1 >0)$, and ${\tilde \delta}_1  = {\tilde Q}_1 \delta$, where
  \begin{equation}
    {\tilde Q}_1 = \frac {s^2_{p,1}}{2 - s^2_{p,1} + s_{p,1} \sqrt{s^2_{p,1}-1}}.
  \end{equation}
  For $0 < F^\e_1 < 3/5$, $s_{p,1}$ varies between $1$ and $2/\sqrt{3}$,
  and
  ${\tilde Q}_1 <1$, as we assumed. For $F^\e_1 =
  3/5
  $, we have $s_{p,1} = 2/\sqrt{3}$ and ${\tilde Q}_1=1$, i.e., the pole merges with the branch cut. For larger $F^\e_1$, we have  ${\tilde Q}_1 >1$, violating our assumption that the pole is above the branch cut, so that Eq.~(\ref{n_2}) has no solution.
A more careful analysis shows that the pole has moved to the unphysical Riemann sheet on which $\sqrt{1-(a + i b)^2}$  near the branch cut is defined as $\sqrt{1-(a + i b)^2} = i \sqrt{(a+ib)^2-1}$ instead of $\sqrt{1 - (s+i \delta)^2} = -i \text{sgn} s \sqrt{(s+ i \delta)^2 -1}$,  which we used to search for the poles on the physical Riemann sheet.

\begin{figure}
    \centering
    \includegraphics[width=\hsize]{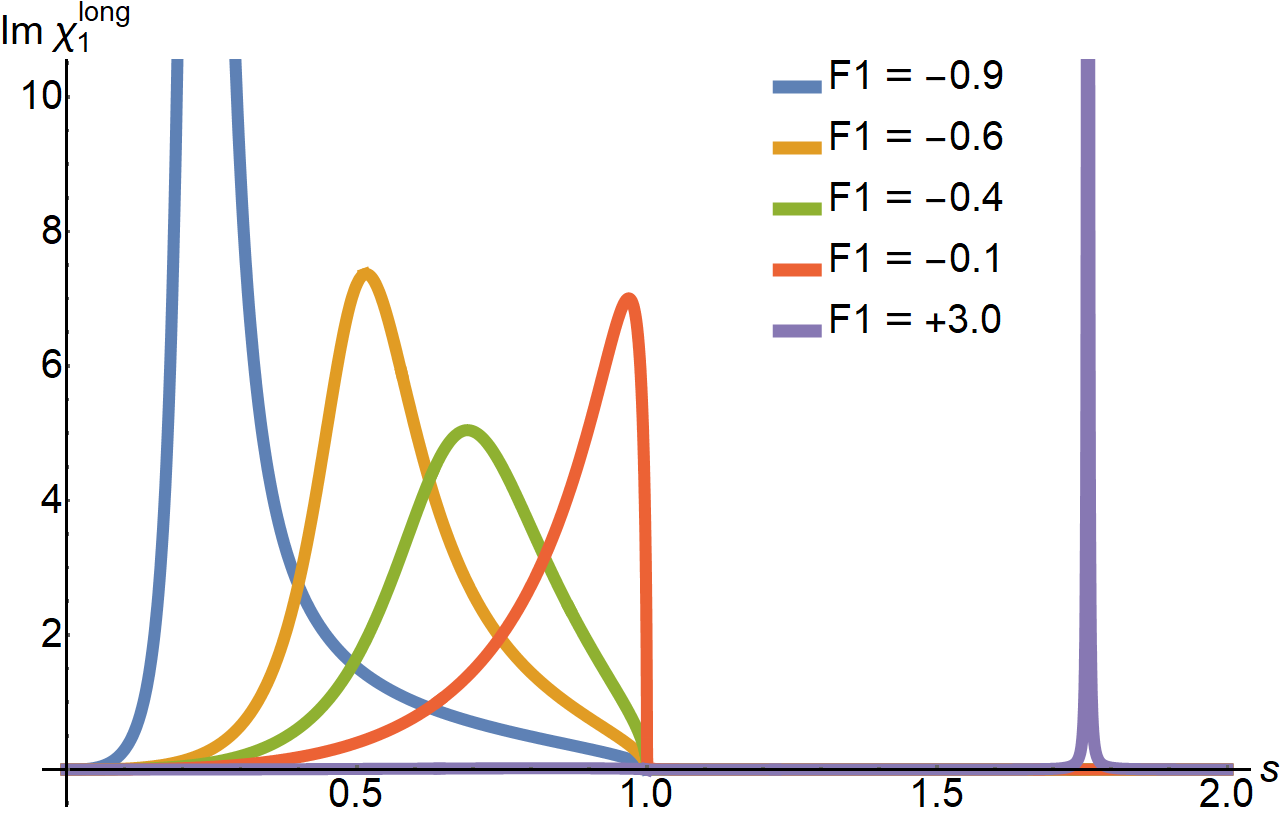}
    \caption{
      (color online)  $\text{Im}\chi_{\text{qp},1}^{\text{long}}(s)$ for different $F_1$. (A small impurity scattering $\delta = 10^{-3}$ was added to make the pole at positive $F_1$ visible.
    )
      \label{fig:im-chi1-clean}}
  \end{figure}

  The absence of the zero-sound pole for  $F^\e_1  \geq 3/5$ is a surprising result,
 but it has little effect on the form of $\chi_{\text{qp},1}^{\text{long}}(s)$ for real $s$. The latter has a conventional form for all positive $F^\e_1$:
  \begin{equation}
 \chi^{\text{long}}_\text{qp,
 1} (s) \propto \left(\frac{1}{s + s_{p,1} +i{\tilde \delta}_1} - \frac{1}{s-s_{p,1} + i {\tilde \delta}_1}\right),
 \label{naa_6}
\end{equation}
and $\text{Im}\chi^{\text{long}}_\text{qp,1} (s)$ has peaks at $s=\pm s_{p,1}$, as shown in Fig.~\ref{fig:im-chi1-clean}.

There also exists another solution for $F^\e_1 >0$, which is purely imaginary:
$s = -is_{i,1}$. Assuming that $s_{i,1}\gg
\delta$, we obtain an equation for
 $s_{i,1}$ from Eq.~(\ref{n_2}):
\begin{equation}
 \frac{1+ F^\e_1}{2 F^\e_1}=s^2_{i,1} \left(1 + \frac{s_{i,1}}{\sqrt{1+s^2_{i,1}}}\right).
\label{last_1}
\end{equation}
For small positive $F^\e_1$, $s_{i,1} \approx 1/
2\sqrt{F^\e_1}
\gg 1$. As $F^\e_1$ increases, $s_{i,1}$ decreases and eventually saturates at $s_{i,1}=1/\sqrt{3}$. This additional solution will be relevant for the case of finite damping, analyzed in Sec.~\ref{sec:gamma}. Note that Eq.~(\ref{last_1}) has a solution only for positive $s_{i,1}$, i.e., the pole is in the lower half-plane, as it should be.

For negative $F^\e_1$,  we again search for complex solutions in the form
 \begin{equation}
s_{1,2} = \pm a -i b,\;  b >0 \label{n_3}
\end{equation}
Right above the Pomeranchuk instability, i.e, at
$ F^\e_1 \approx -1$ but $ F^\e_1
>-1$, we find
\begin{equation}
 a= \left(\frac{1 + F^\e_1}{2}\right)^{1/2},~~b = \frac{1+F^\e_1}{4}.
 \label{F1-1}
\end{equation}
In contrast to the $l=0$ case, the collective modes are almost propagating because
$a
\gg b$. Below the Pomerachuk transition, i.e., for $ F^\e_1<-1$, both poles become purely imaginary and split away from each other along the imaginary $s$ axis:
\begin{equation}
s_{1,2} \approx \pm i  \left(\frac{|1+ F^\e_1|}{2}\right)^{1/2}.
\end{equation}
One of these poles is now in the upper frequency half-plane, i.e., a perturbation with the structure of the longitudinal $l=1$ order parameter grows exponentially (see Sec. \ref{sec:gamma}).  This indicates a Pomeranchuk instability.

\begin{figure*}
   \centering
   \begin{minipage}{0.3\hsize}
     \centering
     \fbox{\includegraphics[height=0.6\hsize,clip,trim=0 0 0 -0]{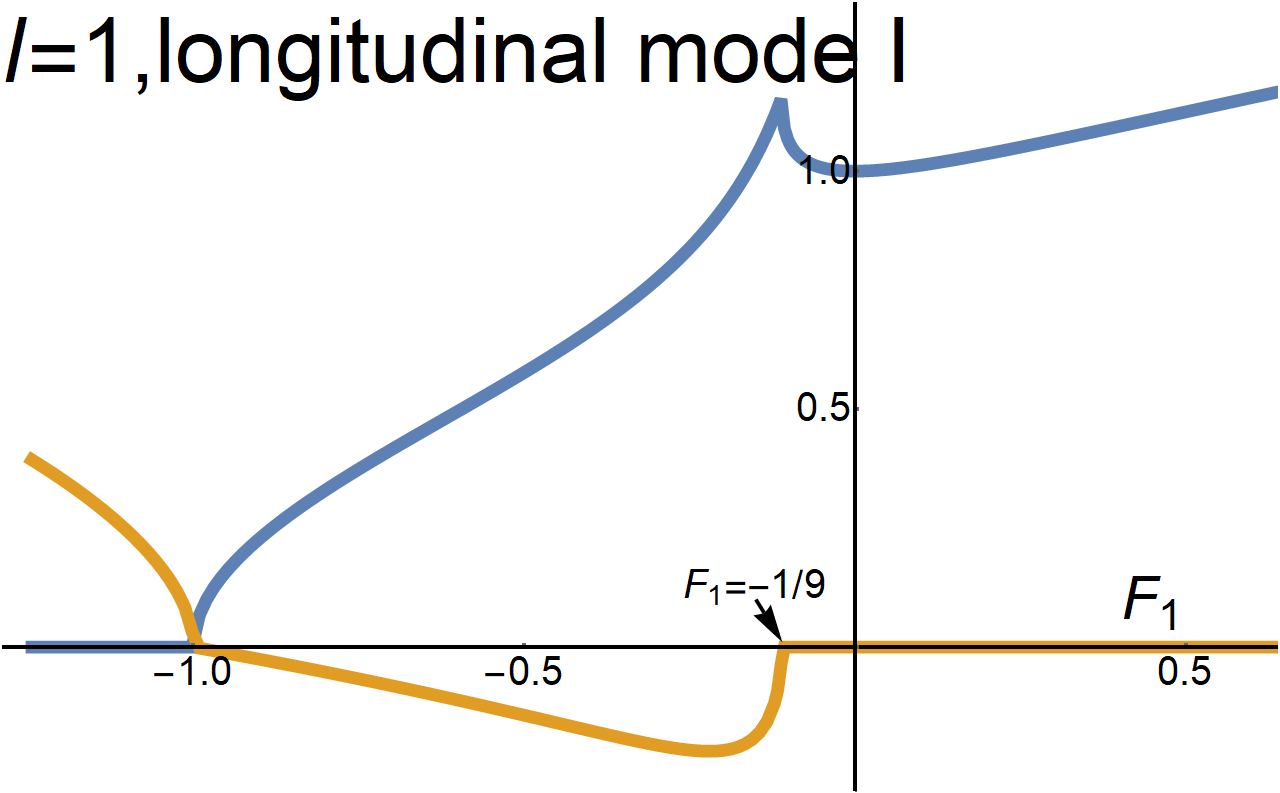}}
     \fbox{\includegraphics[height=0.45\hsize]{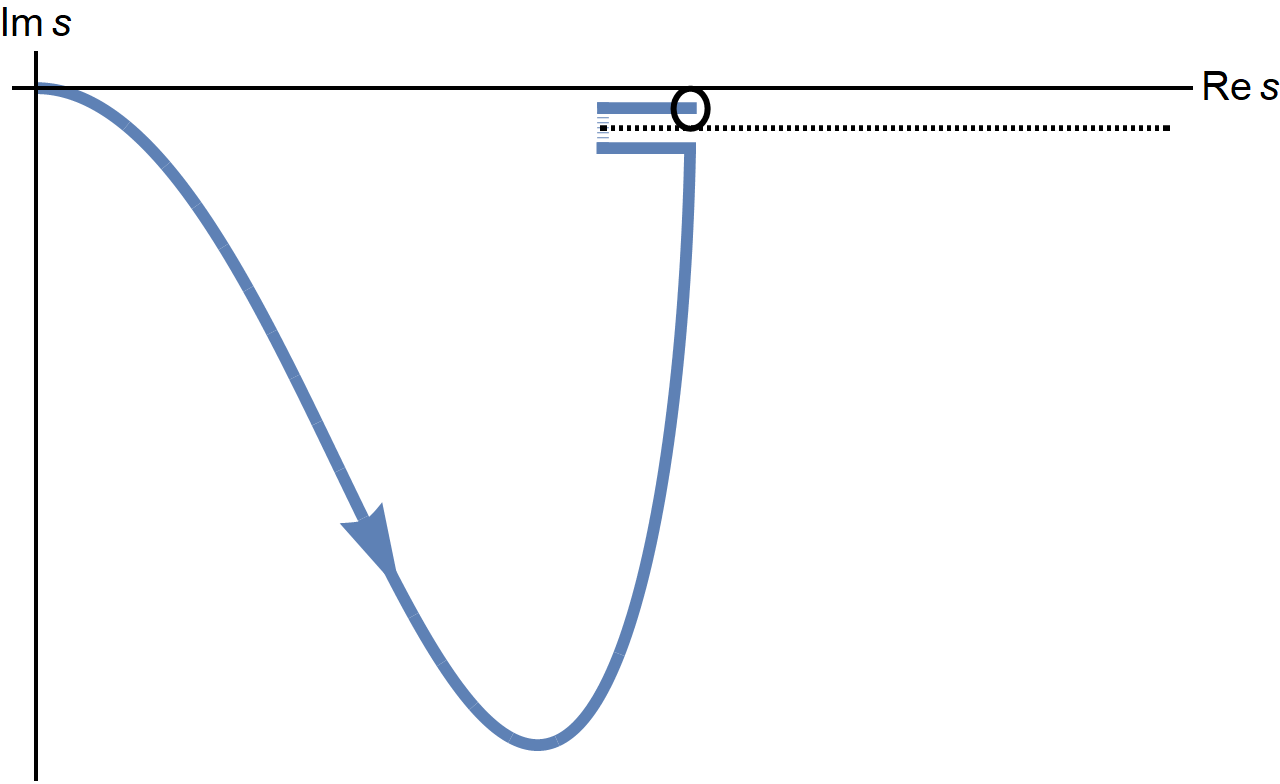}}
   \end{minipage}\hfill
   \begin{minipage}{0.3\hsize}
     \centering
     \fbox{\includegraphics[height=0.6\hsize,clip,trim=0 0 0 -0]{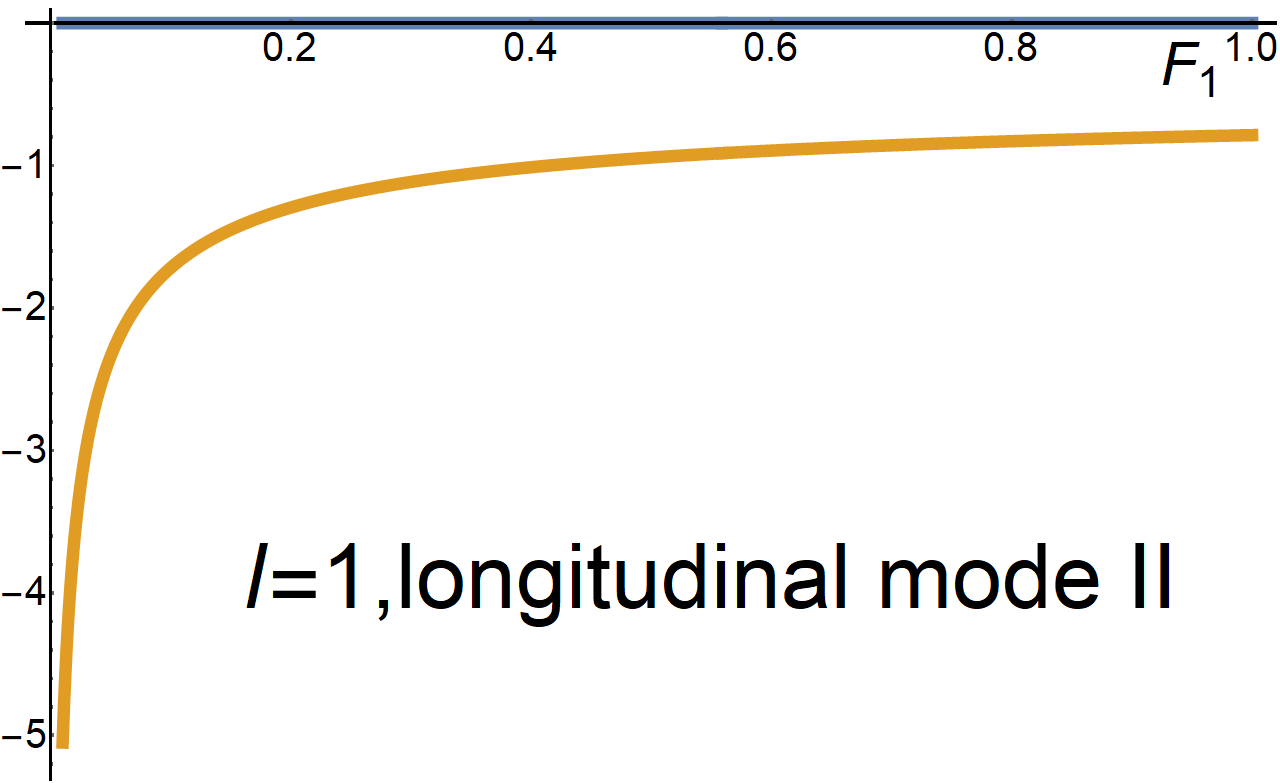}}
     \fbox{\includegraphics[height=0.45\hsize]{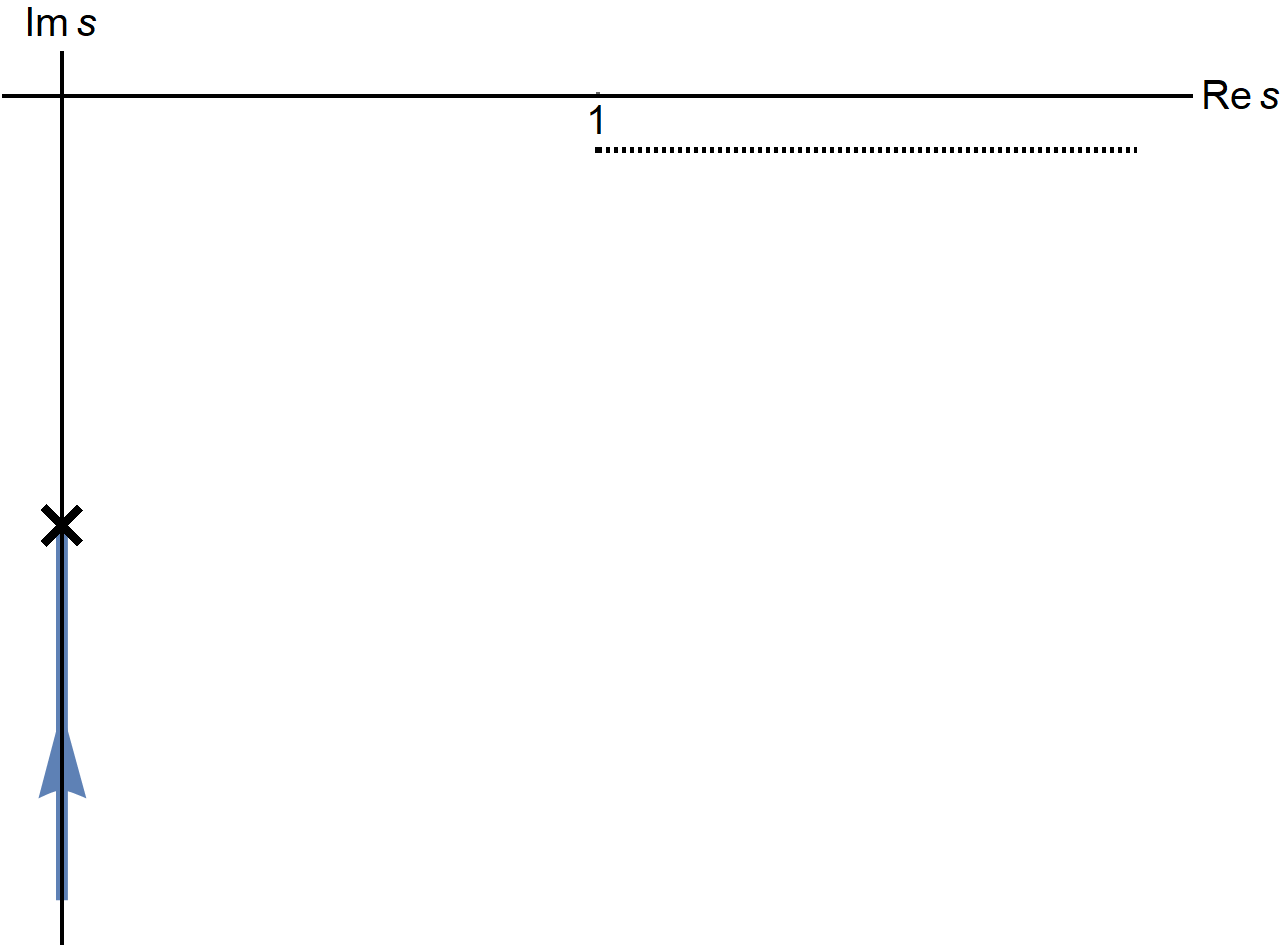}}
   \end{minipage}\hfill
   \begin{minipage}{0.3\hsize}
     \centering
     \fbox{\includegraphics[height=0.6\hsize,clip,trim=0 0 0 -0]{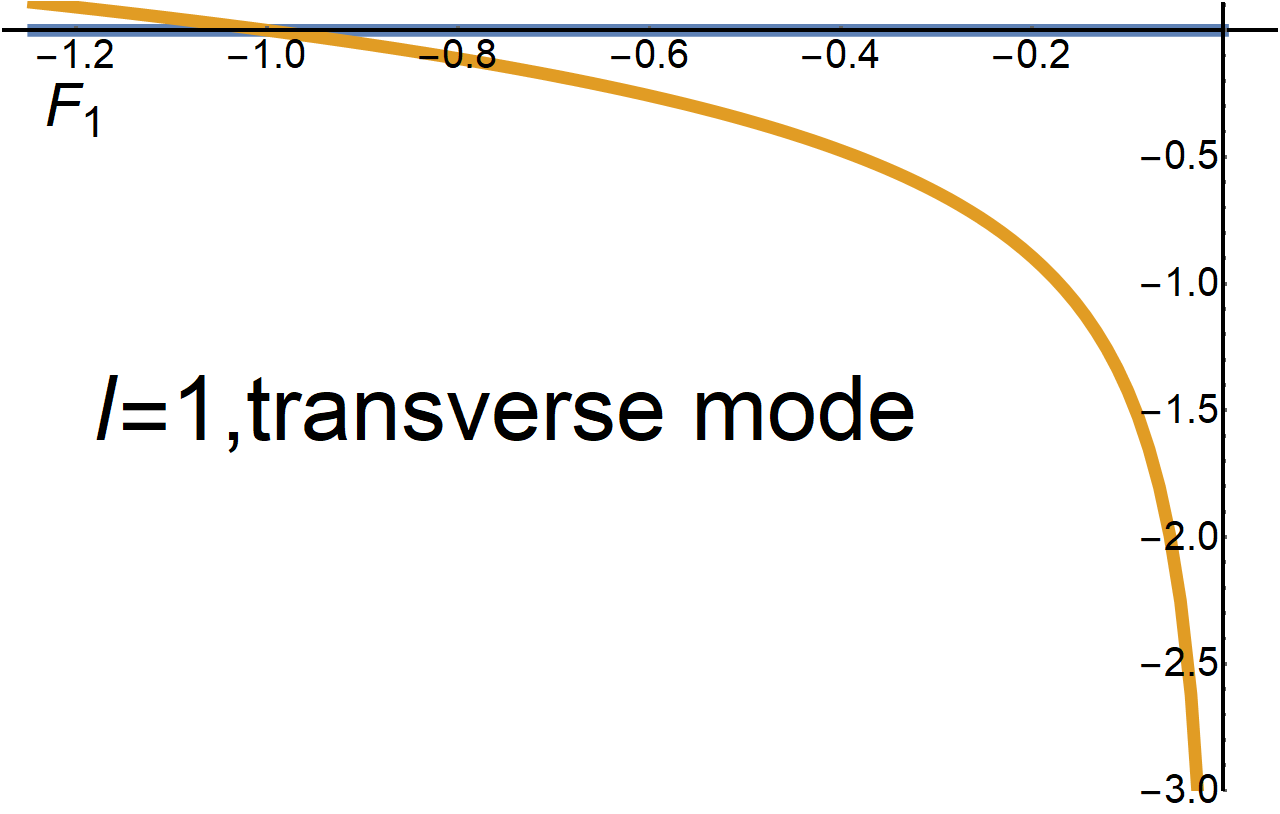}}
     \fbox{\includegraphics[height=0.45\hsize]{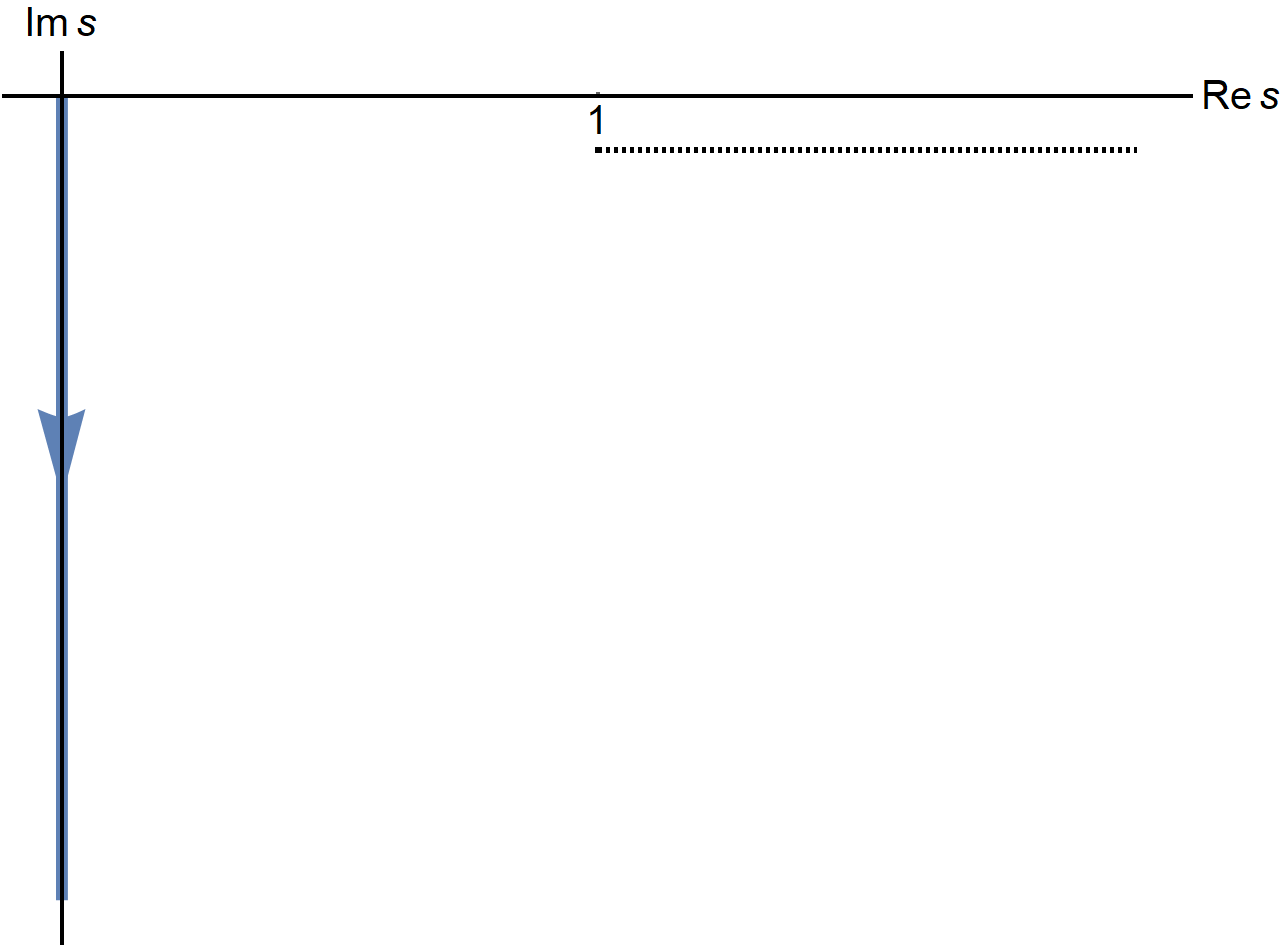}}
   \end{minipage}
   \caption{(color online) The poles of $\chi_{\text{qp},1}^\e(s)$ in the longitudinal and transverse channels. The use of colors and notations is the same as in Fig. \ref{fig:l0-plots}.
     In addition, the circle in the bottom left panel denotes where a pole of the $l=1$ longitudinal mode moves to the unphysical Riemann sheet. See Sec.~\ref{sec:l=1} for a detailed discussion.
   \label{fig:l1-plots}}
 \end{figure*}

In the interval $-1\leq F_1^{c(s)} \leq F^\e_{1,\text{cr}}$, where $F^\e_{1,\text{cr}} \equiv-1/9$, we find $s_{1,2} = \pm a_1 -i b_1$, where
 \begin{align}
  a_1&=
       \frac{1 + \sqrt{|F^\e_1|}}{4\sqrt{|F^\e_1|}} \left[\left(1 -  \sqrt{|F^\e_1|}\right)\left(1 + 3  \sqrt{|F^\e_1|}\right)\right]^{1/2},\nonumber\\
   b_1 &=
 \frac{1 -\sqrt{|F^\e_1|}}{4\sqrt{|F^\e_1|}} \left[\left(1 + \sqrt{|F^\e_1|}\right)\left(3  \sqrt{|F^\e_1|}-1\right)\right]^{1/2}.
\label{na_2}
 \end{align}
 As $|F_1^{c(s)}|$ decreases, $a_1$ monotonically increases, while $b_1$ first increases and then changes trend and start decreasing (see Fig. \ref{fig:l1-plots}). The poles reach the lower edges of the branch cuts at
 $F^\e_{1,\text{cr}}$. At this critical value of $F^\e_1$, $a_{\text{cr},1} = 2/\sqrt{3} >1$ and $b=0$ (up to a term of order $\delta$). For $|F^\e_1|$ slightly below $F^\e_{1,\text{cr}}$,
 $a_1$ and $b_1$ are approximately given by
 \begin{eqnarray}
 b_1 && = \sqrt{\frac{3}{2}} \left(|F^\e_1| - |F^\e_{1,\text{cr}}|\right)^{1/2}, \nonumber \\
 a_1 && = \frac{2}{\sqrt{3}} - \frac{243 \sqrt{3}}{2} \left(|F^\e_1| - |F^\e_{1,\text{cr}}|\right) \nonumber \\
     && = \frac{2}{\sqrt{3}} - O(b^2_1).
\end{eqnarray}
When $ F^\e_1
$ approaches  $ F^\e_{1,\text{cr}}
 $, the poles approach the real axis along the paths that are almost normal to it.

The existence of the solution with $a_1 >1$ but finite $b_1$ for $ F^\e_1
\leq F^\e_{1,\text{cr}}
$  is at first glance questionable, because conventional wisdom suggests that a mode with $\text{Re}s >1$ is located outside the particle-hole continuum and thus should be purely propagating. However,
as for the $l=0$ case,
these poles are located below the branch cuts,
 cannot be accessed from the real axis and do not lead to a peak in $\text{Im}\chi^{\text{long}}_1(s)$ for real $s$.

For $F^\e_{1,\text{cr}}<F^\e_1<0$, the poles are located at $s_{1,2} = \pm {\bar s}_{p,1} - i
{\bar \delta}_1$, where
${\bar s}_{p,1}$ is determined from
\begin{equation}
{\bar s}^2_{p,1} \left(1 + \frac{{\bar s}_{p,1}}{\sqrt{{\bar s}^2_{p,1} -1}}\right) = \frac{1- |F^\e_1|}{|F^\e_1|}
\end{equation}
and
${\bar \delta}_1  = {\bar Q}_1 \delta$ with
 \begin{eqnarray}
 {\bar Q}_1 = \frac{{\bar s}^2_{p,1}}
  {2 - {\bar s}^2_{p,1} - {\bar s}_{p,1} \sqrt{{\bar s}^2_{p,1}-1}}.\label{ors_1}
 \end{eqnarray}
 The magnitude of ${\bar s}_{p,1}$ varies between ${\bar s}_{p,1}= 2/\sqrt{3}$ at $F^\e_1 = F^\e_{1,\text{cr}}$ and  ${\bar s}_{p,1} =
   1 + 2 (F^\e_1)^2
   $ for  $-F^\e_1
   \ll 1$, , i.e., at vanishing $F^\e_1$ the poles approach the end points of the branch cuts. As follows from Eq.~(\ref{ors_1}), $ {\bar Q}_1 \geq 1$ for ${\bar s}_{p,1}$ in this interval, hence  ${\tilde \delta}_1 \geq \delta$, i.e., the poles are located below the lower edges of the cuts, as expected. This is very similar to what we found in the $l=0$ case for $ -1/2<F^\e_0
     <0 $. Like in that case, the $l=1$ susceptibility for $ F^\e_{1,\text{cr}} < F^\e_1 <0$  has poles at $s = \pm {\bar s}_{p,1} - i {\bar \delta}_1$:
      \begin{equation}
 \chi^{
 \text{long}}_{\text{qp},1} (s) \propto \left(\frac{1}{s + {\bar s}_{p,1} +i{\bar \delta}_1} - \frac{1}{s-{\bar s}_{p,1} + i {\bar \delta}_1}\right).
 \label{naa_4}
 \end{equation}
However, Eq.~(\ref{naa_4}) is again only valid for complex $s$ in the lower half-plane below the branch cut, and cannot be extended to real $s$. These poles correspond to silent modes, and the susceptibility does not have peaks above the particle-hole continuum. We plot $\text{Im}\chi^{
   \text{long}}_{\text{qp},1} (s)$ for real $s$ in Fig.~\ref{fig:im-chi1-clean}.
 The evolution of the real and imaginary parts of the poles with $F^\e_1$  is shown in Fig.~\ref{fig:l1-plots}.

  \subsubsection{$l=1$, transverse channel}

We next consider the transverse quasiparticle susceptibility in the $l=1$ channel. The retarded susceptibility of free fermions with the
    $\sqrt{2} \sin{\theta}$ form-factors at the vertices is
\begin{eqnarray}
  \chi^{\text{tr}}_{\text{free},1}
  (s)
=1 -2s^2  + 2 i   \left(1- s^2\right)    \frac{s}
  {\sqrt{1 - (s+i\delta)^2}}.\nn                                                                                                                                                                                                                                            \\
\label{ssss}
\end{eqnarray}
For real  $|s|>1$ Eq.~(\ref{ssss}) is reduced to
\begin{equation}
 \chi^{\text{
 tr}}_{\text{free},1}(s) =  1 - 2s^2 + 2 |s| \sqrt{s^2-1}.
\end{equation}
Substituting this form into Eq. (\ref{new_2}), we find that the positions of the poles on the real frequency axis and outside the particle-hole continuum are determined by
\begin{equation}
\frac{1+F^\e_1}{F^\e_1} = 2s^2 - 2|s| \sqrt{s^2-1}.
\label{n_1_a}
\end{equation}
In contrast to the longitudinal case, the solutions of this equation
 $s_{1,2} = \pm s_{p,1}$ exist not for any positive $F^\e_1
 $ but only for
$F^\e_1
\geq 1$ (Ref.~\onlinecite{Sodemann2018}). Slightly above the threshold,
 $s_{p,1} =
 1 + (F^\e_1-1)^2/8
 $. For large positive $F^\e_1$,
  $s_{p,1} \approx  \sqrt{F^\e_1}/2$.

To obtain the solutions in the complex plane $s$, we introduce impurity scattering in the same way as in the previous cases. Equation (\ref{ssss}) is then replaced by
\begin{widetext}
\begin{equation}
  \chi^{\text{tr}}_{\text{free},1} (s) =
   1 -2s \left(s+i \delta  -
    i \sqrt{1 - (s+i\delta)^2} \right).
\label{ssss1}
\end{equation}
\end{widetext}
There are no additional terms due to vertex corrections because the form-factor is an odd function of the angle $\theta$ and thus vertex corrections vanish upon angular integration.

Substituting Eq.~(\ref{ssss1}) into Eq.~(\ref{new_2}), we find that for $F^\e_
1
>1$,  where Eq.~(\ref{n_1_a}) has a solution for real $s$, there is actually no solution for the pole of $ \chi^{\text{tr}}_{\text{qp},1}(s)$  in the complex plane of $s$, above the branch cut.
Still, for real $s$,   $ \text{Im} \chi^{\text{tr}}_{\text{qp},1}(s)$ displays sharp peaks even for $F_1^\e > 1$.

For  $0 < F^\e_1 <1$ we assume that $s$ is below the branch cut and re-write the square root in Eq.~(\ref{ssss1}) as $\sqrt{1- (s+i \delta)^2} = i \text{sgn} s \sqrt{(s+i \delta)^2 -1}$. If we just neglect $\delta$ after that, we find another propagating mode, located at $s_{1,2} = \pm {\bar s}_{p,1}$, where ${\bar s}_{p,1} \geq 1$ is the solution of
 \begin{equation}
\frac{1+F^\e_1}{F^\e_1} = 2
{\bar s}_{p,1} \left(
{\bar s}_{p,1} + \sqrt{
(({\bar s}_{p,1})^2-1}\right).
\end{equation}
However, if $\delta$ is treated as a small but finite quantity, we find that there is no solution of $(\chi^{\text{tr}}_
{\text{qp},1}
(s))^{-1} =0$ with $|\text{Im}{\bar s}_{p,1}| > \delta$, i.e., there is no pole below the branch cut.  Combining this with the absence of the pole for $ F^\e_1>1$, we conclude that the $l=1$  transverse susceptibility does not have a pole on the \emph{physical} sheet for $F^\e_1>0$.
However, as was the case for the longitudinal mode, the poles do exist on the unphysical sheet.

For negative $F^\e_1$ the pole of $ \chi^{\text{tr}}_{\text{qp},1}(s)$ is on the imaginary axis:
   $s
   = - i s_{i,1}$. The value of $s_{i,1}$ is determined by
\begin{equation}
 \frac{1-|F^\e_1|}{|F^\e_1|} =
 2 (s_{i,1})^2 +2s_{i,1} \sqrt{1+ (s_{i,1})^2}.
\end{equation}
The solution exists for all negative $F^\e_1$. When $F_1^\e$ approaches zero from below,
$s_{i,1} \approx 1/
2|F^\e_1|
^{1/2}
$.  Near $1 + F^\e_1 =0$, we have $s_{i,1} \approx (1+F^\e_1)/2$, i.e., $s
= -i  (1+F^\e_1)/2$. As before, when $1 +F^\e_1$ changes sign and becomes negative, the pole moves from the lower to the upper frequency half-plane, i.e. an $l=1$  perturbation in the shape of the FS grows with time exponentially. This behavior is similar to the one for $l=0$. Yet, a purely relaxational collective mode in the
  $l=1$ transverse channel exists for all $ -1<F^\e_1
  <
  0$, i.e., it appears without a threshold.

\subsection{$l=2$}
\label{sec:l=2}

\subsubsection{
  $l=2$, longitudinal channel}

\begin{figure*}
  \centering
  \fbox{\includegraphics[width=0.22\hsize]{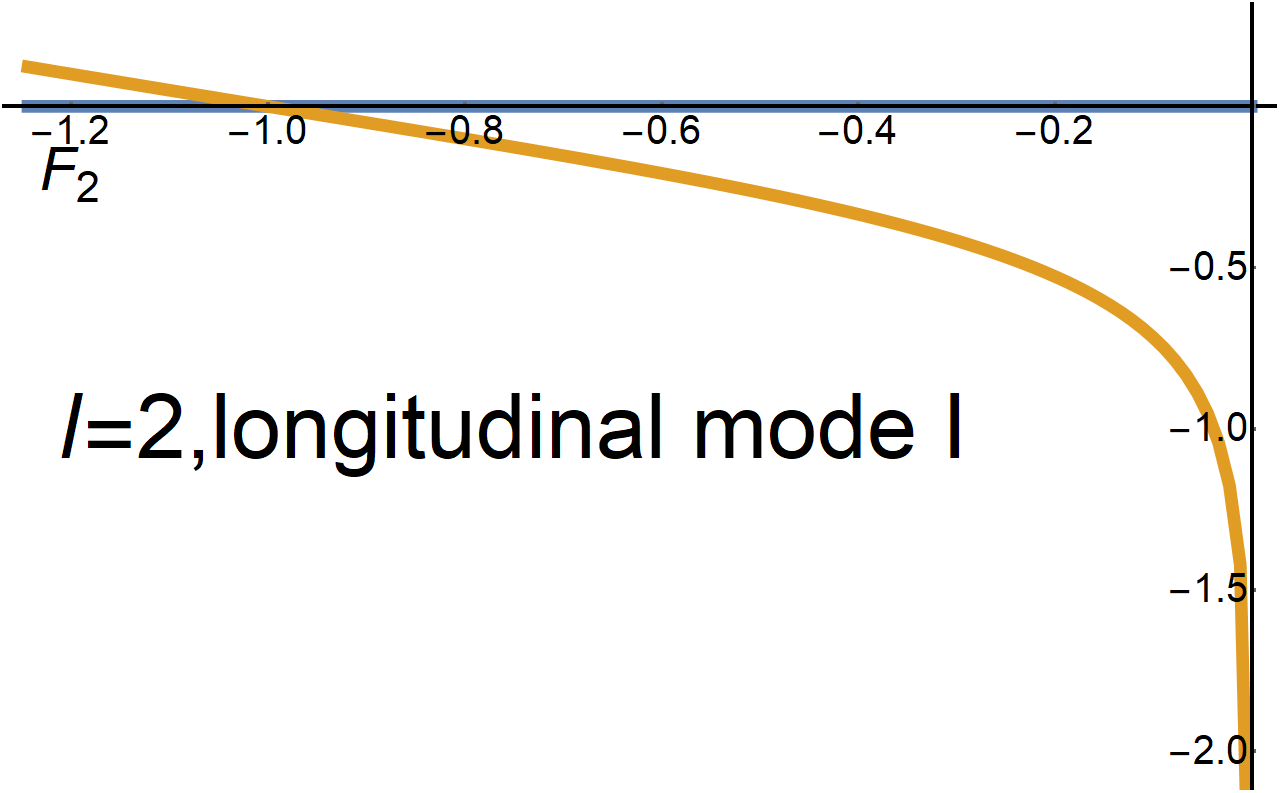}}
  \fbox{\includegraphics[width=0.22\hsize]{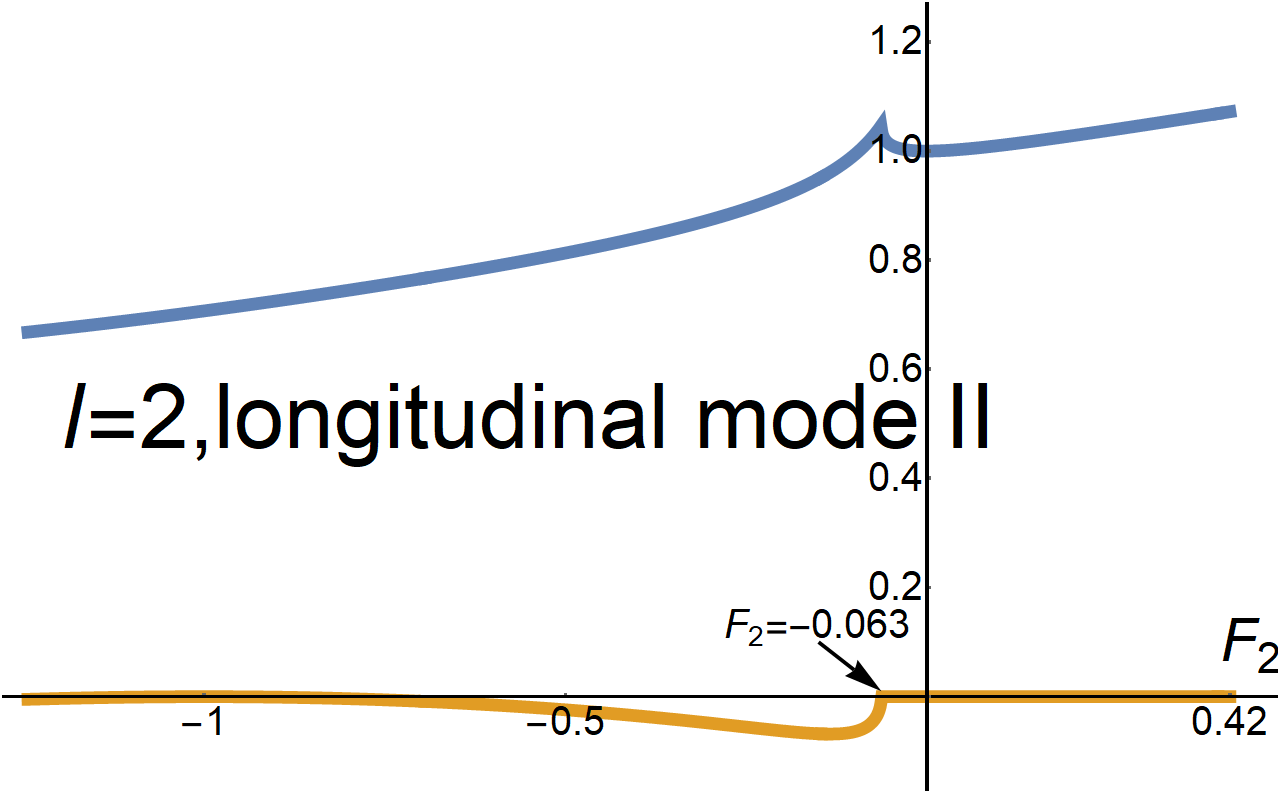}}
  \fbox{\includegraphics[width=0.22\hsize]{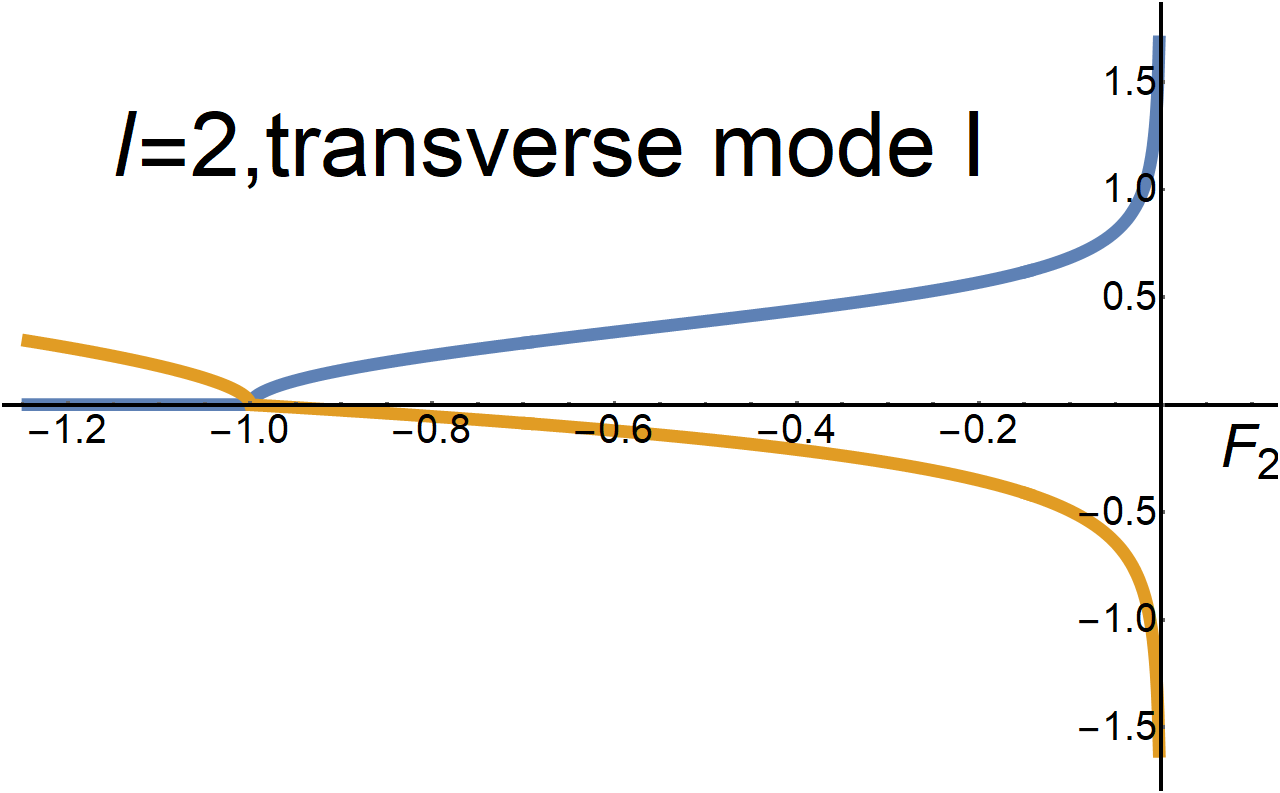}}
  \fbox{\includegraphics[width=0.22\hsize,clip,trim=0 0 0 -30]{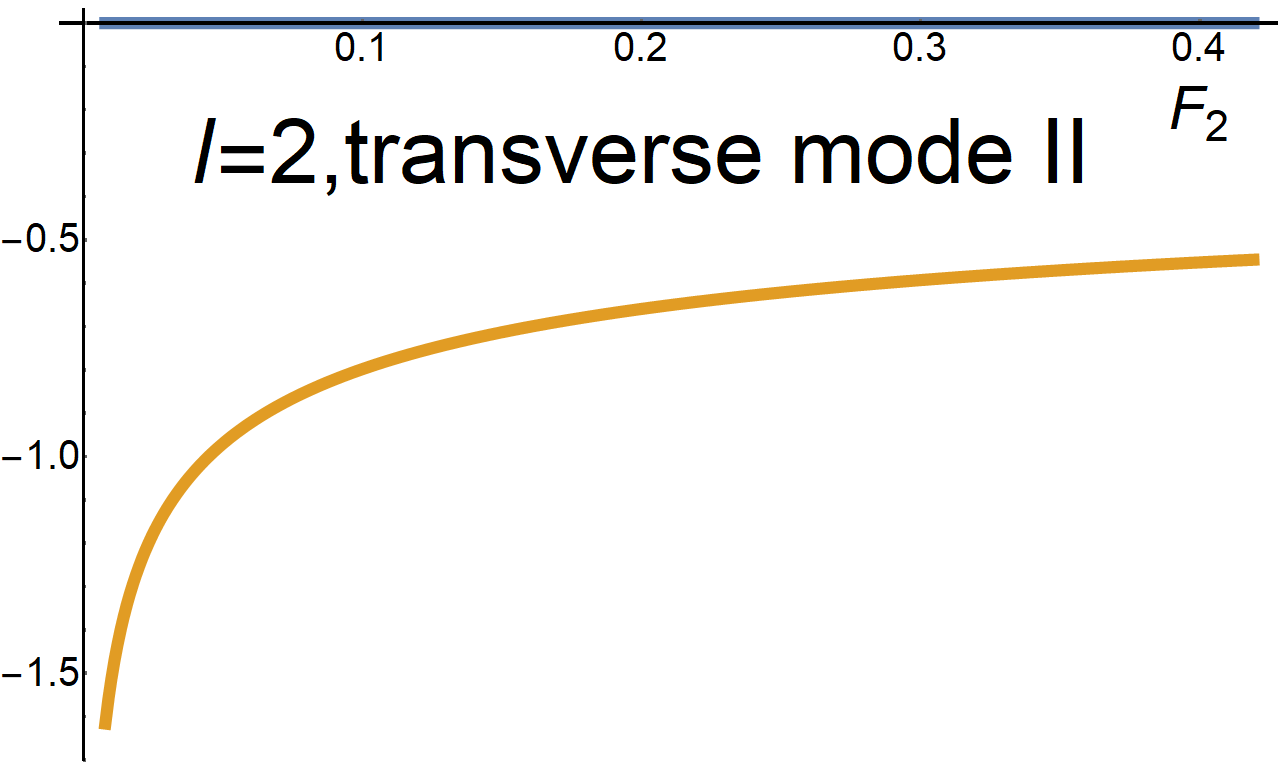}}
  \caption{(color online) The poles of $\chi_{\text{qp},2}^\e(s)$ in the complex plane. The use of color and notation is the same as in Fig. \ref{fig:l0-plots}. See Sec. \ref{sec:l=2} for a detailed discussion.
  }\label{fig:l2-plots}
\end{figure*}

The retarded free-fermion susceptibility with the
$\sqrt{2} \cos{2\theta}$ form-factors at the vertices is
  \begin{eqnarray}
    \chi^{\text{long}}_{\text{free},2}
    (s)
    =
    1 - 4s^2 + 8 s^4 + 2i \left(2s^2-1\right)
    ^2 \frac{s}{\sqrt{1-(s+i \delta)^2}}.\nn                                                                                                                                                                                                                                \\
    \label{l2long}
    \end{eqnarray}
The equation for the poles of
  $ \chi^{\text{long}}_{\text{qp},2} (s)$ outside the continuum, i.e., for $s$ real and
  $|s| >1$, now reads
\begin{equation}
  - \frac{1 + F^\e_2}{F^\e_2} = 8 s^4 -4 s^2 -2 (2 s^2-1)^2 \frac{|s|}
  {\sqrt{s^2-1}} =0 \label{n_6}.
\end{equation}
Similarly to the cases of
 $l=0$ and of the longitudinal channel for $l=1$, the propagating solutions
 $s_{1,2} = \pm s_{p,2}$ exist for all positive $F^\e_2$. For small $F^\e_2$, $s_{p,2} \approx  (1 + 2 (F^\e_2)^2)$;
 for large $F^\e_2$, $s_{p,2} \approx  (F^\e_2/2)^{1/2}$.

To obtain the solutions in the complex plane, we introduce impurity scattering in the same way as before. Combining the self-energy and vertex corrections, we obtain after some algebra
 \begin{widetext}
  \begin{equation}
    \chi^{\text{long}}_{\text{free},2}
    (s) =
    1 - 2i \frac{s}{\sqrt{1-(s+i \delta)
    ^2}-\delta} \left(s+ i \delta-i\sqrt{1-(s+i \delta)^2}\right)^2 \left(1 - 2 s (s + i \delta)\right).
    \label{l2long_1}
    \end{equation}
The equation for the pole becomes
  \begin{equation}
\frac{1 +F^\e_2}{2F^\e_2} =
 \frac{is}{\sqrt{1-(s+ i \delta)^2} -  \delta} \left(s + i \delta -i \sqrt{1-(s+ i \delta)^2}\right)
 ^2
  \left(1 -2 s (s + i \delta)\right)
\label{n_6_1}.
\end{equation}
\end{widetext}
Solving for the pole at small but finite $\delta$, we find $s_{1,2} = \pm s_{p,2} - i {\tilde \delta}_2$, where ${\tilde \delta}_2 = {\tilde Q}_2 \delta$. Evaluating ${\tilde Q}_2$, we find that it is smaller than $1$ for $F^\e_2 < 0.420$, when $s_{p,2} < 1.072$. For these $F^\e_2$, the pole is located above the branch cut, as it should be.  For larger $F^\e_2$ there are no poles near the real axis. This is similar to the behavior in the longitudinal channel for $l=1$. We re-iterate that the absence of a true pole in the complex plane does not affect the behavior of $\chi^{\text{long}}_{2} (s)$ for real $s$;
in particular, $\text{Im}\chi^{\text{long}}_{2} (s)$ still displays sharp peaks at $s= \pm s_{p,2}$.
In  mathematical terms, the pole moves to a different Riemann sheet at $F^\e_2 > 0.420$.

For negative $F^\e_2$ Eq.~(\ref{n_6_1}) has two solutions. One of them is purely imaginary: $s = -i s_{i,2}$. For $ F^\e_2 \approx -1$, $s_{i,2} =(1-|F^\e_2|)/2$; for small negative $F^\e_2$, $ s_{i,2} \approx 1/(2 |F^\e_2| ^{1/4})$.
Another solution does not become critical at the Pomeranchuk transition. To detect this mode, we notice that,
for $F^\e_2 =-1$,
 Eq.~(\ref{n_6_1})
is satisfied not only by $s=0$, but also by     $s_{1,2} = \pm 1/\sqrt{2}$. The latter solutions are on the real axis, but away from the branch cut. At small deviation from the critical value $F^\e_2 =-1$, these solutions evolve into $s_{1,2} =\pm a_2 - i b_2$, where
\begin{eqnarray}
  &   & a_2 = \frac{1}{\sqrt{2}}\left(1 +\frac{1 + F^\e_2}4\right) \nonumber\\
  &   & b_2 = \frac{\left(1 + F^\e_2\right)^2}{8\sqrt{2}}.
\end{eqnarray}
Observe that $b_2$ remains positive even when $1+ F^\e_2 <0$. As $ F^\e_2$ gets larger, the solutions first move away from the real axis but then reverse the trend and, at the threshold value $F^\e_{2,\text{cr}}= -0.0632$, reach the lower edge of the branch cut at $a_{2,\text{cr}} \approx \pm 1.046$. As   $F^\e_{2}$ is varied from $F^\e_{2,\text{cr}}$ to    $0$, the solutions
``slide'' along the lower edge of the branch cut towards $s=\pm 1$. This is very similar to what we found in the $l=0$ case, for
$ -1/2<  F^\e_0    < 0$, and in the    $l=1$ longitudinal channel, for    $ F^\e_{1,\text{cr}}    <    F^\e_1    <    0$. At a small but finite $\delta$, the two sliding solutions are $s_{1,2} = \pm {\bar s}_{p,2} - i{\bar \delta}_2$ with
${\bar \delta}_2 \geq \delta$, i.e., the pole does exist but is located below the branch cut. The evolution of the poles with $F^\e_2$ is shown in Fig. \ref{fig:l2-plots}.

\subsubsection{$l=2$, transverse channel}
The retarded free-fermion susceptibility with the $\sqrt{2} \sin {2\theta}$ formfactors at the vertices is given by
\begin{widetext}
  \begin{eqnarray}
    \chi^{\text{tr}}_{\text{free},2} (s)
    =
    1 + 4s^2 - 8 s^4
    -  \frac{8i s^3(s^2-1)}
    {\sqrt{1-(s+i \delta)^2}}.
    \label{l2long_2}
  \end{eqnarray}
\end{widetext}

The equation for the poles outside the continuum, i.e., for $s$ real and $|s| >1$, reads
\begin{equation}
  \frac{1 + F^\e_2}{F^\e_2} =
  8 s^4
  -
  4 s^2
  -
  8 s^2
  |s| \sqrt{s^2-1}.
   \label{n_6_a}
\end{equation}
For positive $F^\e_2$, the solutions $s_{1,2} = \pm s_{p,2} $  exist for $F^\e_2 >1/3$.  For $F^\e_2$ just slightly above $1/3$, $s_{p,2} \approx
1 + (81/128) (F^\e_2-1/3)^2
$. For large $F^\e_2$,  $s_{p,2} \approx  (F^\e_2/2)^{1/2}$.

To obtain the solutions in the complex plane $s$, we introduce impurity scattering in the same way as before. There are no additional terms due to vertex corrections because the form-factor is an odd function of the angle $\theta$. Equation (\ref{l2long_2}) is then replaced by
\begin{equation}
  \chi^{\text{tr}}_{\text{free},2} (s) = 1 - 4s (s+i \delta) \left( s+i\delta-i\sqrt{1-(s+i \delta)^2}\right)^2.
  \label{ssss1_n}
\end{equation}
For  $F^\e_1 >1/3$,  we assume that $s$ is above the branch cut and use $\sqrt{1- (s+i \delta)^2} = - i \text{sgn} s
\sqrt{(s+i \delta)^2 -1}$. If we neglect $\delta$ after that, we obtain the same solution $s_{p,2}$ as in (\ref{n_6_a}). For $0 <F^\e_2 <1/3$, the same procedure but with an assumption that the pole is below the branch cut, yields another propagating mode, which slides along the lower edge of the branch cut, much like it happens for the pole of the transverse susceptibility for $l=1$.  Once we take into account that $\delta$ is small but finite, we find that the solution along the real axis does not survive in either of the cases, i.e., there is no pole close to the real axis on the physical Riemann sheet. This is similar to the situation for the $l=1$ transverse channel.

As for the $l=1$ case,
there also exists another mode for $F^\e_2 >0$,  at $s = - i s_{i,2}$ on the imaginary axis. The value of $s_{i,2}$ is determined from
\begin{equation}
  \frac{1 + F^\e_2}{4 F^\e_2} =(s_{i,2})^2 \left( s_{i,2} + \sqrt{1 + (s_{i,2})^2} \right)^2.
\end{equation}
For small $F^\e_2 >0$, $s_{i,2} \approx 1/(2 F^\e_2)^{1/4}$; for large $F^\e_2$, $s_{i,2} \approx
1/2\sqrt{2}$.

For negative $F^\e_2$ there is no solution on either real or imaginary frequency axes, and we search for the solutions in the form
$s = \pm a_2 -ib_2$, where both $a
_2$ and $b
_2$ are finite. In this situation, one can safely neglect $\delta$ and write the equation for the poles as
\begin{equation}
\frac{1 -|F^\e_2|}{4 |F^\e_2|} = s^2 \left(1-
2
 s^2 +2 i s \sqrt{1-s^2}\right).  \label{n_6_1_a}
\end{equation}
An analysis of this equation shows that the solution exists for all $F^\e_2 <0$. Near $F^\e_2 =-1$,
\begin{equation}
a_2  \approx \left(\frac{1 - |F^\e_2|}{4}\right)^{1/2}, ~b_2 \approx   \frac{1 -|F^\e_2|}{4}.
\end{equation}
For small negative $F^\e_2$,$ a_2  \approx b_2 \approx 1/(2\sqrt{2}|F^\e_2|^{1/4})$. The evolution of this pole with $F^\e_2$
is shown in Fig. \ref{fig:l2-plots}.
\subsection{arbitrary $l$}
\label{sec:analysis-general-l}
\subsubsection{Equations for the poles}

We now focus in more detail on negative $F^\e_l$ and, in particular, on the behavior of collective modes near a Pomeranchuk instability. Comparing the results for for the $l=0,1,2$ modes, we see a difference between even and odd $l$. Namely, near a Pomeranchuk instability the critical mode in the longitudinal channel is purely imaginary for even  $l=0,2$ and almost real for odd $l=1$. For transverse channels the situation is the opposite -- the mode near a Pomeranchuk instability is purely imaginary for $l=1$ and almost real for $l=2$.   In this section we analyze whether this trend persists for other values of $l$.

The retarded longitudinal and transverse susceptibilities of free fermions can be obtained analytically for any $l$. We have
\begin{equation}
  \label{eq:chi-0-l}
  \chi^{\text{long},\text{tr}}_{\text{free},l>0}(s) =
  K_0
  \pm K_{2l}, ~~
\end{equation}
where
\begin{align}
  \label{eq:K-n-form}
  K_{2l}                                    & = - \int \cos{2l \theta} \frac{\cos\theta}{s+i\delta - \cos\theta}                                                                                                                                                            \\
                                            & = \delta_{l,0} + i\frac{s}{\sqrt{1-(s+i\delta)^2}}(s
                                            -
 i\sqrt{1-(s+i\delta)^2})^{2l}.
\end{align}
The equation for the pole on real frequency axis outside the continuum, i.e., for
$|s|>1$, is
\begin{equation}
  \label{eq:fl-pole-l}
  \frac{1+F^\e_l}{F^\e_l} =    \frac{|s|}{\sqrt{(s+i\delta)^2-1}}\left(1\pm (|s| - \sqrt{s^2-1})^{2l}\right).
\end{equation}
The upper and lower signs correspond to the longitudinal and transverse channels, respectively. One can easily verify that, for any $l$, a solution with real $|s| >1$ exists only for positive $F^\e_l$.

For negative  $F^\e_l$, we search for complex solutions. In this case, we re-write (\ref{eq:fl-pole-l}) as
\begin{equation}
  \label{eq:fl-pole-l_1}
  \frac{1  -
  |F^\e_l|}{|F^\e_l|} =    \frac{is}{\sqrt{1-s^2}}\left(1\pm (s -i \sqrt{1-s^2})^{2l}\right).
\end{equation}

In what follows, we consider the longitudinal and transverse channels separately, first for even $l$ and then for odd $l$. We consider separately the limits of
$F^\e_l \approx -1$ and $|F^\e_l| \ll 1$, and then interpolate between the two limits. We show that there are multiple solutions with complex $s$ in each channel. The structure of the solutions in the longitudinal channel for even $l$ are very similar to those in the transverse channel for odd $l$. We do not discuss here the solutions in the transverse channel for positive $F^\e_l$, but below the threshold on the solution with real $s$ and $s>1$.

\subsubsection{even $l$, longitudinal channel}

For $F^\e_l \approx -1$, we first search for a solution with small $
|s|$.  Expanding Eq.~(\ref{eq:fl-pole-l_1}) in $s$, we find a pole on the imaginary axis
\begin{equation}
s = s_i \approx -i \frac{1+F^\e_l}{2}.
\label{new_4_1}
\end{equation}
There exist additional non-critical solutions for which $s$ remains finite at $F^\e_l =-1$.  To obtain these solutions, we choose the plus sign in Eq.~(\ref{eq:fl-pole-l_1}), set $F^\e_l =-1$, and solve the resultant equation
$1 + (s -i \sqrt{1-s^2})^{2l} =0$. There are $l$ solutions
$s_m = \arccos
\left[\pi(
2m+1)/
2l
\right]
$,  where $0 \leq m <l$ is an integer. They form $l/2$ pairs of solutions with $s_{1,2;p} = \pm a_p$, $a_p <1$, $0 \leq p <l/2$.   For $l=2$, we have a single pair $s_{1,2;0} = \pm 1/\sqrt{2}$, consistent with what we found earlier. At small deviations from  $F^\e_l =-1$, in any direction, these solutions become complex $s_{1,2;p} = \pm a_p -i b_p$, $b_p \propto (1 + F^\e_l)^2$. The imaginary part of these solutions remains negative even for $F^\e_s<-1$.

Next, consider the interval
  $0<-F^ \e_l \ll 1
 $.  In this limit, the magnitude of $s$ must be large for the right-hand side of Eq.~(\ref{eq:fl-pole-l_1})
 to match
  $1/|F^\e_l|\gg 1$ on the left-hand side of the same equation. Using $\sqrt{1-s^2} \approx is$ for $s$ in the lower half-plane, we reduce (\ref{eq:fl-pole-l_1}) at small $|F^\e_l|$ to
 \begin{equation}
  \label{eq:fl-pole-l_2}
  \frac{1}{|F^\e_l|} = (2s)^{2l}
\end{equation}
This equation has $l-1$ solutions with
$ s_m =
e^{-i\pi m / l}/2|F^\e_l|^{1/2l}$, where $0<m<l$  is an integer. The solution with $m =l/2$ is purely imaginary, and the other $l-2$ solutions form $p= (l-2)/2$ pairs of $s{1,2;p} = \pm a_p -ib_p$. The purely imaginary solution,
$s_{l/2}$, evolves towards $s_{l/2} =0$ as $F^\e_l$ approaches $-1$ and moves into the upper half-plane when $|F^\e_l|>1$, signaling a Pomeranchuk instability. The other solutions,
 $s_{1,2;p}$, evolve towards finite values at  $F^\e_l=-1$. Comparing the number of solutions with  $s_{1,2;p} = \pm a_p -ib_p$ at $F^\e_l=-1$  and
 $0<-
 F^\e_l\ll 1
 $, we see that they differ by one pair, which exists for the former case but not for the latter. From the analysis of the
 $l=2$ case, we know that the solution $s_{1,2} = \pm a -ib $ with a non-zero $b$ emerges when $|F^\e_l|$ exceeds a threshold value. At the threshold,  $s_{1,2} = \pm a - i
{\tilde \delta}$ with $a >1$ and ${\tilde \delta} \leq \delta$.

For $|F^\e_l|$ smaller than the threshold, the poles remain
 below the
branch cut at $s = \pm a - i \delta, a \geq 1$. For vanishingly small $\delta$, which we consider in this section, the poles are glued to the lower edge of the branch cut and slide along the branch cut towards its lower end at $|s| =1$ as $|F^\e_l|$ decreases. We see that for any even $l$  there exists exactly one such threshold solution, while other solutions appear already for infinitesimally small negative $F^\e_l$.

\subsubsection{even $l$, transverse channel}

We start again with $F^\e_l \approx -1$ and consider the solution with vanishingly small
 $s$. For the transverse channel
[for which we have to choose the minus sign in Eq.~(\ref{eq:fl-pole-l_1})], the leading, linear-in-$s$ term on the right-hand side of Eq.~(\ref{eq:fl-pole-l_1})
 is absent, and one needs to include the subleading terms.  A straightforward analysis then shows that the poles of the transverse susceptibility are located near the real axis, at
\begin{equation}
s_{1,2} \approx \pm \left(\frac{1+F^\e_l}{2l}\right)^{1/2} - i \frac{1+F^\e_l}{4}.
\label{new_5}
\end{equation}
When $|F^\e_l|$
becomes larger than 1,
 this solution moves into the upper half-plane, signaling an instability towards the development of a Pomeranchuk order.

 There also exist other solutions that remain finite at $F^\e_l\to -1$. These solutions are obtained by setting $F^\e_l =-1$ in Eq.~(\ref{eq:fl-pole-l_1})
 and solving the resultant equation
 $1 - (s -i \sqrt{1-s^2})^{2l} =0$.  There are $l-2$ solutions $s_n = \arccos (\pi n/l)
 $, where $0<n<l$ and $n \neq l/2$. Such solutions do not exist for $l=2$, i.e., there is only a solution that vanishes at $F^\e_l\to -1$.

For
$0<-F^\e_l\ll 1$ we need to solve $ 1/|F^\e_l| = -(2s)^{2l}$. The solutions are
$ s_n = e^{-i\pi (1+2n)/(2l)}/2|F_l^e|^{1/2l}$ with $0\leq n <l$. The number of solutions is $l$, and they form $l/2$ pairs with $s_{1,2;p} = \pm a_p -ib_p$, $0 \leq p<l/2$. One pair evolves towards $s_{1,2;p} =0$, as $F^\e_l$ approaches $-1$, while the other $l-2$  solutions tend to finite values $\pm a_p$ at $F^\e_l =-1$. We see that the number of non-critical solutions is the same,
i.e.,
$l-2$,
both for $0<-F^\e_l\ll 1$ and at $F^\e_l=-1$.

\subsubsection{odd $l$, longitudinal channel}

The analysis for odd $l$ proceeds along the same lines. We do not present the details of calculations and just state the results. For $F^\e_l \approx -1$, there are $l+1$ solutions, which form $(l+1)/2$ pairs $s_{1,2;p} = \pm a_p -ib_p$, $0 \leq p <(l+1)/2$. One pair is the same as in Eq.~(\ref{new_5}), the other solutions tend to finite $s_m = \arccos(\pi(1+2m)/(2l)$, with
 $0 \leq m <l$, $m \neq (l-1)/2$, at $F^\e_l =-1$. For $0<-F^\e_l\ll 1$, there are $l-1$ solutions
$s_m =
e^{-i\pi m/l}/2|F_l^e|^{1/2l}$, with
$0< m <l$. They form $(l-1)/2$ pairs $s_{1,2;p} = \pm a_p -ib_p$, $0\leq p < (l-1)/2$.  Comparing the number of solutions at $F^\e_l
 \approx -1$ and for $0<-F^\e_l\ll 1$, we see that there exists one pair of solutions $s_{1,2} =\pm a - i b$ with $b>0$, which emerges once $|F^\e_l|$ exceeds a threshold value. For $|F^\e_l|$ smaller than the threshold, this pair of solutions remains glued to the lower edge of the branch cut immediately below the real axis.

\subsubsection{odd $l$, transverse channel}

For $F^\e_l \approx -1$, there is one purely imaginary solution with vanishing $s$, as in Eq.~(\ref{new_4_1}), and $l-1$ solutions $s_n = \arccos (\pi n/l)
$, $0 <n<l$. For  $0<-F^\e_l\ll 1$, there are $l$ solutions
$ s_m =
e^{-i\pi (1+2m)/(2l)}/2 |F_l^e|^{1/2l}$, with $0 \leq m <l$. One solution, with
$m= (l-1)/2$, is purely imaginary, while the other solutions form $(l-1)/2$ pairs $s_{1,2;p} = \pm a_p -ib_p$, $0 \leq p < (l-1)/2$.   Comparing the number of solutions at $F^\e_l \approx -1$ and for $0<-F^\e_l\ll 1$, we see that the number is the same, i.e., all solutions develop already at infinitesimally small $F^\e_l$.  The purely imaginary solution moves into the upper frequency half-plane when $F^\e_l+1$ becomes negative,
signaling a Pomeranchuk instability,
while
other solutions $s_{1,2;p} = \pm a_p -ib_p$ remain in the lower frequency half-plane
even for  $F^\e_l  <-1$

Comparing the solutions for even and odd $l$, we see that at $F^\e_l \approx -1$, the solutions in the longitudinal channel for even $l$ are quite similar to those in the transverse channel at odd $l$ and vise versa. For smaller negative $F^\e_l$ there is a difference between the longitudinal and transverse channels at any $l$. Namely, there exists one solution in the longitudinal channel which remains glued to the lower edge of the branch cut at $|s|>1$, until $|F^\e_l|$ exceeds a threshold value, while in the transverse channel all solutions with $\text{Im}s<0$ emerge already at infinitesimally small $F^\e_l$.

\subsection{The case of two comparable Landau parameters}
\label{sec:F0F1}

As a more realistic example, we consider the case when two Landau parameters, e.g.,
    $F^\e_0$ and $F^\e_1$, are comparable in magnitude, while the rest of the Landau parameters are negligibly small. In this situation, the relation between the quasiparticle and free susceptibilities is more complicated than in Eq.~(\ref{new_2})
    because the $l=0$ and $l=1$ channels are coupled at finite $s$ via the $F^\e_0 F^\e_1$ term. Resumming the coupled RPA series for  $\chi^\e_{\text{qp},
  0}$ and $\chi^{\text{long}}_{\text{qp},
  1}$ or, equivalently, solving the FL kinetic equation, one arrives at
\cite{Zyuzin2018,Wu2018}
\begin{align}
  \label{extra_2}   \chi^\e_{\text{qp},0}
  (s)
                                            & =
  \nu_F
   \frac{K_0  - \frac{2 F^\e_1 K^2_1}{1  +  F^\e_1 \left(K_0 + K_2 \right)}}{1 + F^\e_0 K_0 - \frac{2 F^\e_0 F^\e_1 K^2_1}{1  +  F^\e_1 \left(K_0+ K_2\right)}}, \nonumber                                                                                                  \\
  \chi^
  {\text{long}}
  _{\text{qp},1}
  (s)
                                            & =
  \nu_F
   \frac{K_0+K_2  - \frac{2 F^\e_0 K^2_1}{1  +  F^\e_0 K_0}}{1 + F^\e_1 \left(K_0+K_2\right) - \frac{2 F^\e_0 F^\e_1 K^2_1}{1  +  F^\e_0 K_0}},
\end{align}
 where $K_{n}$ are given by Eq.~(\ref{eq:K-n-form}). Explicitly,
 \begin{align}
  K_0                                       & =1 + i \frac{s}{\sqrt{1-(s+i \delta)^2}},\nonumber                                                                                                                                                                            \\
  K_0 + K_2                                 & = 1 +2s^2 + 2 i \frac{s^3}{\sqrt{1-(s+ i \delta)^2}}, \nonumber                                                                                                                                                               \\
  \text{and}\;
  K_1                                       & = s + i  \frac{s^2}{\sqrt{1-(s+i\delta)^2}}.
        \label{new_2_2}
\end{align}

The denominators of $\chi^\e_{\text{qp},0}$ and $\chi^{\text{long}}_{\text{qp},1}$ vanish when
\begin{equation}
(1+ F^\e_0 K_0) (1+ F^\e_1 (K_0 + K_2)) = 2 F^\e_0 F^\e_1 K^2_1. \label{new_6}
\end{equation}
Suppose that $F^\e_1$ is negative and close to $-1$ while
$1 + F^\e_0 >0$ ($F^\e_0$ can be of either sign). In the previous sections, we saw that a critical zero-sound mode corresponds to small $s$. Substituting the forms of $K_n$ into Eq.~(\ref {new_6}) and assuming that $s$ is small, we obtain
\begin{eqnarray}
(1+F^\e_0)(1+F^\e_1)=2s^2+2is^3+iF^\e_0(1+F^\e_1)s.\nn                                                                                                                                                                                                                      \\
\end{eqnarray}
Iterating this equation in $1+F_1^\e\ll 1$, we obtain its approximate solution as
\begin{eqnarray}
s                                           & = & \pm \left(\frac{1+ F^\e_1}{2}\right)^{1/2} (1 + F^\e_0)^{1/2} \nonumber                                                                                                                                                   \\ &  & -\frac{i}{4}(1+F^\e_1).
\end{eqnarray}
This form does not differ qualitatively from Eq.~(\ref{F1-1})
for the case $F^\e_0 =0$, i.e., both the real and imaginary parts of the zero-sound velocity vanish when $ F^\e_1 $ approaches $-1$, and the imaginary part vanishes faster. The only effect of non-zero
  $F^\e_0$ is to renormalize the prefactor of $\text{Re}s$.

  In the opposite case, when $F^\e_0$ is close to $-1$ while $F^\e_1$ is not close to $-1$ but otherwise arbitrary, we find from (\ref{new_6})
\begin{eqnarray}
s \approx  -i \left[1 + F^\e_0 +\left(1+F^\e_0\right)^2\frac{1-F_1^\e}{1+F_1^\e}
\right].
 \end{eqnarray}
 We see that the pole remains on the imaginary axis and moves from the lower to upper frequency half-plane when $1 + F^\e_0$ changes sign. The Landau parameter $F^\e_1$ affects only the subleading term. We expect this behavior to hold when $F^\e_l$ with
$l>1$ are also present, as long as
$F^\e_{l>1}$ are not close to $-1$.

The simultaneous presence of $F^\e_0$ and $F^\e_1$, however, changes the threshold for the existence of a propagating zero-sound mode. ( For 3D systems, this effect was noticed in ~Ref.~\onlinecite{Orso1999}.
)
 For example, if only $F_0^\e$ is non-zero, a propagating mode exists only for positive $F^\e_0$. If $F^\e_1$ is also non-zero and positive, a propagating mode exists also for negative $F^\e_0$. Moreover, for large enough
  $F^\e_1>0$, a propagating mode exists even at the $l=0$ Pomeranchuk instability, i.e., when $F^\e_0 =-1$. Namely, setting $F^\e_0 =-1$ and varying $F^\e_1 >0$, we find the solution of Eq.~(\ref{new_6}) in the form of a propagating zero-sound mode for $F^\e_1 >1$. The mode frequency is
\begin{equation}
s = \pm \frac{1 +F^\e_1}{2\sqrt{F^\e_1}},  ~~|s| >1.
\end{equation}
 For large $F^\e_1$, $s \approx \sqrt{F^\e_1}/2$.

 \subsection{3D systems}
 \label{sec:3D}

For comparison, we also briefly discuss the behavior of zero-sound excitations near a Pomeranchuk instability in a 3D system. We present the results for $l=0$ and $l=1$ and, in each case, consider only one non-zero Landau parameter $F^\e_l <0$.

\subsubsection{$l=0$}
Zero-sound modes in the
 $l=0$ channel were analyzed in Refs.~\onlinecite{Pethick1988,Orso1999}. The free-fermion susceptibility with the form factor  $f^\e_0 ({\bf k}_F) =1$  is
\begin{equation}
\chi_{\text{free},0} (s) = 1 - \frac{s}{2} \ln{\frac{s+ i \delta +1}{s+i \delta-1}}.
\end{equation}
The equation for the pole reads
\begin{equation}
\frac{1}{|F^\e_0|}-1 = - \frac{s}{2} \left[-i \pi + \ln{\frac{1+s+ i \delta}{1-s - i \delta}}\right]. \label{n_7}
\end{equation}
The pole is completely imaginary: $ s = -ib$ (hence, $i\delta$ in (\ref{n_7}) is irrelevant). In contrast to the 2D case, such solution exists for all negative $F
^\e
_0$, i.e., there is no threshold. For $ F^\e_
0
\approx -1$,  $b \approx (2/\pi) (1-|F
^\e
_0|)$.  For
$0<
-F
^\e_0
\ll 1
$, $b \approx 1/(\pi |F^\e_0|)$.

\subsubsection{$l=1$}

The eigenfunctions of angular momentum $l=1$ are spherical harmonics $Y^m_1 (\theta, \phi)$. We normalize $Y^m_1
$ as $Y^0_1 (\theta) = \sqrt{3} \cos{\theta}$ and $Y^{\pm 1}_1 (\theta, \phi) = \mp  \sqrt{3/2} \sin{\theta} e^{\pm i \phi}$. Then the critical value of $ F
^\e
_1
$ for a Pomeranchuk instability is $ F^\e_1
=
-
1$.

In the longitudinal channel, the form-factor is $f^\e_1 ({\bf k}_F) = Y^0_1 (\theta)$. The free-fermion susceptibility is
\begin{equation}
\chi_{\text{free},0} (s) = 1 + 3s \left(s+ i \delta -  \frac{3 (s+ i \delta)^2}{2} \ln{\frac{s+ i \delta+1}{s+ i \delta-1}}\right)
\end{equation}
The equation for the zero-sound pole is
\begin{equation}
\frac{1
- |F^\e_1|}{3|F^\e_1|} = s \left(s+ i \delta - \frac{(s+ i \delta)^2}{2} \left(-i \pi + \ln{\frac{s+ i \delta+1}{1-s-i \delta}}\right)\right)
\label{n_8}
\end{equation}
When $ F^\e_1\approx -1
$, the solution is
\begin{equation}
s = \pm \epsilon^{1/2} - i \frac{\pi}{4} \epsilon, ~~~ \epsilon = \frac{1- |F
^\e
_1|}{3}.
\end{equation}
This is very similar to the 2D case, cf. Eq.~(\ref{F1-1}). In contrast to
  2D case, however, a complex solution in 3D exists for all negative $ F^\e_1
  $, i.e., there is no threshold.  When $|F^\e
_1|$ is small,
\begin{equation}
s \approx \left(\frac{
1}{3 \pi |F^\e_1|}\right)^{1/3} e^{-i\pi/6}.
\end{equation}

The form-factor in the transverse channel is $f^\e_1 ({\bf k}_F) = Y^{\pm 1}_1 (\theta,
\phi
)$. The free-fermion susceptibility is
\begin{equation}
\chi_{\text{free},0} (s) = 1 - \frac{3}{2} s^2 - \frac{3 s (1-s^2)}{4} \ln{\frac{s+1}{s-1}}.
\end{equation}
The equation for the zero-sound pole is
\begin{equation}
\frac{4 (1 + F^\e_1)}{3 |F^\e_1|} = -2 s^2 - s(1-s^2)  \left(-i \pi + \ln{\frac{s+1}{1-s}}\right).
\label{n_9}
\end{equation}
One can easily verify that the pole is located on the imaginary axis, at $s = -ia$, where $a$ is the solution of
\begin{equation}
\frac{4 (1
- |F^\e_1|
)}{3 |F^\e_1|} = \pi a (1 +a^2) + 2 a^2 + ia \ln{\frac{1 + ia}{1-ia}}
\end{equation}
Near $F^\e_1 =-1$,  $a \approx (4/3\pi) (1
-
|F^\e_1|)$. For small negative $F^\e_1$,  $a \approx  (4/(3\pi |F^\e_1|))^{1/3}$. This is again similar to 2D, except for the solution  $s =-ia$ in 3D exists for all negative $ F^\e_1
$, i.e., there is no threshold.

\section{
Finite disorder}
\label{sec:gamma}
\subsection{General formalism}
In this section we analyze how the results of the previous sections change in the presence of finite disorder.
  As in the previous section, we consider separately the cases of
$l=0,1,2$. Other cases can be analyzed in the same manner as these two.

\begin{figure*}
  \centering
  \includegraphics[width=\hsize]{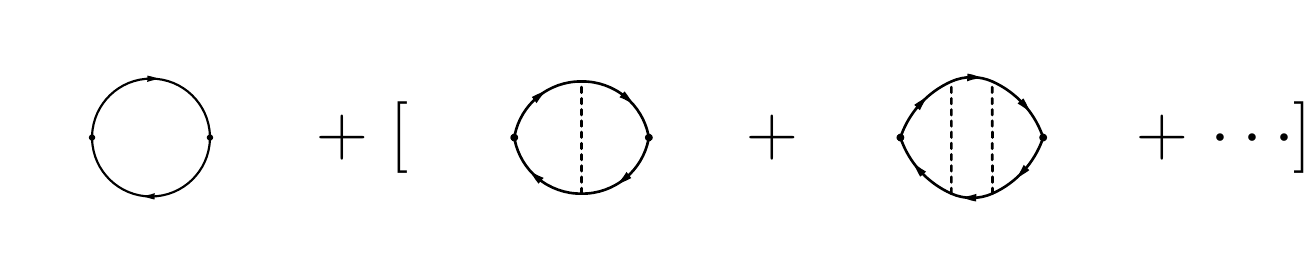}
  \caption{The susceptibility of non-interacting fermions in the presence of impurity scattering. The diagram on the left corresponds to $\chi^B_{\text{free},l}$ in Eq.~(\ref{B}), while the diagrams in the square brackets correspond to $\chi^V_{\text{free},l}$ in Eq.~(\ref{V}). The solid lines represent disorder-averaged fermionic propagators,
   the dashed lines represent the correlation functions of the impurity potential.}
  \label{fig:diffuson}
\end{figure*}

The free-fermion susceptibility in the presence of scattering by short-range impurities consists of two parts: the bubble part and the vertex part:
\begin{eqnarray}
\chi_{\text{free},l}(q,\omega_m)=\chi^{\text{B}}_{\text{free},l}(q,\omega_m)+\chi^{\text{V}}_{\text{free},l}(q,\omega_m).
\end{eqnarray}
(cf. Fig.~\ref{fig:diffuson}).
The bubble part is formed from the (Matsubara) Green's functions $G({\bf k},\nu_m)=(i\nu_m-\epsilon_{k}+i\text{sgn}\nu_m\tilde\gamma/2)^{-1}$, where $\tilde\gamma$ is the impurity scattering rate:
\begin{widetext}
\begin{eqnarray}
\chi^{\text{B}}_{\text{free},l}(q,\omega_m) & = & -
\frac{2}{N_F}
\int \frac{d^2k}{(2\pi)^2}\int\frac{d\nu_m}{2\pi} |f^\e_l({\bf k})|^2 G({\bf k}+{\bf q}/2,\nu_m+\omega_m/2,)G({\bf k}-{\bf q}/2,\nu_m-\omega_m/2).\label{B}
\end{eqnarray}
The vertex part is
\begin{eqnarray}
\chi^{\text{V}}_{\text{free},l}(q,\omega_m) & = & -
\frac{2}{N_F}
 \int \frac{d^2k}{(2\pi)^2}\int \frac{d^2k'}{(2\pi)^2}\int\frac{d\nu_m}{2\pi} f^\e_l({\bf k})\left( f^\e_l({\bf k}')\right)^*G({\bf k}+{\bf q}/2,\nu_m+\omega_m/2)G({\bf k}-{\bf q}/2,\nu_m-\omega_m/2)\nn               \\
                                            &   & \times G({\bf k}'+{\bf q}/2,\nu_m+\omega_m/2)G({\bf k}'-{\bf q}/2,\nu_m-\omega_m/2){\cal D}(\omega_m,q;\nu_m).\label{V}
\end{eqnarray}
\end{widetext}
where ${\cal D}(q,\omega_m;\nu_m)$ is the diffusion propagator~\cite{Zala2001}
\begin{eqnarray}
{\cal D}(q,\omega_m;\nu_m)=\frac{\tilde\gamma}{2\pi N_F}\frac{\theta(\omega_m+|\nu_m|/2)\theta(|\nu_m|/2-\omega_m)}{\sqrt{(v_F^*q)^2+(|\omega_m|+\tilde\gamma)^2}-\tilde\gamma}.\nn                                                                                         \\
\end{eqnarray}
Diagrammatically, ${\cal D}(q,\omega_m;\nu_m)$ is represented by the sum of ladder diagrams in the particle-hole channel
(the sequence of diagrams in the square brackets in Fig.~\ref{fig:diffuson}).

The retarded forms of the susceptibilities are obtained by choosing $\omega_m>0$ and replacing $i\omega_m\to\omega$ in the final results. The vertex part is especially important for the $l=0$ case, because the corresponding order parameters (charge or spin) are conserved quantities, and hence the bubble and vertex parts of the susceptibility must cancel each other at $q=0$. For $l >0$, the corresponding order parameters are not conserved,
 but the vertex parts must be also included in order to obtain the correct positions of the zero-sound poles in the complex plane.

\subsection{l=0}
\begin{figure*}
  \centering
  \begin{subfigure}{0.49\hsize}
    \fbox{\includegraphics[height=0.8\hsize]{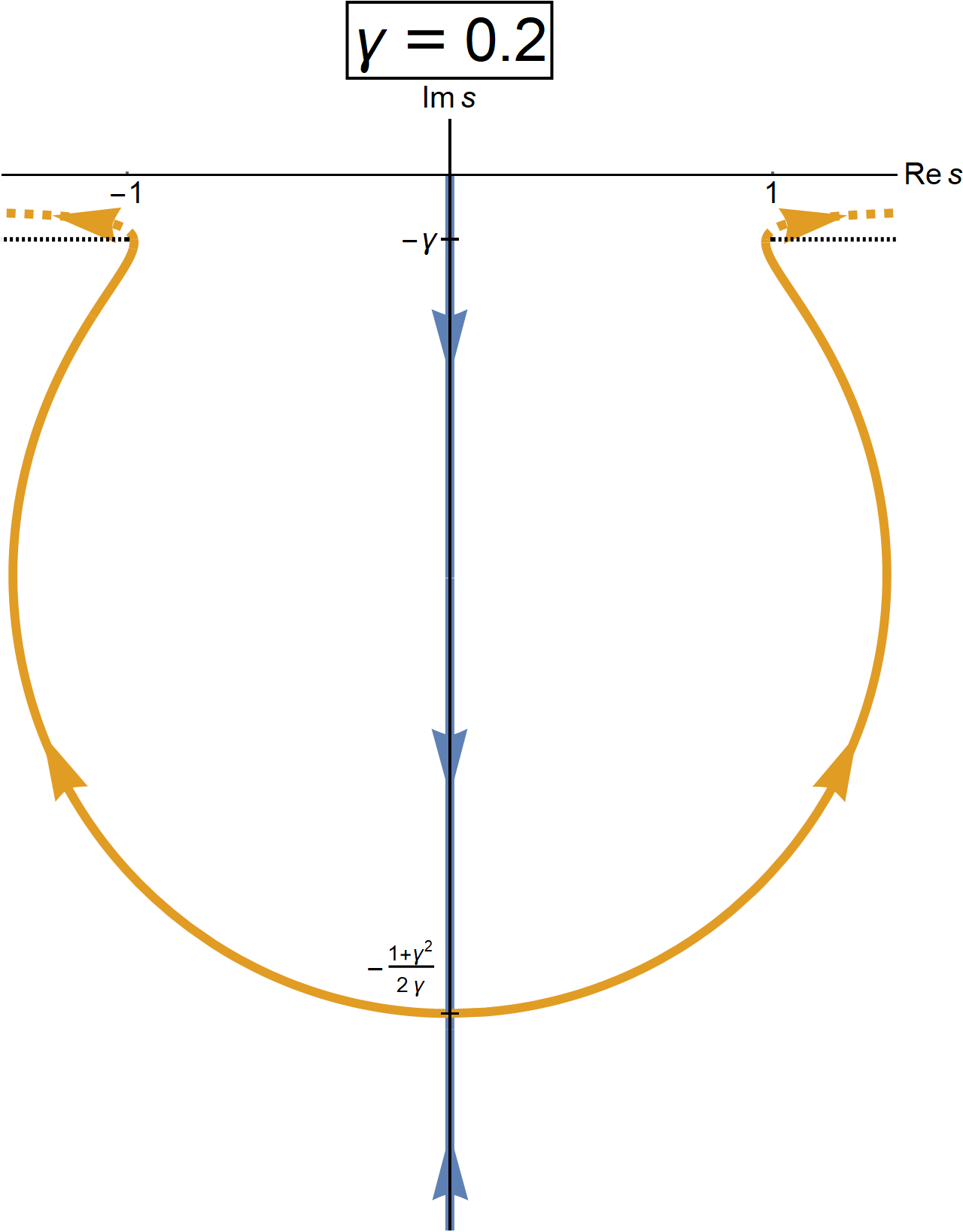}}
  \end{subfigure}\hfill
  \begin{subfigure}{0.49\hsize}
    \fbox{\includegraphics[height=0.8\hsize]{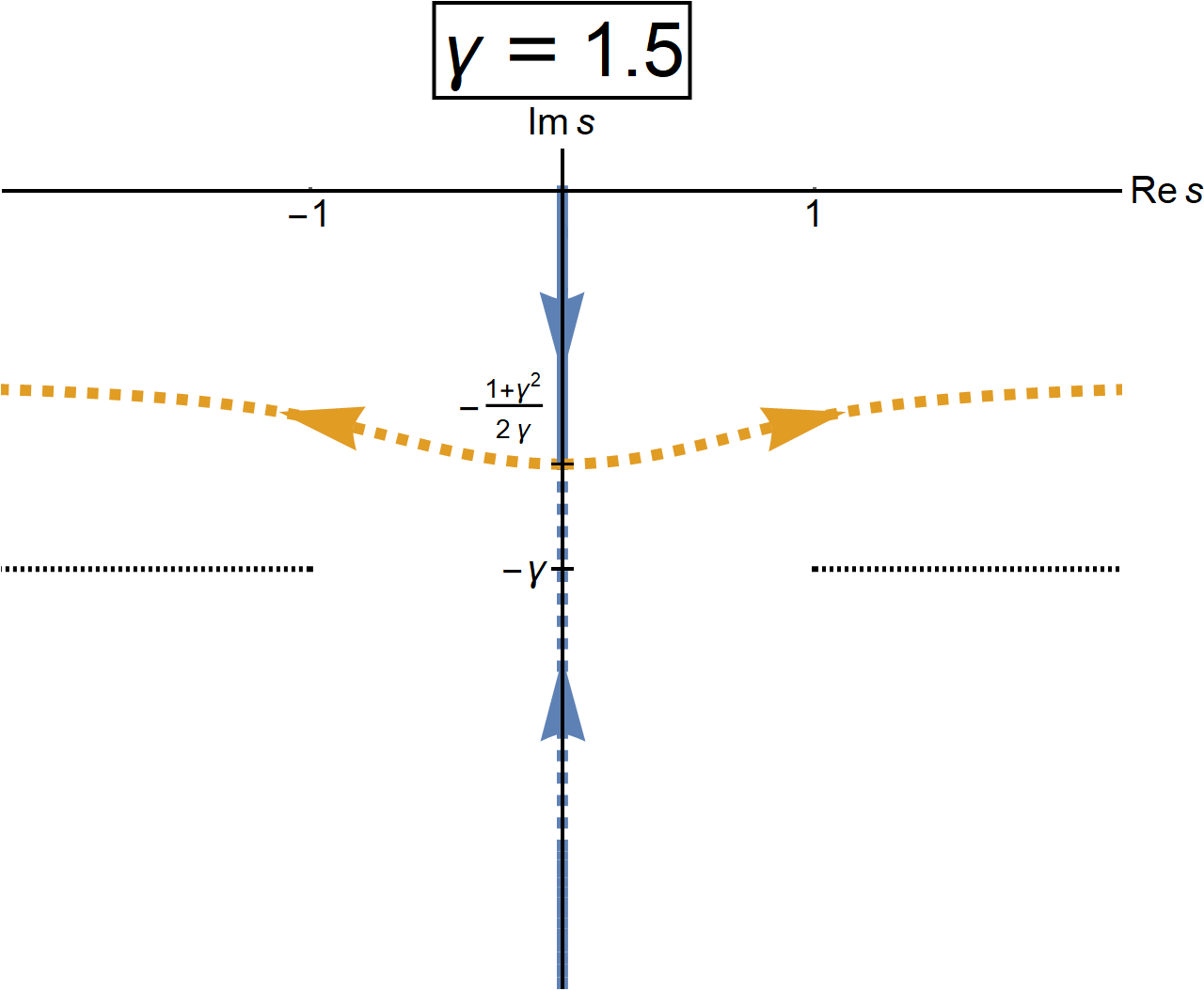}}
  \end{subfigure}
  \caption{(color online)
  Evolution of the poles of $\chi_{\text{qp},0}^\e(s)$ with Landau parameter $F^\e_0$ for finite disorder, parameterized by the dimensionless scattering rate $\gamma$. The blue and yellow lines denote the behavior of the poles of $\chi_{\text{qp},0}^\e (s)
  $ for negative $F_0^\e$ (solid traces) and positive $F_0^\e$ (dashed traces). The horizontal dotted lines denote the square-root branch cuts of $\chi_{\text{qp},0}^\e(s)$ at $\mbox{Im}s = -\gamma$. The arrows identify the direction of the poles' motion with
  $F_0^\e$ increasing from $-1$ to $\infty$.
  We use different colors to show how poles merge and then bifurcate.}
  \label{fig:chi0-poles-gamma}
\end{figure*}

We recall that in a clean Fermi liquid with positive $F^\e_0$ the pole of $\chi^\e_0 (s)$ is on the real axis at $|s| >1$. For negative $F^\e_0$, the pole is on the imaginary axis, at $s = -i (1-|F^\e_0|)/(2 |F^\e_0|-1)^{1/2}$, when $-1 < F^\e_0 <-1/2$. At $F^\e_0 =-1/2$, the pole at $s = - \infty$ splits into two, and the new poles instantly move to the $s =
 \pm
  \infty$ points on the real axis. At larger, but still negative $F^\e_0
  $, the poles move towards $s= \pm 1$ along the lower edge of the branch cut.

 The bubble and vertex part for the $l=0$ case are given by
 \begin{eqnarray}
 \chi^{\text{B}}_{\text{free},0} & = & 1+\frac{is}{\sqrt{1-(s+i\gamma)^2}},\nn  \\
 \chi^{\text{V}}_{\text{free},0} & = &
                                       \frac{is\gamma/\sqrt{1-(s+i\gamma)^2}}{\sqrt{1-(s+i\gamma)^2}-\gamma},
 \end{eqnarray}
 where $\gamma=\tilde\gamma/v_F^*q$. Adding these up, we obtain
\begin{equation}
 \chi_{\text{free},0} (s) =  1 + i\frac{s}{\sqrt{1 - (s+ i \gamma)^2} - \gamma},
 \label{imp}
\end{equation}
which is the result quoted in Eq.~(\ref{retd}), except that we have changed the notations $\delta\to\gamma$ to emphasize that $\gamma$ does not have to be small. This result, as well as a corresponding result for the $l=1$ case, holds for $\tilde\gamma
\ll E_F$ while the ratio $\gamma/s$ can be arbitrary. At $q\to 0$, i.e., at $s\to \infty$, the susceptibility vanishes, which guarantees that the charge and spin are conserved.

For $|s|\ll \gamma$, Eq.~(\ref{imp}) reduces to the well-known diffusive form~\cite{Altshuler1985,Lee1985}
\begin{eqnarray}
\chi_{\text{free},0} (s)=\frac{1}{1-2 i \gamma s}=\frac{Dq^2}{Dq^2-i\omega},
\end{eqnarray}
where
$D=(v^*_F)^2/2\tilde\gamma$
is the diffusion coefficient in 2D.

Substituting Eq.~(\ref{imp}) into Eq.~(\ref{new_2}) and solving for the poles, we find that for $F^\e_0 < -1/2$ the pole is on the imaginary axis, at
\begin{equation}
  s = s_1 = -i \gamma \frac{1- |F^\e_0|}{2 |F^\e_0|-1} \left(\sqrt{1 + \frac{2 |F^\e_0|-1}{\gamma^2}} -1 \right).
  \label{na_1}
\end{equation}
For $\gamma \to 0$, this reduces to Eq.~(\ref{ac1}). For large $\gamma$, $s_1 = -i (1 - |F^\e_0|)/
 2 \gamma
 $. In this limit, we have a diffusion pole at $\Omega = - i D^* q^2$, where $D^* = D (1 - |F^\e_0|)$ is the renormalized diffusion coefficient~\cite{Finkelstein2010}. In the ballistic regime at small $\gamma$, the damping term accounts for a small correction to the result for a clean Fermi liquid, cf. Eq.~(\ref{ac1}).

When $
1+ F^\e_0$ becomes negative, $s_1$ moves into the upper half-plane of $s$, which signals a Pomeranchuk instability. For positive  $
1+ F^\e_0$ the pole $s_1$ moves down along the imaginary axis as $1+ F^\e_0$ increases, but remains finite at $ F_0^\e = -1/2$, in contrast to the behavior in the clean limit. Expanding the square root in Eq.~(\ref{na_1}) in $2 |F^\e_0|-1$, we find $s_1 = - i/(4\gamma)$ at $F^\e_0=-1/2$. At larger $ F^\e_1
 $, another solution,
\begin{equation}
  s_2 =  -i \gamma \frac{1- |F^\e_0|}{1-2 |F^\e_0|} \left(\sqrt{1 - \frac{1-2 |F^\e_0|}{\gamma^2}} +1 \right),
\end{equation}
appears in the lower half-plane, initially at $s_2 = -i \infty$. As $ F^\e_0
   $ keeps increasing, $s_1$ and $s_2$ move towards each other. For $\gamma<1$,  the two solutions merge into a single pole $s_1=s_2 = -i (1+\gamma^2)/(2\gamma)$ at $ F^\e_0 = (\gamma^2-1)/2<0$ (see Fig.~\ref{fig:chi0-poles-gamma}a). For  $F^\e_0$  slightly larger than this value, the double pole bifurcates into two poles with finite real parts. For even larger $F^\e_0$, the two poles move along arc-like trajectories $s_{1,2}=\pm a -i b$ in the complex plane. In contrast to the case of $\gamma =0$, the poles remain at a finite distance from the lower edges of the branch cuts, as long as $F^\e_0$ remains negative.

At  $F^\e_0=0$, the poles reach the points $s_{1,2}  = \pm \sqrt{1-\gamma^2} - i \gamma$.  At positive   $F^\e_0$, $a$ increases and $b$ becomes smaller than $\gamma$. This implies that the poles are now located in between the branch cuts and real axis. At  $F^\e_0 \to \infty$,
   $a \to (F^\e_0/2)^{1/2}$ and $b \to \gamma/2$.
    \begin{figure*}[hbt]
     \centering
     \begin{subfigure}{0.45\hsize}
       \centering
       \includegraphics[width=\hsize]{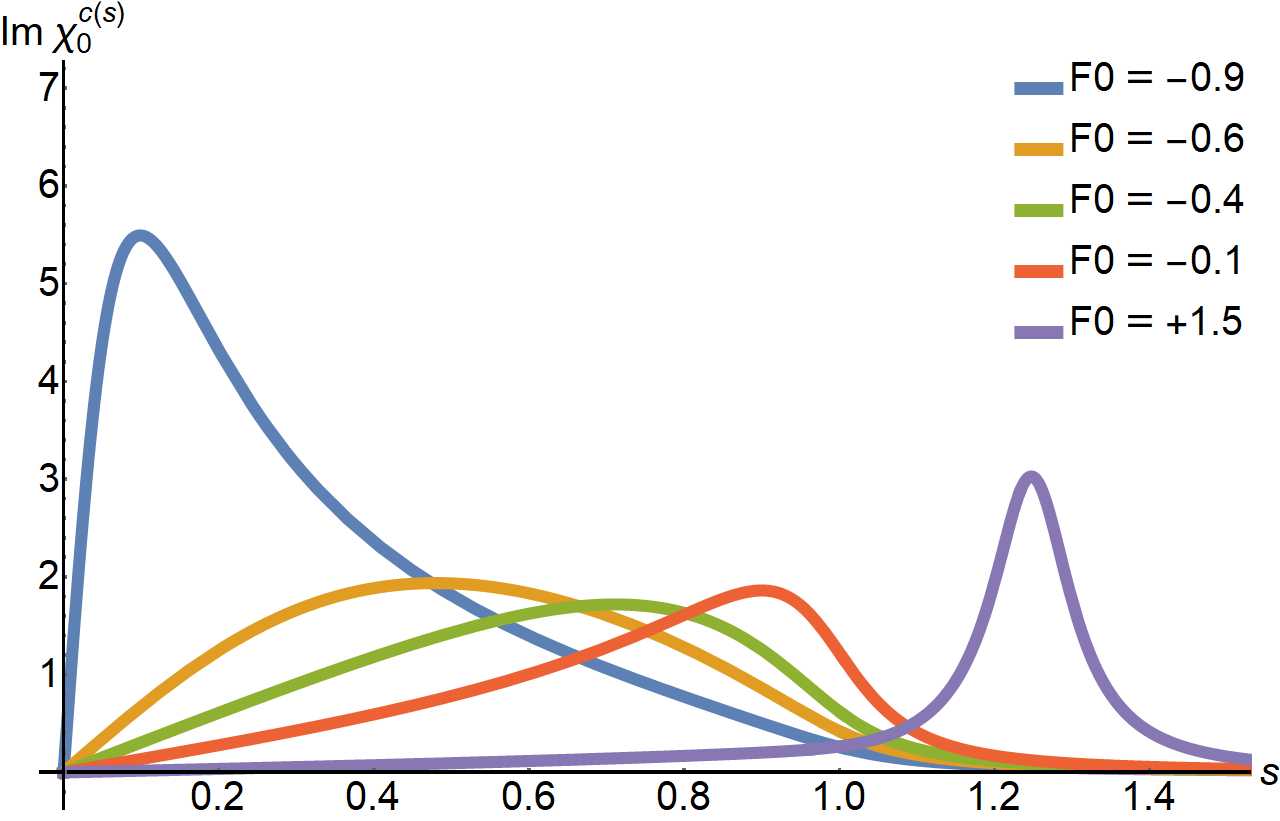}
       \caption{
         \label{fig:im-chi0-gamma-weak}}
     \end{subfigure}
     \begin{subfigure}{0.45\hsize}
       \centering
       \includegraphics[width=\hsize]{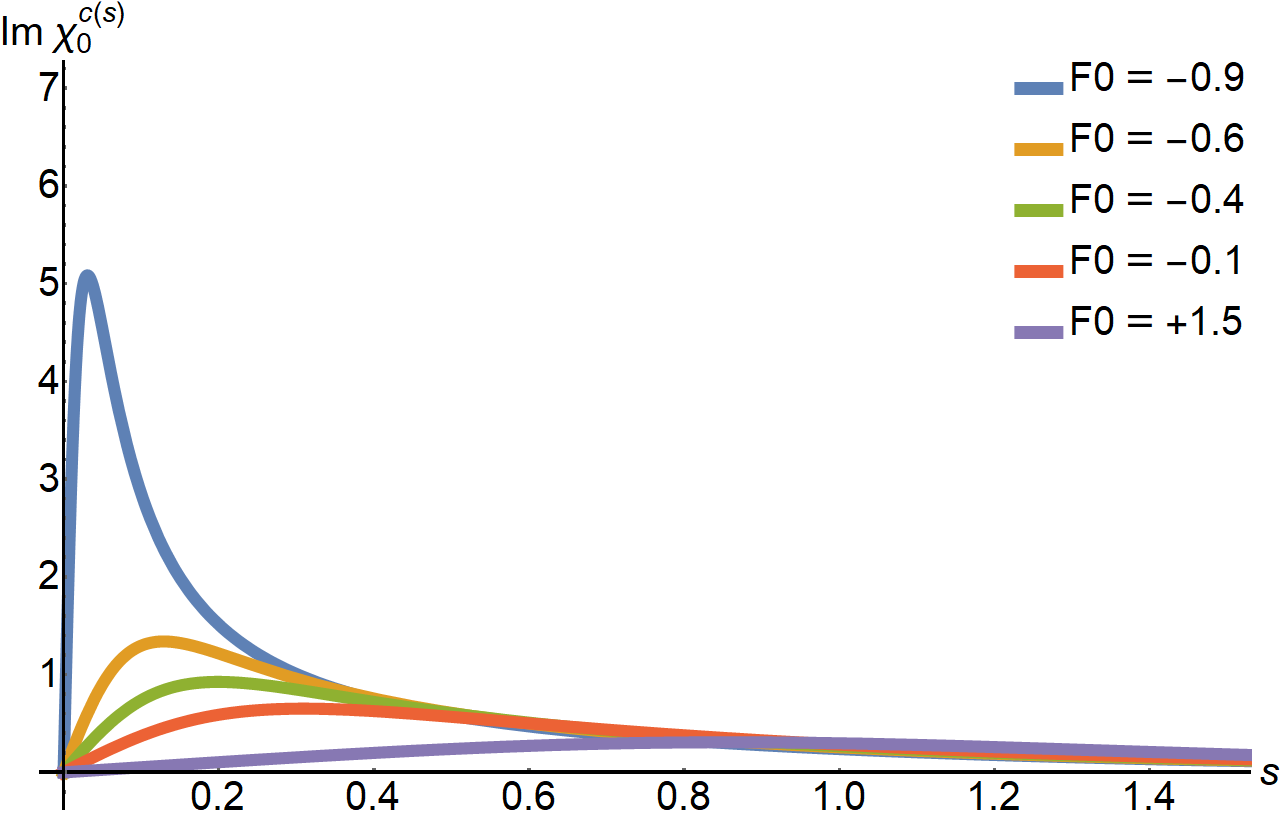}
       \caption{
         \label{fig:im-chi0-gamma-weak}}
     \end{subfigure}
     \caption{(color online)
The imaginary part of $\chi_{\text{qp},0}^\e (s)$
       for finite disorder, characterized by the dimensionless coupling constant
      $\gamma$. (a)
      Weak disorder ($\gamma = 0.2$). For $-1 < F_0^\e < -1/2$,
      the shape of $\mbox{Im}\chi_{\text{qp},0}^\e (s)$ has a characteristic overdamped form.
     As $F_0$ increases from $-1/2$ to $0$,
      the shape changes its form due to the appearance of
      ``hidden'' poles below the branch cut.
     For $F_0^\e > 0$,  $\mbox{Im}\chi_{\text{qp},0}^\e (s)$ has a conventional form of a damped zero-sound mode.
     (b)
      Strong disorder ($\gamma = 1.5$). In this case  $\mbox{Im}\chi
      ^\e_0(s)
      $ has an overdamped shape for all $F_0^\e$,
      negative and positive.
      \label{fig:im-chi0-gamma-new}}
   \end{figure*}

   For $\gamma>1$, the second pole $s_2$ still emerges at $F^\e_0 =-1/2$, but the poles at $s=s_1$ and $s=s_2$ on the imaginary axis remain at finite distance from each other for all negative $F^\e_0$ and merge only at
   $F^\e_0 =
    (\gamma^2-1)/2>0$ (see Fig.~\ref{fig:chi0-poles-gamma}b). At larger  $F^\e_0$, the poles again follow the trajectories
    $s_{1,2} = \pm a -ib$ with $a$ increasing and $b$ decreasing with increasing $F^\e_0$. For $F^\e_0\gg 1$, the poles reach the same values as for
    $\gamma<1$:
   $a
   \approx (F^\e_0/2)^{1/2}$ and $b
    \approx
    \gamma/2$.
 In Fig.~\ref{fig:im-chi0-gamma-new}
   we plot  $\text{Im}\chi^\e_0 (s)$ for real $s$ for a range of
   $F^\e_0$, both for $\gamma <1$ and $\gamma >1$ (panels (a) and (b), respectively).

\subsection{$l=1$}
\subsubsection{$l=1$, longitudinal}
 \begin{figure*}[p!]
  \centering
\setlength{\fboxsep}{0pt}  \fbox{\includegraphics[width=0.45\hsize]{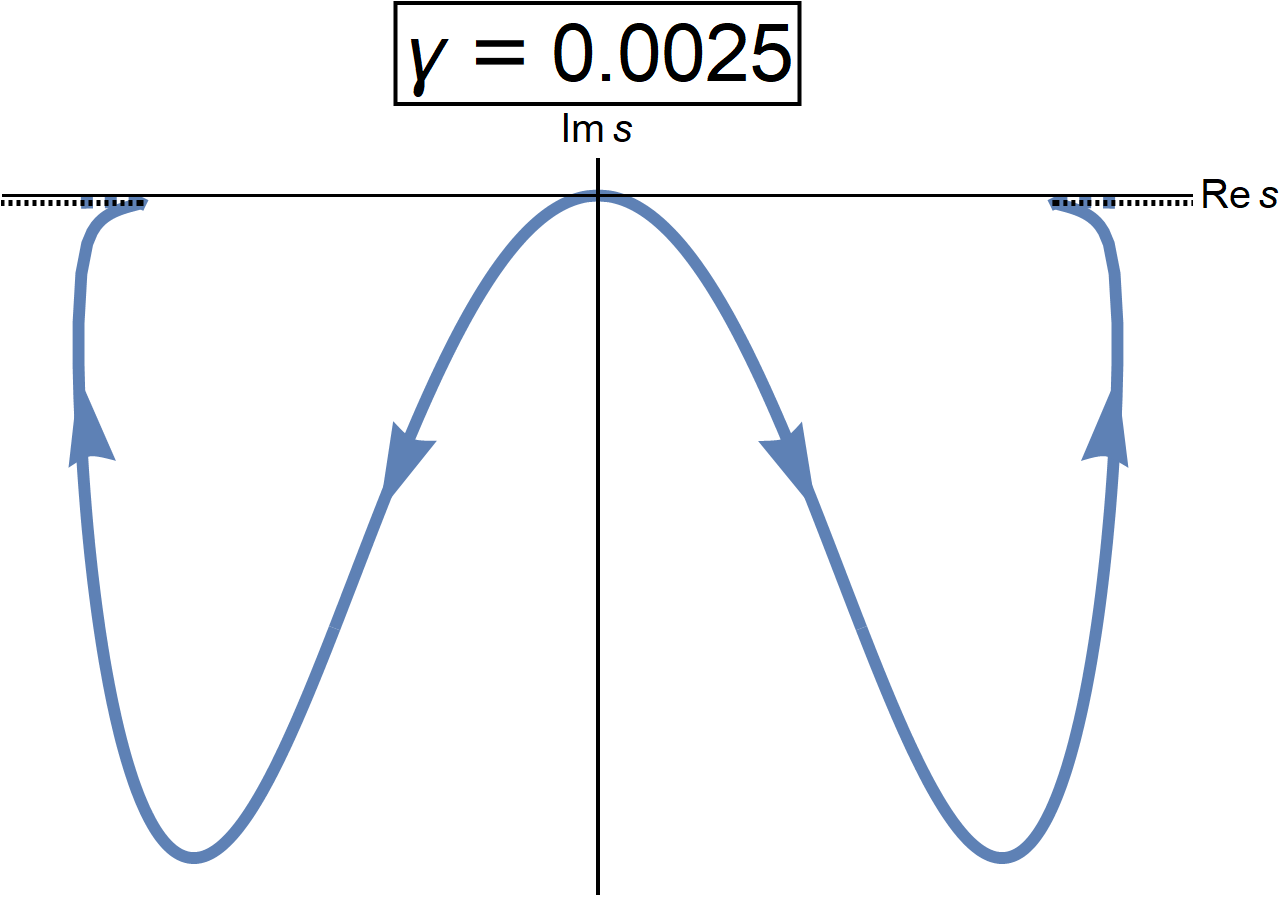}}\llap{\raisebox{0.265\hsize}{\fbox{\includegraphics[width=0.15\hsize,clip,trim=300 0 0 270]{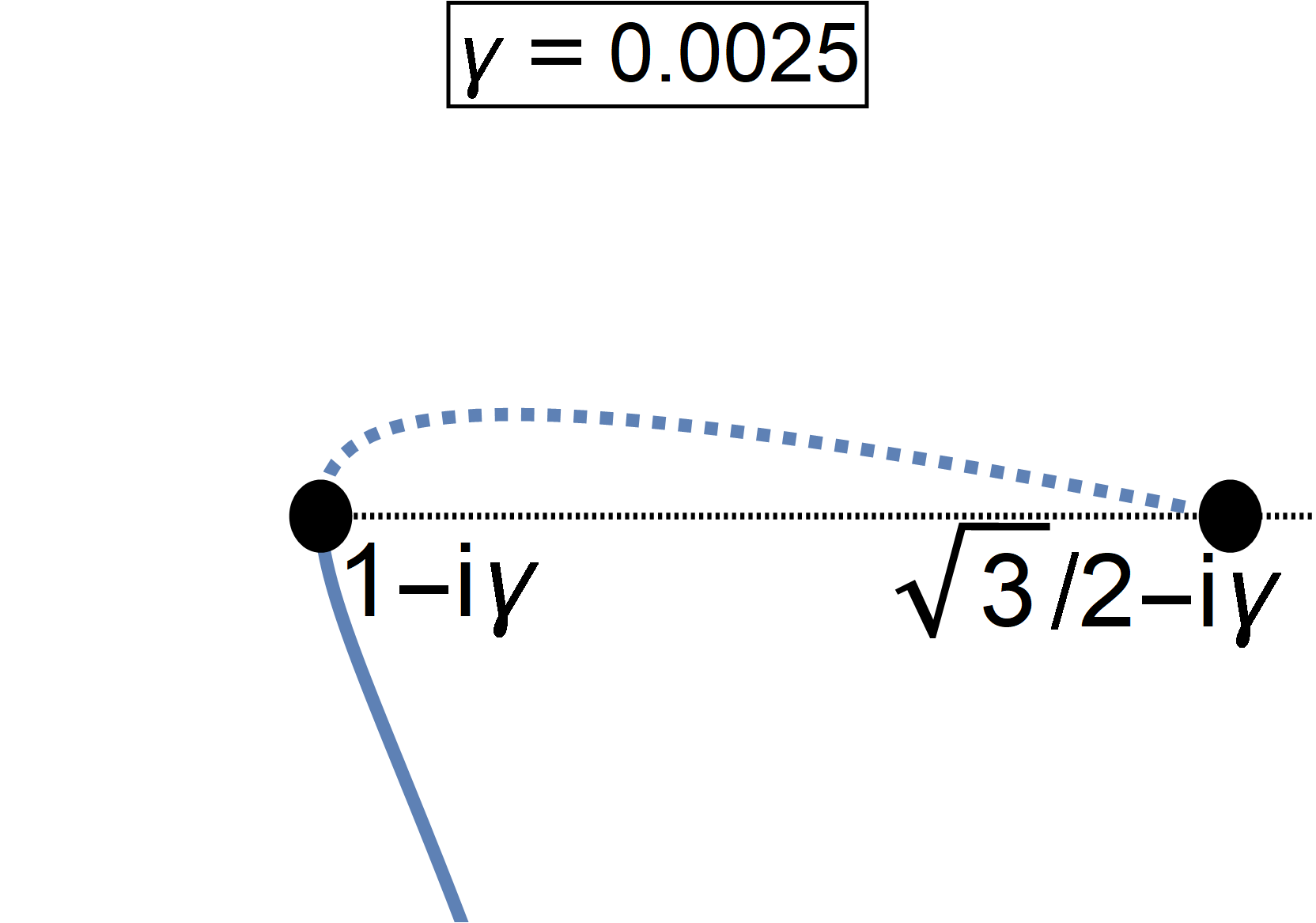}}}}
  \fbox{\includegraphics[width=0.45\hsize]{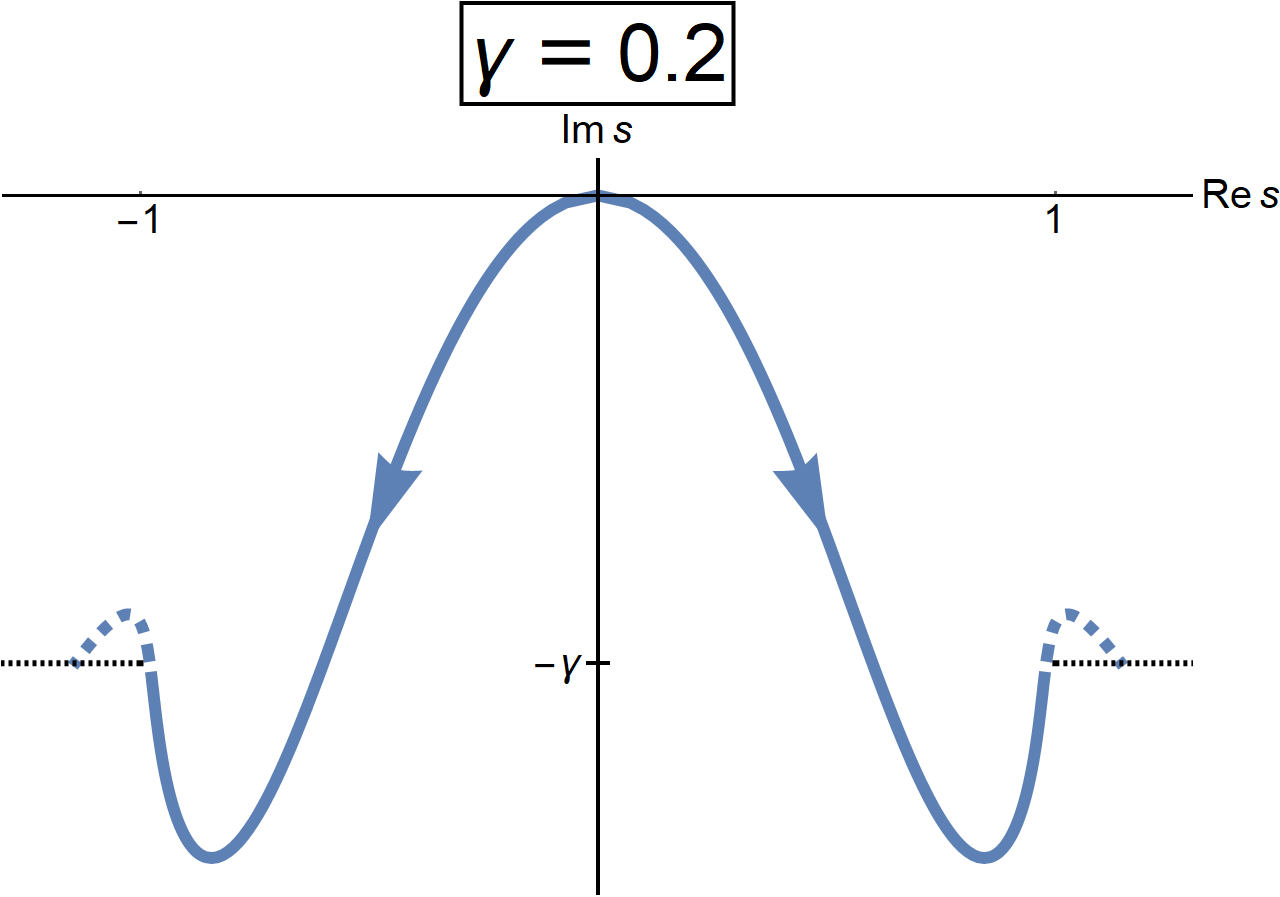}}

  \fbox{\includegraphics[width=0.45\hsize]{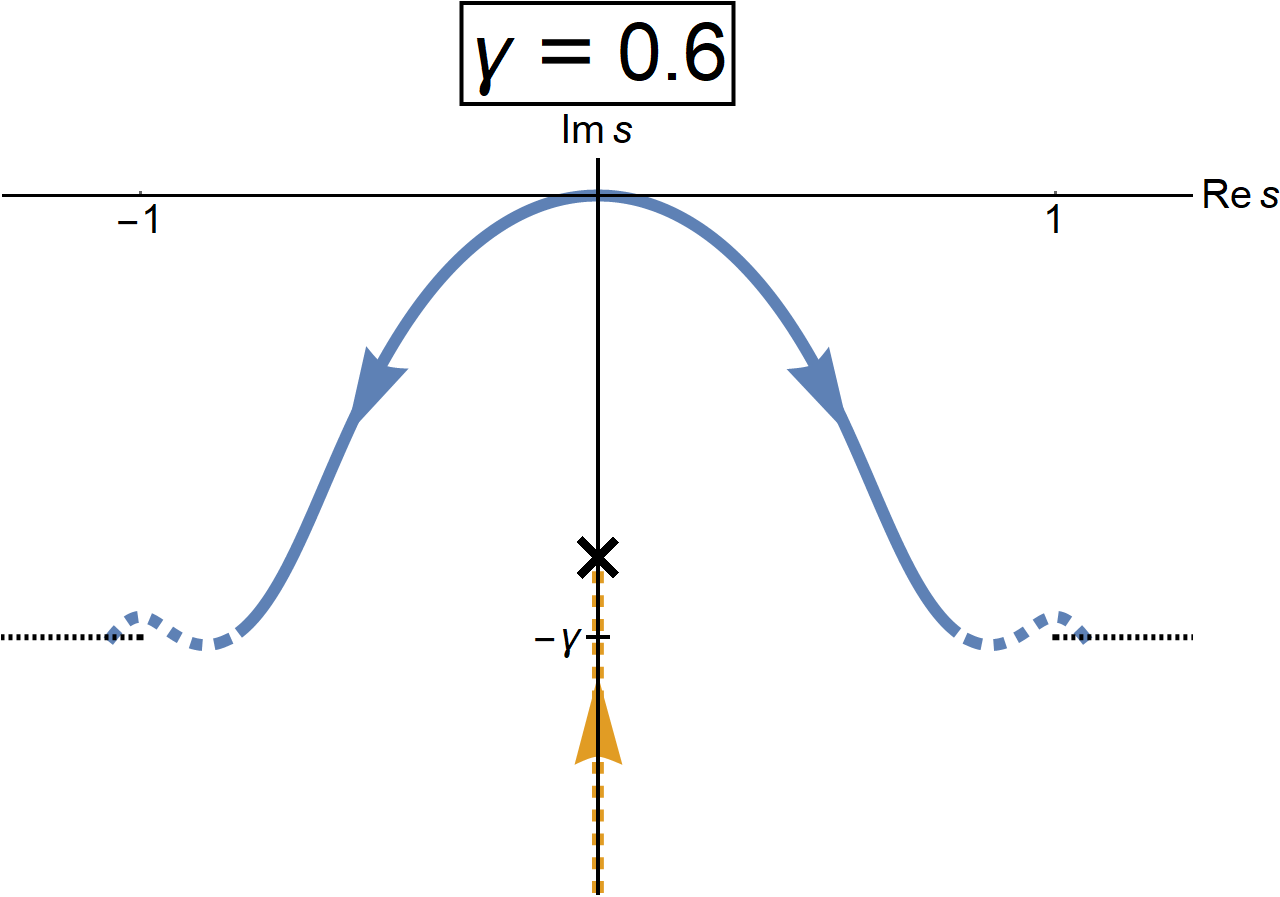}}

  \fbox{\includegraphics[width=0.45\hsize]{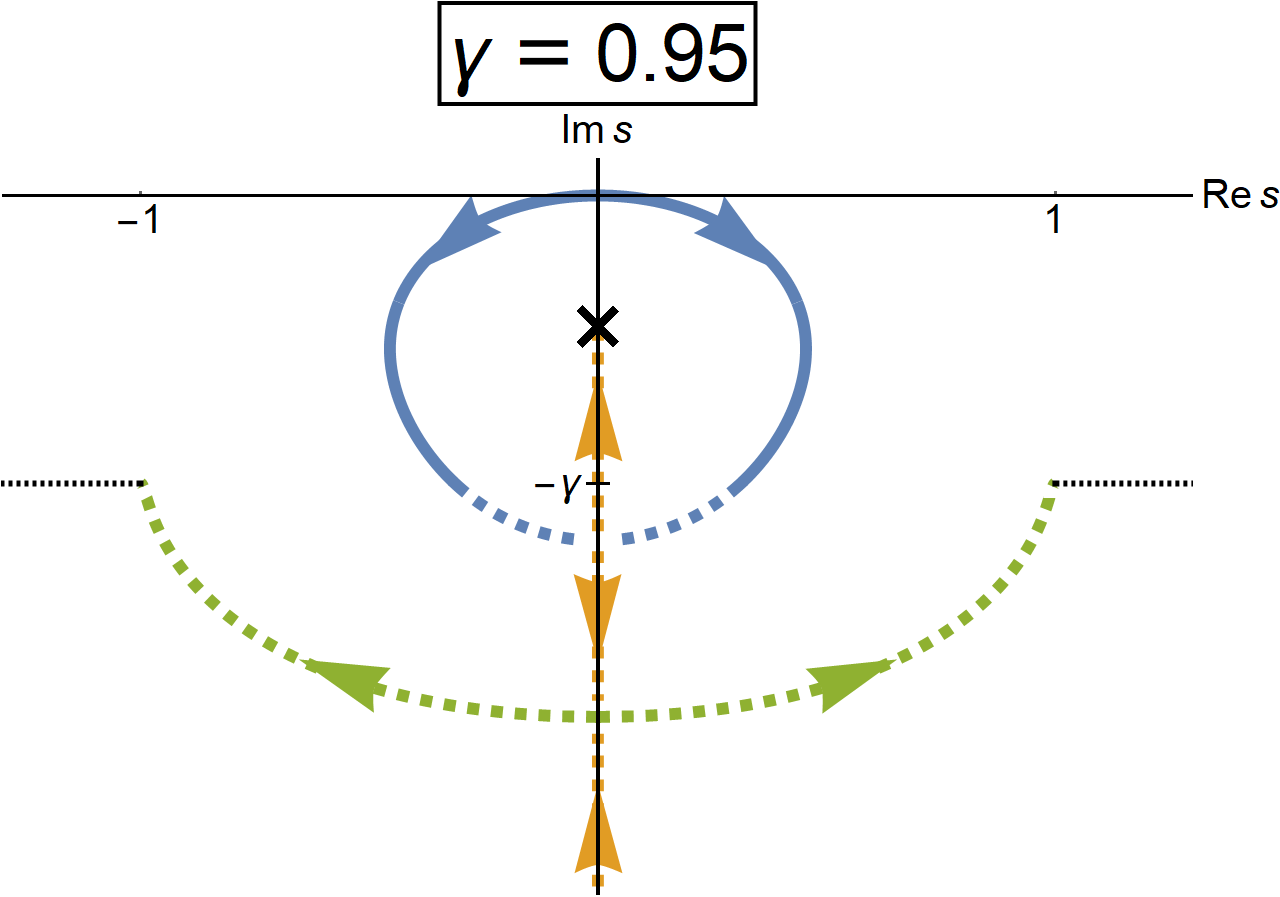}}
  \fbox{\includegraphics[width=0.45\hsize]{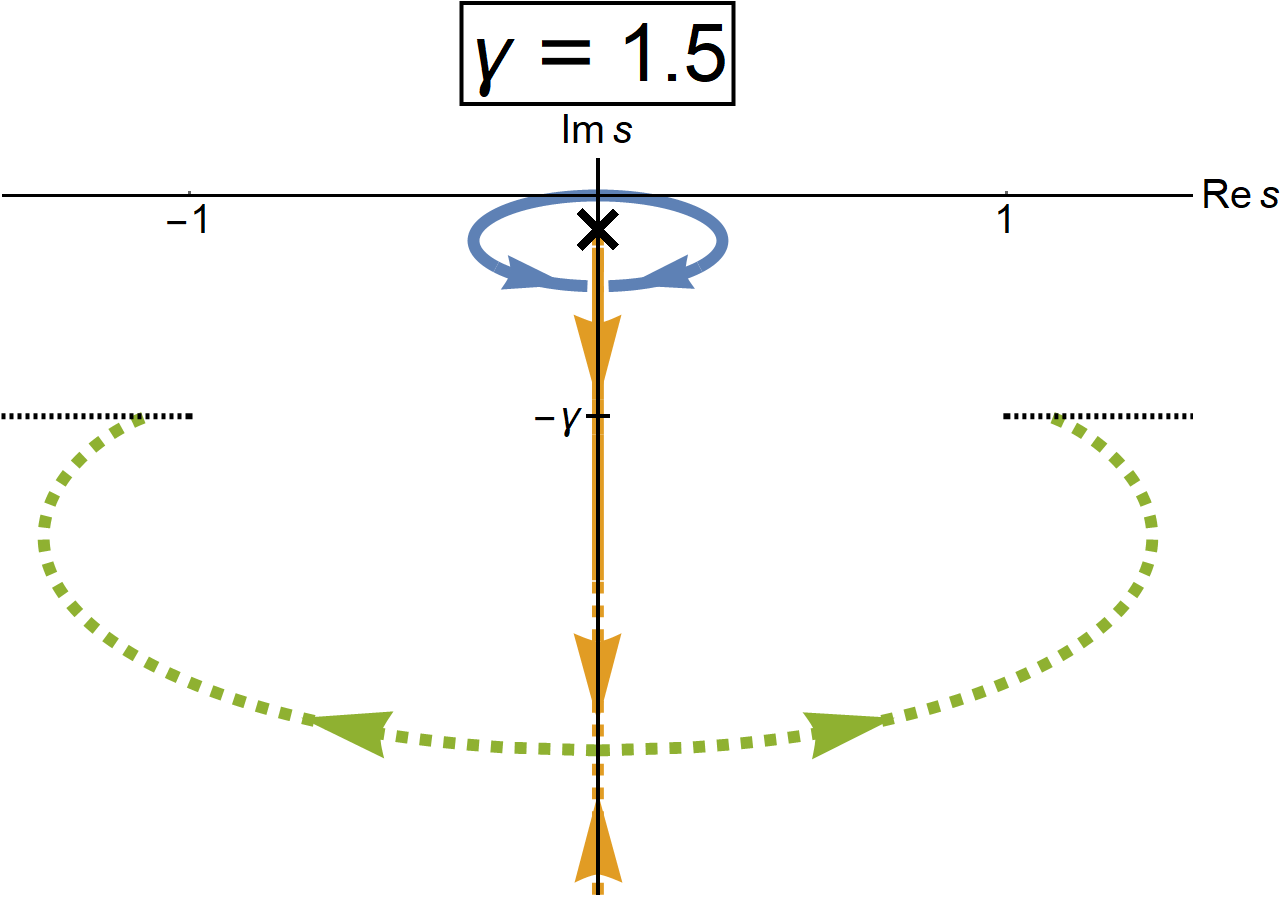}}
  \caption{(color online)
    Evolution of the poles of $\chi_
    {\text{qp},
    1}^{\text{long}}(s)$ with the Landau parameter $F^\e_1$ for finite disorder,
parameterized by the dimensionless coupling constant $\gamma$, as specified in the legend.
       Like in Fig. \ref{fig:chi0-poles-gamma}, we use different colors to show how the poles merge and bifurcate.
    The
    $\boldsymbol\times$
    denotes the limiting position of the pole for $F
    ^\e
    _1 \to \infty$. The inset in the first panel depicts how, for small $\gamma$, the pole
    bypasses the $s = 1$ branching point before finally moving to an unphysical Riemann sheet for  $s > 2/\sqrt{3}$ ($F^\e_1 > 3/5$).}
   \label{fig:chi1-poles-gamma}
\end{figure*}
\begin{figure}
  \centering
  \includegraphics[width=\hsize]{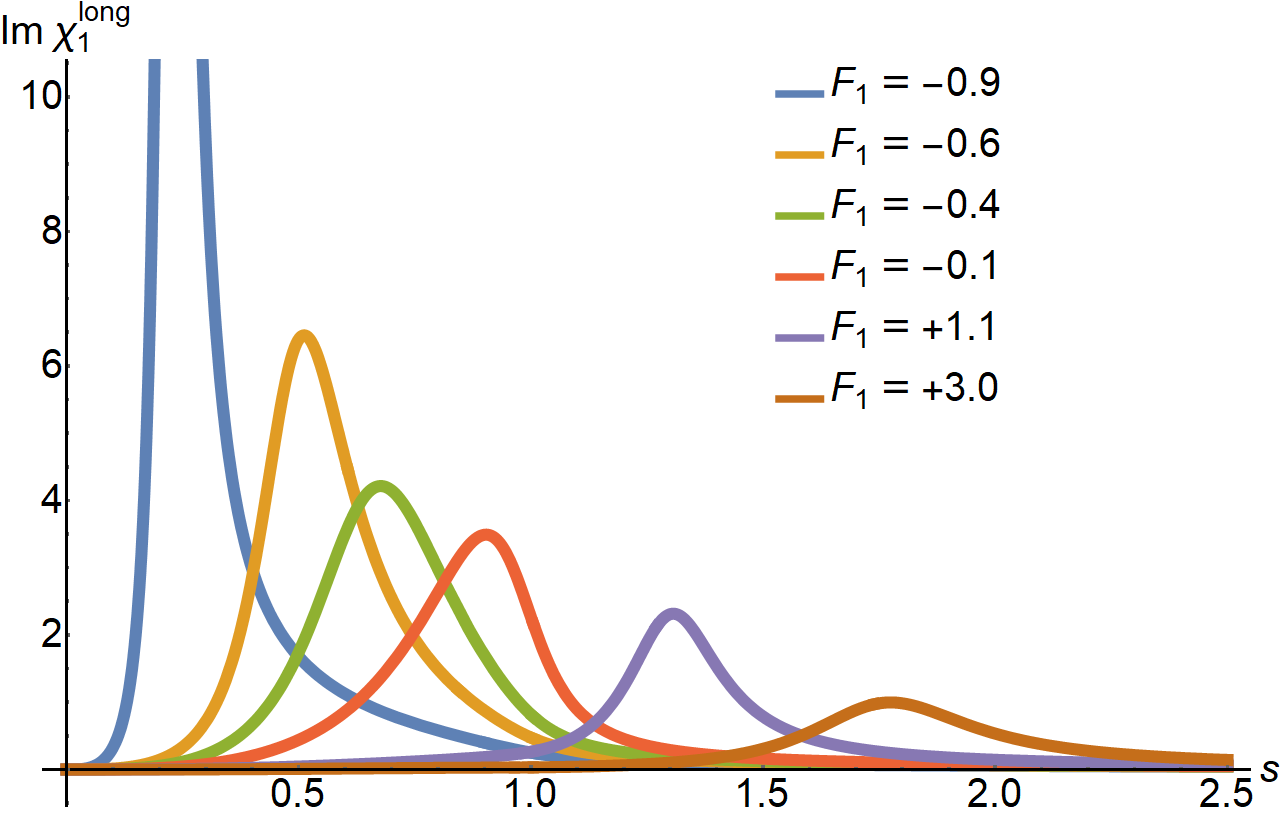}
  \caption{(color online) $\text{Im}\chi_{
  \text{qp},
  1}^{\text{long}}(s)$ for finite disorder $\gamma = 1/10$ and various interaction strengths.
 \label{fig:im-chi1-gamma}}
\end{figure}

We start with recalling the situation at vanishingly small damping. For $-1<F^\e_1 <0$, the poles of $\chi^{\text{long}}_{\text{qp},1} (s)$ are at $s_{1,2} = \pm a_1 -ib_1$, where $a_1$ and $b_1$ are given by Eq.~(\ref{na_2}). For $F_1^\e$ just above $-1$, $a _1  \approx ((1-|F^\e_1|)/2)^{1/2}$ and $b_1 \approx (1-|F^\e_1|)/4$, i.e., the the poles are almost on the real axis. For larger $F^\e_1$ (smaller $|F^\e_1|$), the two poles evolve such that $a_1$ increases monotonically, while $b_1$ first increases and then decreases. The poles approach the lower edges of the branch cuts along the real axis at $ F^\e_{cr,1}  =-1/9$, when $a_1 =  2/\sqrt{3} >1$. For larger $ F^\e_1 $, the poles remain slightly below the lower edge of the branch cut and move towards  $a_1 = \pm 1$. For $  0<  F^\e_1<  3/5$ the poles are located slightly above the branch cuts. 
For $F^\e_1>3/5$ they
 move off
from
 the physical Riemann sheet.
 For all $F^\e_1 >0$, there exists another,  purely imaginary solution  $s= -i s_{i,1}$. For small damping, this solution and the $s_{1,2}$ solution are not connected.

We now show that the behavior of $\chi^{\text{long}}_1$ changes qualitatively in the presence of disorder. The bubble and vertex parts of the free-fermion susceptibility are now given by
\begin{eqnarray}
  \chi^{\text{long,B}}_{\text{free},1} (s)  & = & 1 +  2  s (s+ i \gamma)\left(
  1+i \frac{
  s+i \gamma
  }{\sqrt{1 - (s+i\gamma)^2}}\right),\nn                                                                                                                                                                                                                                    \\
  \chi^{\text{long,V}}_{\text{free},1} (s)  & = & -2i\gamma s\frac{\left[\sqrt{1-(s+i\gamma)^2}+i(s+i\gamma)\right]^2}{\sqrt{1-(s+i\gamma)^2}}\nn                                                                                                                           \\
                                            &   & \times\frac{1}{\sqrt{1-(s+i\gamma)^2}-\gamma}.
\label{naa_3_1}
\end{eqnarray}
Adding these up, we obtain
\begin{eqnarray}
\chi^{\text{long}}_{\text{free},1} (s)=1+2s^2\frac{\sqrt{1-(s+i\gamma)^2}+i(s+i\gamma)}{\sqrt{1-(s+i\gamma)^2}-\gamma},\label{na_3}
\end{eqnarray}
which is the result quoted in Eq.~(\ref{ret1a}), up to a replacement $\gamma\to\delta$. Note that the vertex part vanishes at $q\to 0$, i.e., at $s\to\infty$, while the bubble part is reduced to a form which is identical to the Drude conductivity at finite frequency $\omega$. This indicates that the charge and spin currents are not conserved in the presence of disorder.

The analysis of the evolution of the poles with $F^\e_1$ for different $\gamma$ is straightforward but somewhat involved.  We omit the details of the calculations and present only the results. These results are summarized graphically in the panels of Fig. \ref{fig:im-chi1-gamma}.

 The beginning stage of the evolution is the same for all $\gamma$:
  for
   $1 + F^\e_1 \ll 1$, the poles are located at $s_{1,2} = \pm a_1 - i b_1$, where $a_1 \approx ((1 + F^\e_1)/2)^{1/2}$ is independent of
    $\gamma$, and $b_1 \approx ((1 + F^\e_1)/4) (\sqrt{1 + \gamma^2} + \gamma)$. However, the behavior at larger $F^\e_1$ depends strongly on $\gamma$. We find that there are three values of $\gamma$, at which the evolution of the poles changes qualitatively: $\gamma =1/2$, $\gamma = 0.923$, and $\gamma=1$.

 For
 $\gamma <1/2$ the evolution of the poles is similar to that for vanishingly small $\gamma$ (see Fig. \ref{fig:im-chi1-clean}), although the interval, where the imaginary part of the pole frequency varies non-monotonically with $F_1^\e$, shrinks rapidly with increasing $\gamma$. The pole positions $s_{1,2} = \pm a_1 -ib_1$ cross the line $s = -i\gamma$ first at some negative $F^\e_1 $, when $a_1 = \sqrt{1- (1 -\gamma)^2}$, and then again at $F^\e_1 =0$, when
 $a_1 = \sqrt{1-\gamma^2}$. The pole moves to the other,
  unphysical Riemann sheet at $F^\e_{1,R}$ given by
 \begin{eqnarray}
  \frac{1+ F^\e_{1,R}}{2 F^\e_{1,R}} = \frac{(a^2_{1.R} - \gamma^2)(a_{1,R} - \sqrt{a^2_{1,R}-1})}{\sqrt{a^2_{1,R}-1}},
  \end{eqnarray}
  where
  \begin{eqnarray}
   a_{1,R} = \left(\frac{2 \sqrt{\gamma^4-\gamma^2 +1}+ 2- \gamma^2}{3}\right)^{1/2}.\label{a1R}
  \label{last_2}
  \end{eqnarray}
  The values of $F^\e_{1,R}$ and of $a_{1,R}$ decrease as $\gamma$ increases, but
  $F^\e_{1,R}
   $ remains positive, and $a_{1,R}
   $ remains larger than $1$ as long as $\gamma <1$.
   For large $F^\e_1$ the pole on the unphysical sheet is at $s_{1} \approx \left(\sqrt{3F_1^\e} - i\gamma F^\e_1\right)/2$,
   i.e.
   $\text{Im}s_1$
   increases with $F^\e_1$.

   The purely imaginary pole  $s=-is_{i,1}$, which exists only for $F_1^\e>0$, moves up the imaginary axis from  $s_{i,1} \approx
  1/
  2 (F^\e_1)^{1/2}
  \gg 1
  $ for small $F^\e_1
  >
  0$ towards smaller values for larger $F^\e_1
  $. For  $F^\e_1 \gg 1$, $s_{i,1}$ is determined from the equation
  \begin{eqnarray}
    s^2_{i,1}\left(
    1+ \frac{s_{i,1}}
    {\sqrt{1 + (s_{i,1} -\gamma)^2} -\gamma}
    \right)
    =1/2.
  \end{eqnarray}
  For $\gamma <1/2$, this limiting value $s_{i,1} > \gamma$.

  At $\gamma =1/2$, the points at which the $s_{1,2}$ poles cross the $s = -i\gamma$ line merge at $F^\e_1 =0$, and the region of the non-monotonic evolution of $s_{1,2}$ for negative $F^\e_1$ disappears.
  At this $\gamma$, the limiting value of the purely imaginary pole at $F^\e_1 \gg 1$ becomes $s_{i,1} = \gamma =1/2$.

  For $1/2< \gamma < 0.923$, the poles evolve in the complex plane as shown in Fig \ref{fig:im-chi1-gamma}, third panel.  The $s_{1,2}$ poles cross the line $s = - i \gamma$ first at $F^\e_1 =0$, when
  $a_1 = \sqrt{1-\gamma^2}$, and then again at some positive $F^\e_1$, when $a_1 = \sqrt{1- (1 -\gamma)^2}$.  The limiting value of the purely imaginary pole for  $F^\e_1 \gg 1$ is now smaller than   $\gamma$. This implies that, for large $F^\e_1$, this pole gives the main contribution to
$\chi^{\text{long}}_{1} (t)$ in the time domain.

At $\gamma = 0.923$, the $s_{1,2}$ poles touch the imaginary axis of $s$ at $F^\e_1 \approx 0.031$. The corresponding value of $ a_1
= 1.391$.  At this  $F^\e_1$, the purely imaginary pole is located at the same point on the imaginary axis i.e., there are three degenerate solutions.

For  $0.923 < \gamma < 1$, the poles, which initially move away from the imaginary axis, return to this axis at some positive value of
$F^\e_1
   $, at which the purely imaginary is still located at a higher point on the imaginary axis (see Fig. \ref{fig:im-chi1-gamma}, fourth panel). The subsequent evolution with increasing $F^\e_1$ involves two bifurcations.
   After the second bifurcations, the two solutions approach the upper edge of the branch cut and move to a different Riemann sheet at $F^\e_1 = F^\e_{1,R}$ given by Eq.~(\ref{last_2}).

At $\gamma =1$, the first bifurcation occurs at $F^\e_1 =0$, at the point $s = - i \gamma$. After the bifurcation, one solution moves up along the imaginary axis, while another moves down. The one that moves up eventually reaches the point
$s = - 0.39 i$ at $F^\e_1 \gg 1$. The second bifurcation occurs at $F^\e_1 =0.022$ at the point $s=-2i \gamma$.  After that, the two solutions $s_{1,2} = \pm a_1 - i b_1$ move towards the end point of the branch cut and reach $a_1 = 1, b_1 = \gamma$ at $F^\e_1 = 1/3$.

For $\gamma >1$, the first bifurcation happens at $F^\e_1 <0$ and, after bifurcation, the first (second) solution moves up (down)
the imaginary axis.  At $F^\e_1 =0$, these two solutions are at $s=-i (\gamma \pm \sqrt{\gamma^2-1})$. For $F^\e_1 >0$, the third solution emerges on the imaginary axis and, eventually, it merges with the solution that moves down (see Fig. \ref{fig:im-chi1-gamma}, fifth panel), that, the two solutions bifurcate and move towards the branch cut. In distinction to the case $\gamma <1$, now they merge with the lower edge of the branch cut at $F^\e_1 = {\bar F}^\e_{1,R}$, where
 \begin{eqnarray}
 \frac{1+ {\bar F}^\e_{1,R}}{2 {\bar F}^\e_{1,R}} = \frac{({\bar a}^2_{1,R} - \gamma^2)({\bar a}_{1,R} + \sqrt{{\bar a}^2_1-1})}{\sqrt{{\bar a}^2_{1,R}-1}}
  \label{last_3}
  \end{eqnarray}
and ${\bar a}_{1,R}$ is given by Eq.~(\ref{a1R}.)
For large $\gamma$,
 ${\bar a}_{1,R} \approx \gamma/\sqrt{3}$ and ${\bar F}_{1,R} \approx 3/
 8 \gamma^2
  \ll 1$. For
  $F^\e_1 > {\bar F}^\e_{1,R}$ the poles again move to a different Riemann sheet. The solution that moves up the imaginary axis survives for all $F^\e_1 >0$ and, for $F^\e_1 \gg 1$ and $\gamma \gg 1$, it approaches the point
   $s \approx -
  i/
  2\gamma
  $.

We note that there are certain similarities between the evolution of the poles with
$ F^\e_1
$ and the behavior of the plasmon modes in a 2D electron gas with conductivity exceeding the speed of light. This problem was studied some time ago \cite{FalKo1989} and has recently been re-visited in Ref.~\onlinecite{Oriekhov2019}.

\subsubsection{$l=1$, transverse channel}

   For vanishingly weak damping ($\gamma\to 0$), the pole moves along the imaginary axis ($s = -i s_i$) for $-1 < F^\e_1 <0$,  towards larger $s_i$, as $|F^\e_1|$ decreases. At  $F^\e_1 =0+$, $s_i$ tends to infinity. For positive $F^\e_1$, there is no pole on our Riemann sheet.

For finite $\gamma$, the evolution remains essentially the same. There are still no solutions for
$F^\e_1 >0$, while for
$-1 < F^\e_1 <0$ the pole is on the imaginary axis, at
 $s = - i s_{i,1}$, where
\begin{equation}
s_{i,1} = \frac{S}{1 + 2S}  \left(\sqrt{ \gamma^2 + (1 + 2S)^2} +  \gamma \right),~~ S = \frac{1 - |F^\e_1|}{|F^\e_1|}.
\label{last_5}
\end{equation}
At
$F^\e_1 \approx -1$, $s_{i,1} \approx (1- |F^\e_1|) (\sqrt{1 + \gamma^2} + \gamma)/2$. Note that there is no diffusive behavior for large $\gamma$. In this limit, $s_{i,1} \approx \gamma (1- |F^\e_1|)$, i.e., $
\omega
 \approx - i
 \tilde
 \gamma (1- |F^\e_1|)$, where $\tilde\gamma$ is the dimensionful impurity scattering rate. At $F^\e_0 \to 0$, $s_{i,1} \approx 1/
2 |F^\e_1|^{1/2}
$ for all $\gamma$.

\subsection{$l=2$}
\subsubsection{$l=2$, longitudinal channel}
 For vanishingly weak damping ($\gamma\to 0$)
  and
   $F^\e_2 <0$, one of the poles is on the imaginary axis while the other one is in the complex plane. For small negative $F^\e_2$, the latter pole is at the lower edge of the branch cut. When $F^\e_2
     $ crosses zero, the pole bypasses the end point of the branch cut, moves slightly above it, and continues to stay there as  $F^\e_2$ increases from $0$ up to
     $F^\e_2 \approx 0.4$. For larger $F^\e_2$, the pole is located on an unphysical Riemann sheet.

  For finite $\gamma$, the poles are determined from the equation
  \begin{widetext}
    \begin{equation}
\frac{1 +F^\e_2}{2F^\e_2} =
 \frac{is}{\sqrt{1-(s+ i \gamma)^2} -  \gamma} \left(s + i \gamma -i \sqrt{1-(s+ i \gamma)^2}\right)
 ^2
  \left(1 -2 s (s + i \gamma)\right)
\label{n_6_1_1}.
\end{equation}
\end{widetext}
As for the
$l=1$ case, the behavior of the poles is quite involved, particularly for $\gamma >1$, and we refrain from presenting all the details. We note only that at
$ F^\e_2
\approx -1
$ the purely imaginary pole is located at $s \approx -
i (1
 -
 | F^\e_2|) (\sqrt{1 + \gamma^2} + \gamma)/2
 $. For large $\gamma$, $s \approx -
 i (1
 -|F^\e_2|) \gamma$, i.e., $
 \omega
 \approx -
 i (1
 -
 |F^\e_2|)
 \tilde\gamma
 $. This pole is not a diffusive one, which to be is expected because the
  $l=2$ order parameter is not a conserved quantity.

\subsubsection{$l=2$, transverse channel}
 For vanishingly weak damping ($\gamma\to 0$), the poles $s = \pm a_2 -i b_2$ are in the complex plane of $s$ for negative $F^\e_2$. As $F_2^\e$ increases from $-1$ towards $0$, the poles move from the vicinity of the real axis at $F_1^\e\approx -1$ ($a_2 \sim (1
-
 |F^\e_2|)^{1/2} \gg b_2$ )
 towards $ a_2  \approx b_2 \approx
1/ 2\sqrt{2}|F^\e_2|^{1/4}
$ at
$0<-F^\e_2\ll 1$. For positive $F^\e_2$, there is only a single pole on the imaginary axis.

For finite $\gamma$, the equation for the pole is
\begin{widetext}
\begin{equation}
 s (s+i \gamma) \left(
 s+i\gamma-i\sqrt{1-(s+i \gamma)^2}
\right)^2 = - \frac{1 + F^\e_2}{F^\e_2}.
\label{ssss1_n_1}
\end{equation}
\end{widetext}
It has two solutions. At small $1+F^\e_2$ both
 are on the imaginary axis: one is
$a_2 =0, b_2 \approx (1
-
|F^\e_2|
) (\sqrt{1+ \gamma^2} + \gamma)^2/
4 \gamma
$ and another one is
 $a_2 =0, b_2 \approx - \gamma$. Note that neither mode is diffusive for large $\gamma$.  As $1 + F^\e_2$ increases, the two solutions moves towards each other and merge at some critical value
 $F^\e_2=F^\e_{2,\text{cr}}$. For small $\gamma$,
   $F^\e_{2,\text{cr}} \approx -1 + \gamma^2$ and the solutions merge at  $b_2 \approx \gamma/2$. At $ F^\e_2
   =
   F^\e_{2,cr} + 0$, the poles split and move away from the imaginary axis, i.e.,
  $a_2$ becomes finite. The subsequent evolution is essentially the same as for vanishingly small $\gamma$. For large positive $F^\e_2$, the pole is located on the imaginary axis at $b_2 = 1/
   2/\sqrt{2}
   $ for small $\gamma$ and at $b_2 = \gamma +1/(4\gamma)$ for large $\gamma$.

   \section{
Susceptibility in the time domain}
\label{sec:time}
\subsection{General results}

In this section we study the real-time response of an order parameter on both sides of the Pomeranchuk transition by analyzing the susceptibility in the time domain $\chi_l^\e(q,t)$. For definiteness we consider
  $l=0$ and the longitudinal channel for $l=1$. In both cases,
 \begin{equation}
 \chi^\e_l (q, t) = \int_{-\infty}^\infty  \frac{d\omega}{2\pi}  \chi
 ^\e_l (q, \omega) e^{-i \omega t},
 \end{equation}
 where $ \chi^\e_l (q, \omega)$ is the retarded susceptibility. Introducing $t^* = v^*_F q t$ and going over from integration over $\omega$ to integration over $s = \omega/v^*_F q$, we obtain
\begin{equation}
  \label{eq:fourier-s}
  \chi^{\e}_l (t^*)  \equiv  \frac{1}{v^*_F q}\chi^{\e}_l (q,t)=
  \int_{-\infty}^\infty \frac{d s}{2\pi}  \chi^{\e}_l (s) e^{-i s t^*}.
\end{equation}
The time-dependent
$\chi^{\e}_l (t^*)$ can be measured in pump-probe experiment, by applying an instantaneous perturbation $h_l (t^*) = h \delta (t^*) \Delta^\e_l$  with the symmetry of the Pomeranchuk order parameter, to momentarily move the system away from the FL state without Pomeranchuk order
 (here $\delta (...)$ is the $\delta$-function).
  The order parameter $\Delta^\e_l (t^*)$ will then relax to zero as $\Delta^\e_l (t^*) \propto h \chi^{\e}_l (t^*)$, if
    $1 + F^\e_l >0$, and will grow with time, if
     $1 + F^\e_l <0$.

Causality requires that $\chi^{\e}_l (t^* < 0) = 0$. The vanishing of $\chi^{\e}_l (t^*)$ for $t^* < 0
$ is guaranteed because the poles and branch cuts of the retarded susceptibility
$ \chi^\e_l (s)$ are located in the lower frequency half-plane.
For $t^* <0$, $e^{-i s t^*}$ vanishes at $s\to i\infty$,
  and the integration contour can be closed in the upper half-plane of complex $s$, where $\chi^{\e}_l (s)$ is analytic.
  The integral over $s$ in Eq.~(\ref{eq:fourier-s})
   then vanishes. For $t^* >0$, the integration contour should be closed in the lower-half-plane of $s$, where $\chi^{\e}_l$ has both poles and branch cuts. In this situation,  $\chi^{\e}_l (t^*)$  is finite.

The susceptibility in the time domain can be obtained either by contour integration or directly, by using the form of $\chi^\e_l (s)$ above the branch cut and integrating over real $s$. In a clean system Eq.~(\ref{eq:fourier-s}) can be re-written as
\begin{widetext}
  \begin{eqnarray}
\chi^{\e}_l (t^* >0)                        & = & \frac{1}{\pi} \int_{0}^\infty d s  \left( \text{Re} \chi^{\e}_l (s) \cos{s t^*} + \text{Im} \chi^{\e}_l (s) \sin{s t^*}\right)
= \frac{2}{\pi} \int_{0}^1 d s  \text{Im} \chi^{\e}_l (s) \sin{s t^*}.
\label{naaa_7}
\end{eqnarray}
\end{widetext}
   \begin{figure}
   \centering
   \includegraphics[width=0.6\hsize]{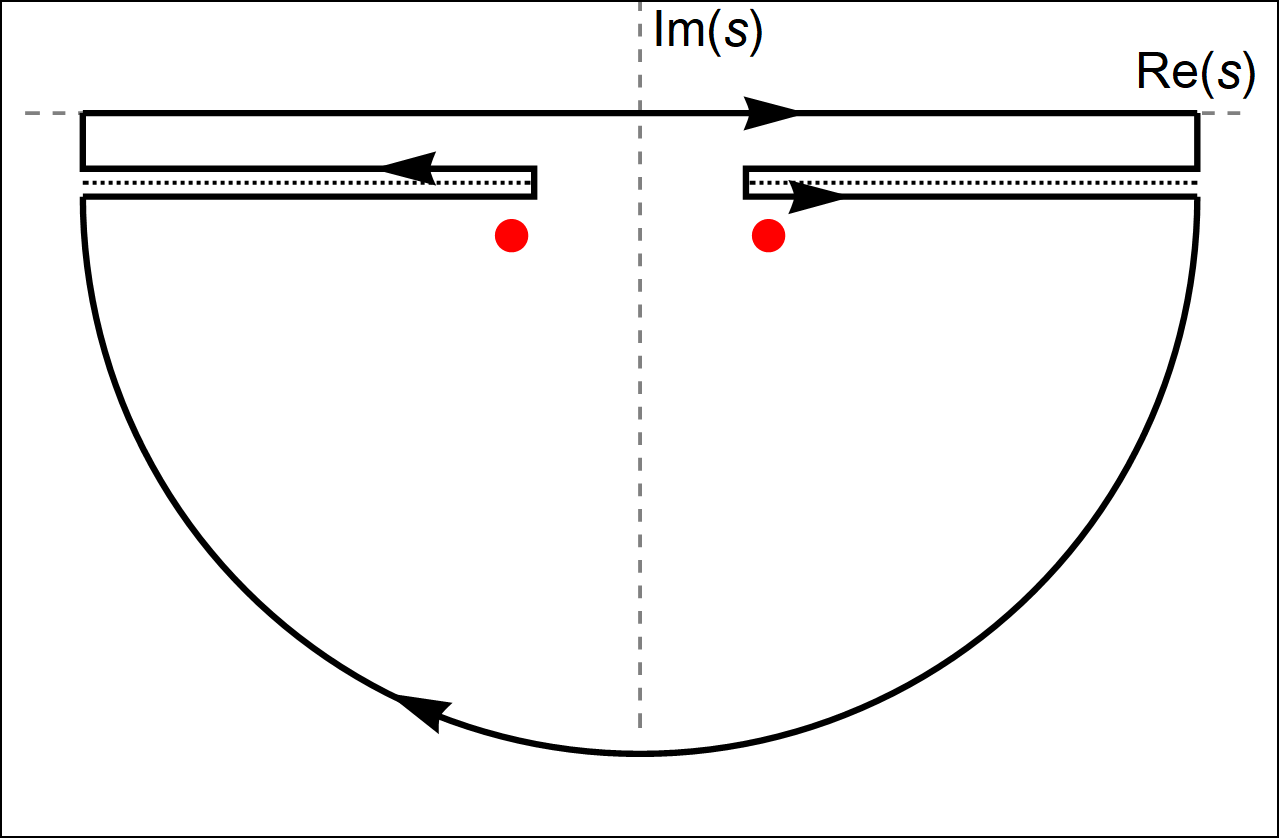}
   \caption{(color online) Integration contour for evaluation of $\chi_l^\e(t^*)$,
   defined in Eq.~(\ref{eq:fourier-s}).
The contour is shown for the case of finite disorder, and the branch cuts are at $s = -i\gamma + x, |x| > 1$. The poles are located at finite distance below the lower edges of the branch cuts.}
   \label{chi1-int-contour}
 \end{figure}

 In the last line we used that $\chi^{\e}_1 (t^* <0) =0$ and that $\text{Im} \chi^{\e}_l (s)$ is non-zero only for $|s| <1$. Equation (\ref{naaa_7}) is convenient for numerical calculations.
 To analyze of the behavior of $\chi^{\e}_1 (t^*)$
analytically,
   it is more convenient to integrate over the contour shown in Fig. \ref{chi1-int-contour}, and evaluate the contributions from the poles and branch cuts. This way, we get
 \begin{equation}
 \chi^\e_l (t^*) =  \chi_{\text{pole}, l
 } (t^*) - \chi_{\text{bcut}, l
 } (t^*),
 \label{naaa_8}
 \end{equation}
  where
$ \chi_{\text{pole},l} (t^*)$ is the sum of the residues of the poles,
 multiplied by $-i$,
  and
  \begin{eqnarray}
 \chi_{\text{bcut},l} (t^*)                 & = &
 \int_1^\infty \frac{dx}{\pi} \cos{x t^*}\label{naaa_8_2}                                                                                                                                                                                                                   \\
                                            &   & \times
 \left(\chi^\e_l (x - i \delta- i \epsilon) - \chi^\e_l (x - i \delta + i \epsilon)\right),\nn
 \end{eqnarray}
where
   $\epsilon=
  0^+$ is the combined contribution from the two edges of the branch cut along $
 |x| >1$. (By $\chi^\e_l (x - i \delta- i \epsilon)$,
  we mean the retarded susceptibility
 $\chi^\e_l (s)$ computed at
  $s = x - i \delta- i \epsilon$, where $x$ is a real variable and $0 < \epsilon \ll \delta$).

 In what follows, we focus on the on the behavior of  $\chi^\e_l (t^*)$ near the Pomeranchuk instability for $l=0$ and $l=1$, when the corresponding $ F^\e_l
  \approx - 1$.  The analysis of
  $\chi^\e_l (t^*)$ for larger $ F^\e_l
  $ requires a separate discussion, particularly when the poles in $\chi^\e_l (s)$ are near the lower edges of the branch cuts, and will be presented elsewhere \cite{KMC2019}.
  We will analyze the behavior of $\chi^\e_l (t^*)$ at large $t^* \gg 1$. For such $t^*$, the dominant contribution to $\chi^\e_l (t^*)$ comes from the quasiparticle part of the susceptibility (the first term in Eq. (\ref{new_1})),  the contribution from the incoherent part of the susceptibility is much smaller.  Accordingly, $\chi^\e_l (t^*) = (\Lambda^\e_l)^2 \chi^\e_{\text{qp},l} (t^*)$, where $\chi^\e_{\text{qp},l} (t^*)$ is the Fourier transform of $\chi^\e_{\text{qp},l} (s)$.   In this section we consider a generic case when $\Lambda^\e_l$ is finite near a Pomeranchuk transition, and focus on $\chi^\e_{\text{qp},l} (t^*)$. In the next section we consider the special case of $l=1$ charge/spin current order parameter, for which $\Lambda^\e_1$ vanishes at a Pomeranchuk transition.  To simplify the expressions, below we write $\chi^\e_{\text{qp},l} (t^*)$ simply as $\chi^\e_{l} (t^*)$.

\subsection{l=0}
\label{sec:chit0}

We recall that near the Pomeranchuk transition the only pole of
  $\chi^\e_0 (s)$ in the lower half-plane is located at $s= s_i  \approx - i (1 -|F^\e_0|) $ (see Eq.~\ref{ac1}). Near this pole,
 \begin{equation}
 \chi^\e_0 (s)
  \approx \nu_F \frac{i
 }{
 s+ i(1-|F^\e_0|)}.
 \label{naaa_9}
 \end{equation}
Evaluating the residue, we obtain
\begin{eqnarray}
\chi^\e_{\text{pole},0} (t^*) = \nu_F e^{-t^* (1 - |F^\e_l|)}.\label{pole}
\end{eqnarray}
To obtain the branch cut contribution, we recall that
\begin{equation}
\chi^\e_l (x - i \delta \mp i \epsilon) = \nu_F \frac{1 \pm \frac{x}{\sqrt{x^2-1}}}{1 -|F^\e_l| \left(1 \pm \frac{x}{\sqrt{x^2-1}}\right)}
\end{equation}
Hence
 \begin{eqnarray}
                                            &   &
\chi^\e_0 (x - i \delta - i \epsilon) - \chi^\e_0 (x - i \delta+ i \epsilon) \nn                                                                                                                                                                                            \\
                                            &   & = -2 \nu_F
\frac{x\sqrt{x^2-1} }{(1-|F^\e_0|)^2 + x^2 (2 |F^\e_0|-1)}
 \label{naaa_10}
 \end{eqnarray}
 For large $x$, the r.h.s. of Eq.~(\ref{naaa_10})
approaches a constant value ($=-2$), and the integral over $x$ in
(\ref{naaa_8_2}) formally diverges. This divergence is artificial and can be eliminated by introducing a factor of
 $\exp(-\alpha x)$  with
$\alpha >0$ and taking the limit of
 $\alpha
 \to 0$ at the end of the calculation.

 For $ F^\e_0
 \approx -1$, the leading contribution to the integral in Eq.~(\ref{naaa_8_2})
  comes from non-analyticity of the integrand at
  $x=1$. For $t^* \gg 1$, we use $\int dy \sqrt{y} \cos{y t^*} =  - \sqrt{\pi}/(2t^*)^{3/2}$, $\int dy \sqrt{y} \sin{y t^*} =  \sqrt{\pi}/(2t^*)^{3/2}$ and obtain
 \begin{equation}
 \label{eq:chit-power}
\chi_{\text{bcut},0} (t^*)
= - \nu_F \sqrt{\frac{2}{\pi }}\frac{\cos(t^* - \pi/4)}{(F_0^\e)^2(t^*)^{3/2}}.
\end{equation}
Comparing Eqs.~(\ref{pole}) and (\ref{eq:chit-power}), we see that the pole contribution is the dominant one
for
$1 \ll t^* \ll (3/2) |
\ln
(1-|F^\e_l|)|/(1 -|F^\e_l|)$, while at longer times the time-dependence of the response function comes from the end point of the branch cut.

\begin{figure*}
  \centering
  \includegraphics[width=0.48\hsize,clip,trim=0 00 0 0]{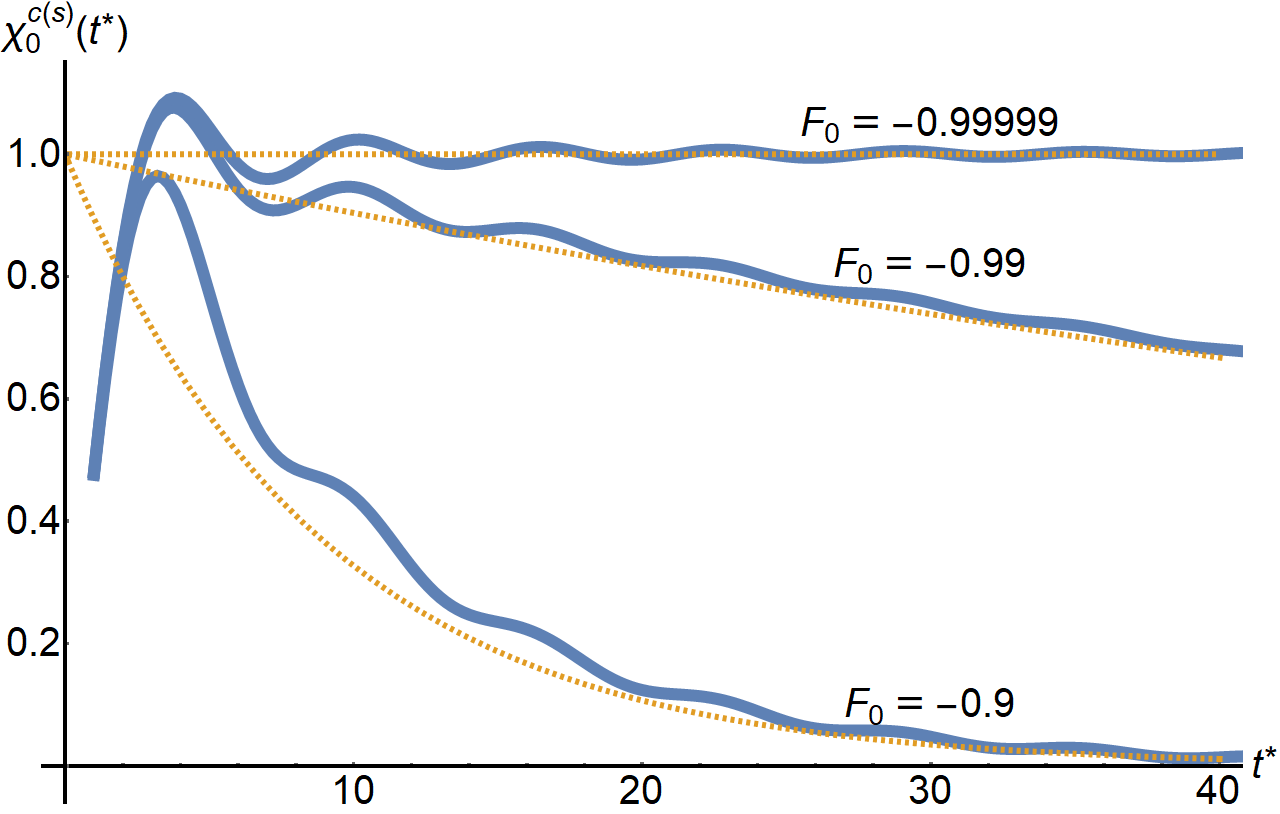}\hfill
  \includegraphics[width=0.48\hsize,clip,trim=0 0 0 0]{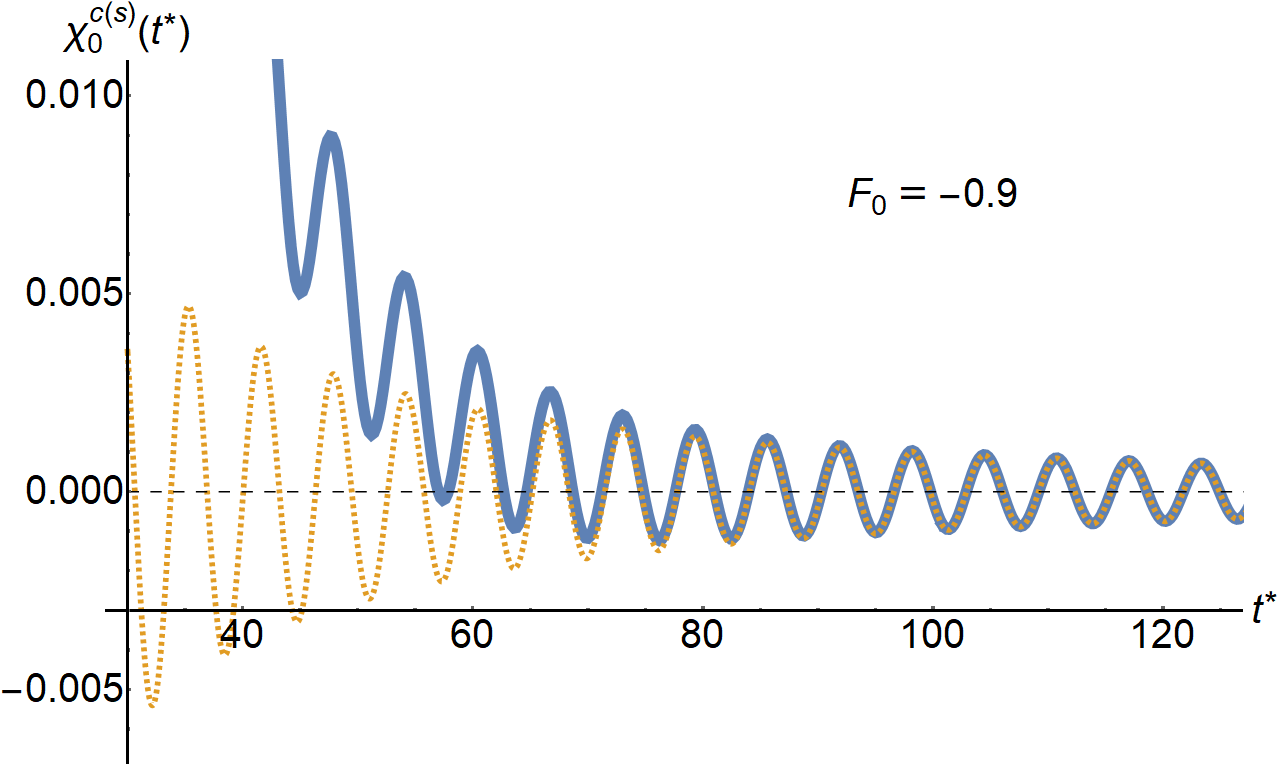}
  \caption{
  (color online) $\chi_{\text{qp},0}^\e(t^*)
$
(with $\nu_F = 1$)
  as a function of the dimensionless time $t^*$, as defined in Eq.~(\ref{eq:fourier-s}),
  for $F_0^\e$ near $-1$. The solid line is the numerically computed response and the dashed lines are the analytic expressions in Eqs.~(\ref{naaa_9})
  and (\ref{eq:chit-power}).
  Left:
  $\chi_{\text{qp},0}^\e(t^*)$
  at short and intermediate times.
  At intermediate time, the time dependence is dominated by the exponentially decaying pole contribution. As $F_0^\e$ approaches $-1$, the decay time goes to infinity. Right: long-time behavior, dominated by the oscillatory and power-law decaying contribution from the branch cut.
 \label{fig:chi1-time-1}}
\end{figure*}

In Fig.~\ref{fig:chi1-time-1} we show
$\chi^\e_0 (t^*)$ computed numerically using Eq.~(\ref{naaa_7}). As is obvious from this equation, $\chi^\e_0 (t^*)$ increases linearly with $t^*$ at short times $t^* \leq 1$ (the pole and branch contributions cancel each other at $t^*=0$). At intermediate times $1 \ll t^* \ll (3/2)
\ln
|{1-|F^\e_l|}|/(1 -|F^\e_l|)$, $\chi^\e_0 (t^*)$ exhibits an exponentially decay augmented by weak oscillations, in agreement with Eqs.~(\ref{pole}) and  (\ref{eq:chit-power}). This behavior is shown in the left panel of Fig.~\ref{fig:chi1-time-1}. At long times,   $\chi^\e_0 (t^*)$ oscillates and decreases algebraically with time, in agreement with Eq.~(\ref{eq:chit-power}). This behavior is shown in the right panel of Fig.~\ref{fig:chi1-time-1}.

As $F_0^\e$ becomes closer to $-1$, the exponential decay of $\chi^\e_0(t^*)$ with $t^*$ becomes slower and the crossover to a power-law behavior shifts to larger $t^*$.   Right at the Pomeranchuk instability, when $F^\e_0=-1$, the form $\chi^\e_0(t^*)$
  can be found directly from Eq.~(\ref{naaa_7}). In this case,
$\text{Im}\chi^\e_0(s)= \nu_F \theta(1-|s|)\sqrt{1-s^2}/{s}$. Substituting into (\ref{naaa_7}) we find that
 $\chi^\e_0(t^*)$ starts off linearly for $t^*\ll 1$, exhibits an oscillatory behavior for $t^*\sim 1$, and approaches the limiting value of $\chi^\e_0(t^*)=\nu_F $ at $t^*\to\infty$.

For $F^\e_0<-1$,
a long-range order develops. Within our approach, we can analyze the initial growth rate of the order parameter
$\Delta^{\e}_0 (t^*)$ induced by an instant perturbation $h (t^*) \propto h \delta (t^*)$ such that $\Delta^\e_0(t^*) \propto h \chi_0^\e(t^*)$.
The computation of $ \chi^{\e}_0 (t^*)$ for
$ F^\e_0
<-1$ requires some care because integrating Eq.~(\ref{eq:fourier-s}) over the same contour as in Fig.~\ref{chi1-int-contour}
we would find that $\chi^\e_0
(t^* <0)$ becomes finite, i.e., that causality is lost. This issue was analyzed in Ref.~\onlinecite{Pethick1988}
(see also, e.g., Ref.~\onlinecite{Wyld1999}), where it was shown that, to preserve causality, one has to modify the integration contour such that it goes above all poles, as shown in Fig.~\ref{fig:upper}. Integrating along the modified contour, we find that $\chi^{\e}_0 (t^* <0)=0$, as required by causality. For $t^*>0$ we now have
\begin{equation}
  \Delta^{\e}_0 (t^*)
  \propto h e^{t^* (|F^\e_
  0
  |-1)},
  \label{t_2_1}
\end{equation}
i.e., a perturbation grows exponentially with time. This obviously indicates that the FL state without Pomeranchuk order becomes unstable. To see how the system eventually relaxes to the final equilibrium state with $\Delta^{\e}_0 (t^*) = \Delta_0$, we would need to re-calculate $\chi^{\e}_0 (t)$ in the broken-symmetry state.

\subsubsection{
$l=1$, longitudinal channel}
For small positive $1 + F^\e_1$,  the poles of $\chi^{\text{long}}_1 (s)$ are given by Eq.~(\ref{F1-1}). Near the poles,
\begin{equation}
  \chi^{\text{long}
  }_1 (s)
   \propto   \frac{1}{(s-s_1)(s-s_2)} + \ldots
\end{equation}
where $\ldots$ stands for non-singular terms. Evaluating the residues, we obtain the pole contribution to $\chi^{
\text{long}}_1 (t^*)$ as
\begin{equation}
 \chi
 ^{\text{long}}
 _{\text{pole},1} (t^*) \propto t^*
 \exp
 \left(
- \frac{1
-|F^\e_1|}{4}t^*\right)
 \frac{\sin\left(\sqrt{\frac{1
 -|F^\e_1|}{2}} t^*\right)}{\sqrt{\frac{1
 -|F^\e_1|}{2}} t^*}.
  \label{t_1}
\end{equation}
The branch-cut contribution has the same structure as for $l=0$, i.e.,
$\chi^{\text{long}}_{\text{bcut},1} (t^*)  \propto
\cos(t^* - \pi/4)
/(t^*)^{3/2}$ for $t^* \gg1$. We see that the pole contribution remains dominant up to
$t^* \sim   |\ln (1-|F^\e_1)|/(1-|F^\e_1|)$,
which becomes progressively larger as $|F^\e_1|$ approaches one.
 We also see that the pole contribution contains two relevant scales
 \begin{equation}
t^*_a = \left[2/(1 -|F^\e_1|) \right]^{1/2} \;\text{and}\; t^*_b =4/(1-|F^\e_1|).
 \label{tab}
 \end{equation}
 Near the Pomeranchuk transition,
 $t^*_b \gg t^*_a
 \gg 1
 $.
 Applying an instant perturbation in the $l=1$ channel
 $h (t^*) \sim h \delta (t^*)$ and analyzing the behavior of
 $\Delta^{
 \text{long}}_1 (t^*) \propto h \chi^{
 \text{long}}_1 (t^*)$, we find that it grows linearly with $t^*$ for
   $t^*
  \ll t^*_a$,   i.e., the system initially tends to move further away from equilibrium. For $t^*_a \ll t^* \ll t^*_b$, the order parameter oscillates between the quasi-equilibrium states with
    $\Delta^{
    \text{long}}_1 (t^*) =\pm \Delta_{
    \text{q-eq}}$, where $\Delta_{
    \text{q-eq}}
     \propto h
    \left[2/(1+F^\e_1)
    \right]^{1/2}$. Finally, for $t^*
    \gg t^*_b$,   $\Delta^{
    \text{long}}_1 (t^*)$ decays exponentially towards zero. At $ F^\e_1 =-1
   $, both $t_a$ and $t_b$ diverge and, following an instant perturbation at $t=0$,    the order parameter
  $\Delta^{
    \text{long}}_1 (t^*) \propto h$ increases linearly with $t^*$
until the perturbation theory in $h$
breaks down.

The difference with the
$l=0$
case,
when  $\Delta^\e_0
(t^* \to\infty)$
 at $ F^\e_0 =-1$
is finite,
 can be understood by noticing that the behavior of $\chi^{\text{long}}_{\text{free},1}(t^*)$ for large  $t^*$ is determined by that of
 $\text{Im}\chi^{\text{long}}_{\text{free},1}(s)$ for small $s$. Equation (\ref{na_3}) shows that
 $\chi^{\text{long}}_{\text{free},1}(s)=1+2s(s+i
 \delta)$ for $|s|\ll 1$. At $F^\e_1=1$, therefore, we have  $\chi^{\text{long}}(s)
 _1
 \approx -1/2s(s+i
 \delta)$, and, at vanishingly small $\delta$,
 $\text{Im}\chi^{\text{long}}(s)$ goes over to $(\pi/2)\delta(s)/s$. Substituting this into Eq.~(\ref{naaa_7}), we find that $\chi^{\text{long}}_1(t^*)\propto t^*$.

  \begin{figure*}
  \centering
  \begin{subfigure}{0.3\hsize}
    \includegraphics[width=\hsize]{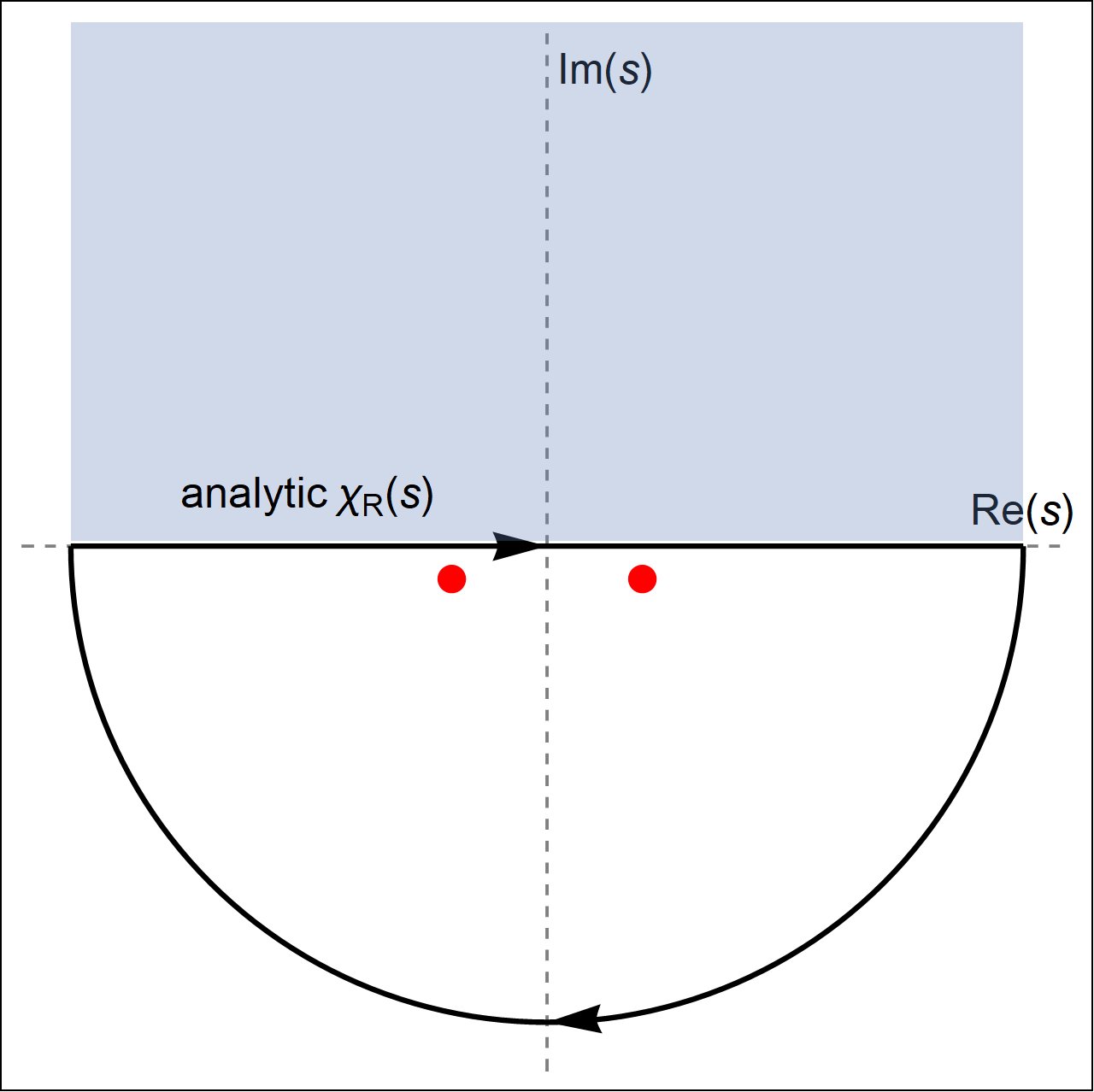}
    \caption{\label{fig:lower}}
  \end{subfigure}
    \begin{subfigure}{0.3\hsize}
      \includegraphics[width=\hsize]{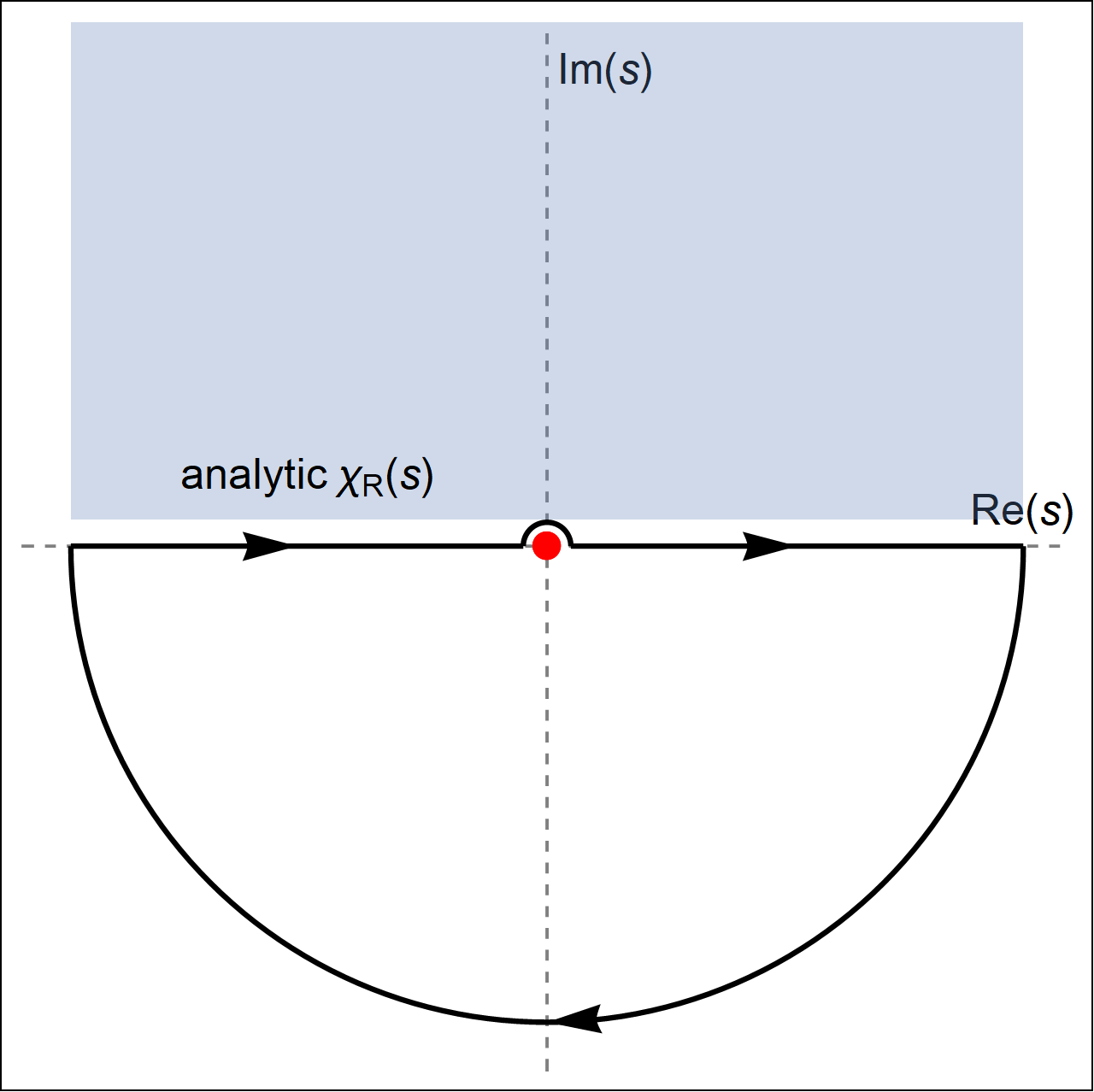}
      \caption{\label{fig:axis}}
  \end{subfigure}
    \begin{subfigure}{0.3\hsize}
      \includegraphics[width=\hsize]{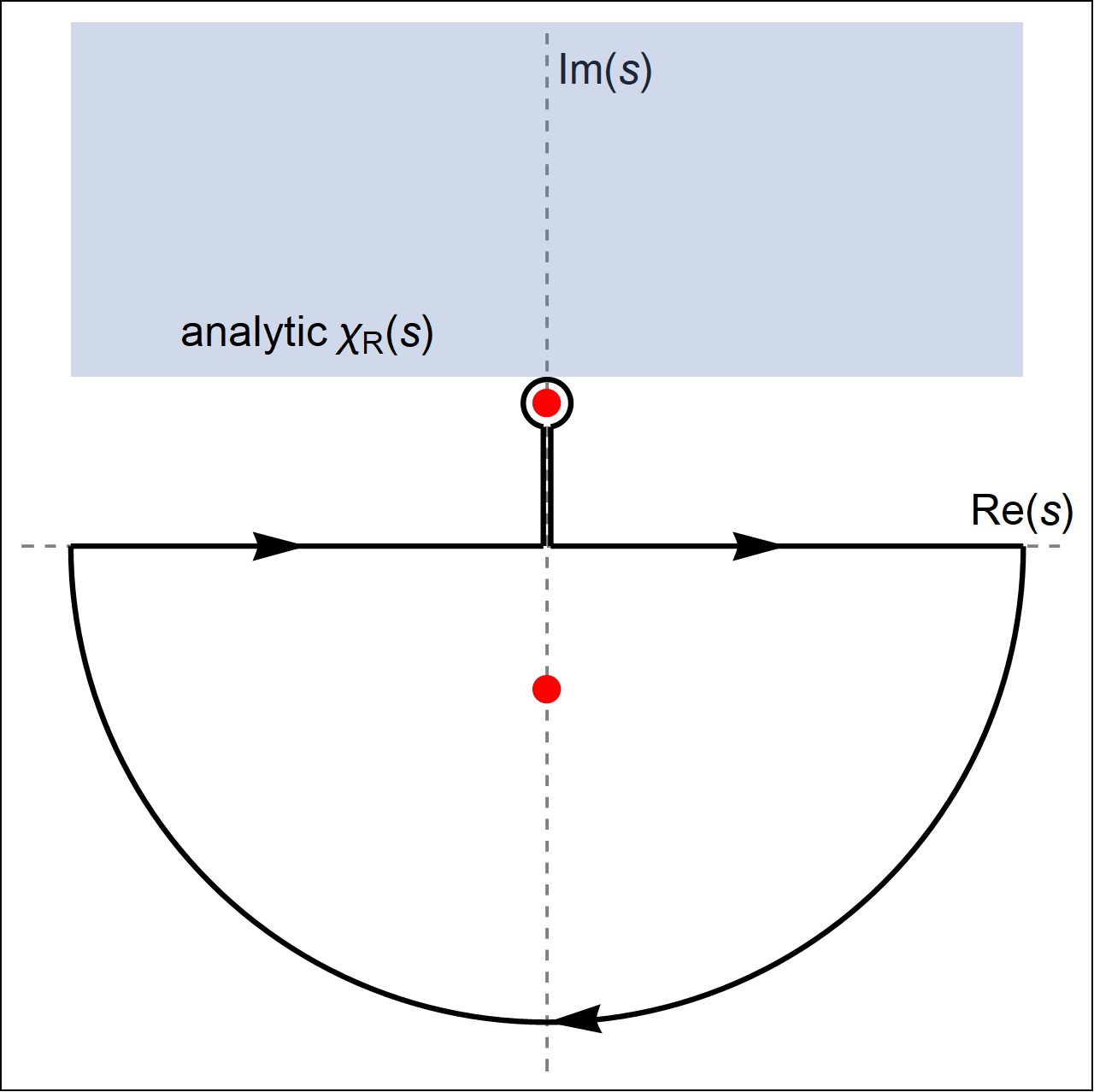}
      \caption{\label{fig:upper}}
  \end{subfigure}
  \caption{(color online)
  Positions of the poles and analyticity regions of $\chi_1^{\text{long}
  }(s)$ in the Fermi-liquid phase without Pomeranchuk order
   [$1+F_1^\e > 0$, panel (a)], at the transition point [$1+F_1^\e = 0$, panel (b)], and in the ordered phase [$1+F_1^\e < 0$, panel (c)]. In the ordered phase, the susceptibility in the time domain
     $\chi_1^{\text{long}
     }(t^*)$ is an increasing function of time, and so its Fourier transform is only analytic at finite distance above the real axis. }
  \label{fig:contours}
\end{figure*}

For $F_1^\e<-1$ both poles of
  $ \chi^{\text{long},
  }_1 (s)$ are located on the imaginary axis, at $s = \pm  (|1 + F^\e_1|/2)^{1/2}$.  One of the poles is now in the upper half-plane of complex $s$. Modifying the integration contour the same way as for $l=0$ to preserve causality, we obtain for
    $t^* >0$
\begin{equation}
  \chi^{\text{long}
  }_1 (t^*)  \propto t^*  \frac{\sinh{\sqrt{\frac{|F^\e_1|-1}{2}} t^*}}{\sqrt{\frac{|F^\e_1|-1}{2}} t^*}.
  \label{t_2}
\end{equation}
For $t^*
\ll t^*_a = (2/|1 + F^\e_1|)^{1/2}$, both
$\chi^{\text{long}}_1 (t^*)$ and $\Delta^{\text{long}}_1 (t^*) \propto h \chi^{\text{long}}_1 (t^*)$ increase linearly with $t^*$. For $t^*
\gg t^*_a$, the perturbation grows exponentially,
indicating
that the FL state becomes unstable.

We note in passing that the need to bend the integration contour around the pole for
$ F^\e_1 <-1$  can be also understood by considering the behavior of $\chi_1^{\text{long}
 }(t^*)$ at $ F_1^\e
 $ approaching $-1$ from above. In the limit $ F_1^\e
  \to -1
  +0^+$, the two poles of $\chi_1^{\text{long}}(s)$ coalesce into a single double pole at the origin, as shown in Fig.~\ref{fig:axis}. Had we tried to compute $\chi_1^{\text{long}}(t^*)$ by integrating along the real axis of $s$, we would have intersected a divergence. To eliminate the divergence, one needs to bend the integration contour and bypass the double pole along a semi-circle above it. The extension of this procedure for $ F^\e_1
   <-1$ yields the contour shown in Fig.~\ref{fig:upper}.

\subsubsection{
$l=1$, transverse channel}

For small $1 + F^\e_1$, the pole in the transverse susceptibility for $l=1$ is on the imaginary axis. The behavior of $\chi^{\text{ tr}}_1 (t^*)$ is then the same as for the $l=0$ case.

\subsection{
Response in the time domain in the presence of disorder}
 Near Pomeranchuk instabilities in the
$l=0$
and transverse
 $l=1$
channels,
 the poles are on the imaginary axis, and adding weak impurity scattering will not change the results obtained in Sec.~\ref{sec:chit0}. Namely, the pole's contribution to $\chi^\e_0 (t^*)$ still decays exponentially with $t^*$ for $ F^\e_0
>-1$, becomes independent of $t^*$ at  $ F^\e_0
 =-1$, and increases exponentially with $t^*$ for $ F^\e_0
<-1$.

\begin{figure}
      \centering
      \includegraphics[width=\hsize]{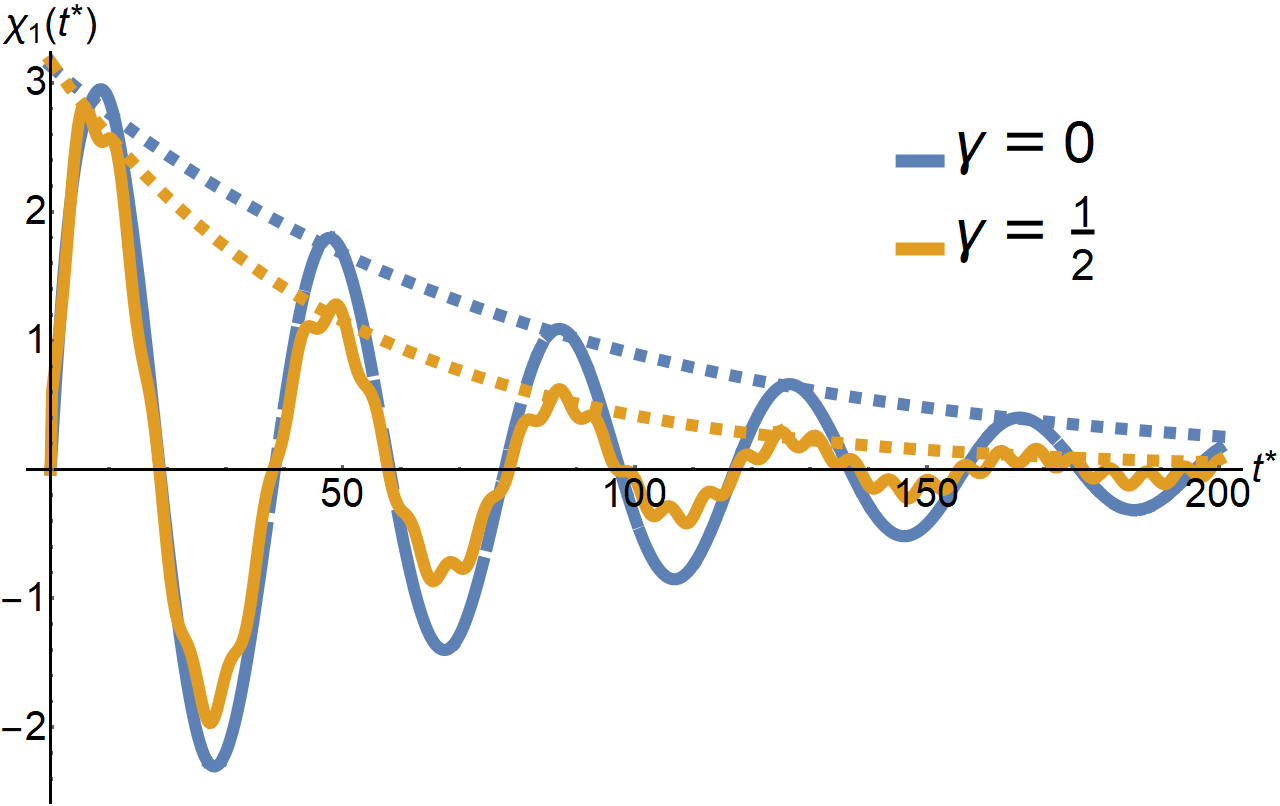}
      \caption
      {(color online) $\chi_
{\text{qp},
1}^{\text{long}}
     (t^*)$
with and without impurity scattering. Solid:
      numerical calculation for $\gamma = 0$ and $\gamma =
1/2
$, both for $F^\e_1 = -0.95$. Dashed:
      asymptotic expressions describing contributions that decay exponentially with characteristic times $t^*_b$ (Eq.~(\ref{tab})
        for $\gamma =0$ and Eq. (\ref{tab-gamma}) for $\gamma =1/2$) . }
      \label{fig:chi-1-time}
    \end{figure}

In the longitudinal
$l=1$,
the poles remain near the real axis also in the presence of $\gamma$. Finite damping changes the time scale  $t^*_b$ from Eq.~(\ref{tab})
to
 \begin{equation}
  t^*_b  = 4(\sqrt{1+ \gamma^2} - \gamma)/(1 -|F^\e_1|).
  \label{tab-gamma}
\end{equation}
After this change, the results for $\chi^{\text{long}}_1 (t^*)$ remain the same as in the absence of disorder. The time dependence of $\chi^{\text{long}}_1$ in the presence of impurity scattering is shown in Fig.~\ref{fig:chi-1-time}.

In all cases, finite $\gamma$ modifies the branch-cut contribution, so that in addition to the algebraic decay, there is also an exponential decay. For $l = 0$ we find,
    \begin{equation}
      \label{eq:chi-t-bcut-gamma}
      \chi_{\text{bcut},0}(t^*) \propto e^{-\gamma t^*} \frac{\cos(t^*-\pi/4)}{(t^*)^{3/2}}.
    \end{equation}
  Similar expressions holds for $l >0$.

  The presence of the exponentially decaying terms due to damping is particularly relevant for $l>0$ and $F^\e_l >0$,
 as it allows one to distinguish between the cases of smaller
 $F^\e_l$.
For the former,
the zero-sound pole is present and located above the branch cut, at $s = \pm a -i b \gamma$, where $a >1$ and $b <1$.
For the latter,
the zero-sound pole is located on the unphysical Riemann sheet.
 In both cases,  $\text{Im}\chi^\e_l (s)$ at real $s$ has a peak at $s = \pm a$,  but for larger $F^\e_l$ its width is ${\bar b} \gamma$ with
   ${\bar b} >1$,
      i.e., it is larger than $\gamma$. Accordingly, for smaller $F^\e_l$, the dominant contribution to $\chi_{l}(t^*)$ at large $t^*$ comes from the pole, and $\chi_{l}(t^*) \propto e^{-b \gamma t^*} \cos(a t^*)$. For larger $F^\e_l$,
 $\chi_{l}(t^*)$ at large $t^*$ comes from the branch cut, and
   $\chi_{l}(t^*) \propto e^{-\gamma t^*} \cos(t^*-\pi/4)/(t^*)^{3/2}$.
Because this property holds only for $l > 0$
and in the presence of disorder,
it was not discussed in previous works,
~\cite{Beal-Monod1994,Anderson2011,Li2012,Oganesyan2001,Zyuzin2018,Sodemann2018,Lucas2018,Torre2019}
which studied collective modes of a 2D Fermi liquid either in the $l=0$ channel or in the absence of disorder.

\section{
Special
 cases of charge-current and spin-current order parameters}
\label{sec:current}
\subsection{Ward identities and static susceptibility in the $l=1$ channel}
In previous sections, we assumed that the behavior of the full susceptibility is at least qualitatively the same as that of the quasiparticle susceptibility, i.e., the collective modes, present in $\chi^\e_{\text{qp},l} (q,\omega)$, are also present in the full $\chi^\e_l (q,\omega)$.   The full and quasiparticle susceptibilities differ by the factor $(\Lambda^\e_l)^2$ [see Eq.  (\ref{new_1})], which accounts for renormalizations from high-energy fermions. For a generic order parameter with a form-factor $f^\e_l ({\bf k})$, the vertex
$
\Lambda^\e_l$  is assumed to be finite for all $F^\e_l$, including $F^\e_l =-1$. The pole structure of $\chi^\e_{l} (q,\omega)$ is then fully determined by that of  $\chi^\e_{\text{qp},l} (q,\omega)$.

We now consider the special case of order parameters with $l=1$, for which $f^\e_{1} ({\bf k}
) =
\left(\begin{array}{c}
\cos\theta                                                                                                                                                                                                                                                                  \\
\sin\theta
\end{array}
\right)
\partial \epsilon_k/\partial k$, up to an overall factor. These order parameters correspond to charge or spin currents.  The special behavior of a FL under perturbations of this form has been discussed in recent studies of the static susceptibility in the $l=1$ channel \cite{Kiselev2017,Wu2018,Chubukov2018}. Namely, for the spin or charge current order parameter, the vertices
$\Lambda^\e_1$ satisfy the Ward identities which follow from conservation of the total number of fermions (the total
``charge'') and total spin.  In the static limit, the Ward identities read \cite{Engelsberg1963,Chubukov2018}
\begin{equation}
\frac{m^*}{m} Z \Lambda^\e_1=
1 + F^\e_1.
 \label{new_7}
\end{equation}
Under certain assumptions, these identities allow one to decide which of three factors on the left vanishes at the instability. First, we assume that the $Z$-factor, being a high-energy property of the system, remains finite at the instability. Therefore, the product $(m^*/m)\Lambda_1^\e$ should vanish at $F_1^\e\to -1$. Next, we divide the versions of Eq.~\eqref{new_7} for the charge and spin channels by each other and obtain
\begin{eqnarray}
\frac{\Lambda_1^c}{\Lambda_1^s}=\frac{1+F_1^c}{1+F_1^s}.
\end{eqnarray}
We then rule out a very special case, when both $F_1^c$ and $F_1^s$ reach the critical value of $-1$ simultaneously, and also assume the charge (spin) vertex remains finite at an instability in the spin (charge) channel.  Then $\Lambda_1^\e$ vanishes as $1+F_1^\e$, which implies that $m^*/m$ remains finite.

Given that $m^*/m$ remains finite while $(\Lambda_1^\e)^2$ vanishes as $(1+F_1^\e)^2$, the full static susceptibility
 $ \chi^\e_1 (q, \omega =0)
 =
 (\Lambda^\e_1)^2 \chi^\e_{\text{qp},1} (q, \omega =0)
 +\chi^\e_{\text{inc},1}
$ does not diverge at  $F^\e_1 =-1$, despite the fact that the quasiparticle susceptibility
 $ \chi^\e_{\text{qp},1} (q,\omega=0)$ diverges
 as $1/(1+F_1^\e)$.

What was said above does not apply to the special case of a Galilean-invariant system.
 In this case,
 the
charge current is
equivalent
 to
 the
 momentum
  and
  thus
 is
 conserved.
The Ward identity for the momentum
implies
 that $Z \Lambda^c_1 =1$, i.e., $\Lambda^c_1$ remains finite at
 $1+F^c_1 =0$.
 Equation (\ref{new_7}) then implies that
  $m^*/m = 1 +F^c_1$, which is
  the standard result for a Galilean-invariant FL.
  The static susceptibility still remains finite
  at $1 +F^c_1  \to 0$, this time because
  the factor
  of $m^*/m$
  in the numerator
  of $\chi^c_{\text{qp},1}$
  cancels
  out with
   $1+F^c_1$ in
   its denominator.
  Furthermore,
   in the Galilean-invariant case the
   static  $l=1$ charge
   and
   spin
susceptibilites
are not renormalized at all by the electron-electron interaction \cite{Leggett1965}.
On the other hand,  $m^*/m=1+F_1^\e$ does vanish at the transition in the $l=1$ charge channel.
  We believe
  that the vanishing mass indicates a global instability of a non-Pomeranchuk type, which is not associated with the $l=1$ deformation of the FS.

\subsection{Dynamical susceptibility in the $l=1$ channel}
Now, let us look at the dynamics. Consider for definiteness the
$l=1$ longitudinal susceptibility.   Near  $1 + F^\e_1 =0$, the quasiparticle susceptibility has poles given by Eq.~(\ref{F1-1}). One of the poles moves into the upper frequency half-plane when $1 + F^\e_1$ becomes negative. To relate the full and quasiparticle dynamical susceptibilities, we need to know $\Lambda^\e_1$ in the dynamical case. The FL theory assumes that
 $\Lambda^\e_1$ can be computed by setting both $\omega$ and $q$ to zero. The argument is that
 $\Lambda^\e_1$ is renormalized only by high-energy fermions, hence its frequency and momentum dependences come in a form of regular functions of $q/k_F$ and $\omega/E_F$. If so, then the relation between
   $\Lambda^\e_1$ and $1 + F^\e_1$, Eq.~(\ref{new_7}),
    holds in the dynamical case as well. We will verify this statement explicitly via a perturbative calculation in Sec.~\ref{sec:perturbation-theory}.

Taking $\Lambda^\e_1$  from Eq.~(\ref{new_7}) and substituting it along with the dynamical quasiparticle susceptibility into Eq.~(\ref{new_1}), we obtain the full susceptibility for $F_1^\e\approx -1$
\begin{flalign}
  \chi^{
  \text{long}}_1 (q, \omega) \approx  -
N_F \frac{m}{m^*} \frac{\left(1 + F^\e_1\right)^2     (v_F^*q)^2}{\omega^2 -
\left (1 + F^\e_1\right)
 (v_F^* q)^2/
 2} \nonumber                                                                                                                                                                                                                                                               \\
  + \chi_{\text{inc},1}.
\label{eq:chi-stat-false}
\end{flalign}
where we recall that the last term represents the contribution from high-energy fermions.
The static limit of the first term in the equation above, i.e., $N_F(1+F_1^\e)(m/m^*)$, in indeed non-singular at the transition, in agreement with the conclusions of the previous section.
Nevertheless, one of the poles of
$\chi^{
\text{long}}_1 (q,\omega)$  moves into the upper frequency half-plane when
 $
1 + F^\e_1$ becomes negative, i.e., a
 {\it dynamical} perturbation with the structure of spin or charge current grows exponentially with time, which is an indication of a Pomeranchuk instability.
The peculiarity of the $l=1$ case in that the residue of the pole vanishes right at the transition, but it is finite both above and below the transition.

\subsection{The case of more than one non-zero Landau parameters}
It is instructive to derive an analog of Eq.~(\ref{eq:chi-stat-false})
for a more general case of several non-zero Landau parameters.
We remind the reader that in this situation the pole structure of $\chi^{
  \text{long}}_1 (q,\omega)$ is more complex than when only $F^\e_1$ is present (see Eq. (\ref{extra_2})
for the case when $F^\e_1$ and $F^\e_0$ are non-zero).  The issue we address is whether  $\chi^{
  \text{long}}_1 (q,\omega)$ for charge/spin current still has $(\Lambda^\e_1)^2 \propto (1 +F^\e_1)^2$ as the overall factor.  We argue that it does.

To demonstrate this, we need to express the full susceptibility via vertices
$\bar\Lambda^\e({\bf q},\omega)$, which include both high-
and low-energy renormalizations.
Vertices $\bar\Lambda
^\e({\bf q},\omega)$ can be expanded into a series of partial harmonics:
$\bar\Lambda
^\e({\bf q},\omega)=\sum_l a_l \Lambda^\e_l(s)\cos l\theta
$, where $\theta$ is the angle between the momentum of the incoming fermion and ${\bf q}$,
$a_0=1$, and $a_{l\neq 0}=\sqrt{2}$.
With this definition, the full longitudinal susceptibility in the $l=1$ channel can be written as
\begin{align}
  \label{eq:susc-full-consistent}
                                            & \chi_1^{c(s),\text{long}}(s) =  \nonumber                                                                                                                                                                                     \\
                                            & -\nu_F
  \int \frac{d\theta}{\pi} \sum_{l} a_l \bar\Lambda_l^{c(s)} (s) \cos l\theta
  \frac{\cos^2 \theta}{s-\cos\theta + i\delta
  }\Lambda_1^{c(s)}.
\end{align}
The vertex
 ${\bar \Lambda}^\e_l ({\bf q},\omega)$ is given by a series of diagrams which contain momentum and frequency integrals of the product
$G_{{\bf p}+\frac{{\bf q}}{2}, \omega_p + \frac{\omega}{2}}
G_{{\bf p}-\frac{{\bf q}}{2}, \omega_p -\frac{\omega}{2}}$,  convoluted with fully renormalized four-fermion vertices. The diagrammatic series can be represented as the sum of subsets of diagrams, each with a fixed number $n =0,1,2, 3...$ of cross-sections which contain contributions from the regions where the poles of
  $G_{{\bf p}+\frac{{\bf q}}{2}, \omega_p + \frac{\omega}{2}}$ and  $G_{{\bf p}-\frac{{\bf q}}{2}, \omega_p -\frac{\omega}{2}}$ are in the opposite half-planes of complex frequency. This constraint binds the internal ${\bf p}$ and  $\omega_p$  to the FS. The subset with $n=0$
       is non-zero only for
$l=1$ and gives $\Lambda^{c(s)}_1$, while the sum of contributions with different
$n>0$
 gives
$\bar\Lambda_l^\e (s)$.
 Combining the contributions from all $n$, we find that the vertex ${\bar \Lambda}^\e_l(s)$ satisfies an integral equation with $\Lambda^{c(s)}_1$ as the source term:
\begin{widetext}
\begin{equation}
  \label{eq:lambda-lambda-relation}
  {\bar \Lambda}^\e_l (s)
  = \Lambda^\e_1
  \delta_{l,1}  +
  \frac{Z^2 m^*}{
  4\pi^3}
   \sum_{l'}
   {\bar \Lambda}^\e_{l'} (s) a_{l'}
  \int ^{2\pi}_0 d \theta  \int ^{2\pi}_0 d \theta'
   \cos l\theta
   \cos{l'\theta'} \frac{ \cos{\theta'} }{s- \cos \theta' + i \delta} \Gamma^\e (\theta, \theta'),
\end{equation}
\end{widetext}
where $\Gamma^\e (\theta, \theta')$ is the four-fermion (four-leg) vertex with external fermions right on the FS. By construction,
$\Gamma^\e (\theta, \theta')$ contains only renormalizations from high-energy fermions (in the FL theory, such a vertex is called $\Gamma^\omega$, see Ref.~\onlinecite{Landau1980}). Landau parameters $F^\e_l$ are related to the angular harmonics of $\Gamma^\e (\theta, \theta')$ via
$F^\e_l=(Z^2 m^*/\pi)\Gamma^\e_l
$. When only  $F^\e_1$ is non-zero, i.e.,
 $(Z^2 m^*/\pi) \Gamma^\e (\theta, \theta') = 2 F^\e_1
  \cos{\theta'} \cos{\theta}$, only ${\bar \Lambda}^\e_1 (s)$ is non-zero as well. Then
\begin{align}
  {\bar \Lambda}^\e_1 (s)                   & = \frac{\Lambda^\e_1}{1 - F^\e_1 \frac{1}{\pi} \int_0^{2\pi} d \theta  \frac{ \cos^3\theta}{s- \cos \theta + i \delta}} \nonumber                                                                                             \\
                                            & = \frac{\Lambda^\e_1}{1 + F^\e_1 \chi
                          ^{\text{long}}
                          _{\text{free},1}
                          (s)
                          }.     \label{new_9}
\end{align}
Substituting Eq.~(\ref{new_9})
into Eq.~(\ref{eq:susc-full-consistent}) and expanding near $F^\e_1 =-1$,  we reproduce Eq.~(\ref{eq:chi-stat-false}).  When, e.g.,
$F_0^{c(s)}$ and $F_1^{c(s)}$ are non-zero, the solution of Eq.~(\ref{eq:lambda-lambda-relation}) is
\begin{align}
  \label{eq:ansatz-F0-F1}
  \bar\Lambda^\e_1(s)                       & = \Lambda^\e_1 \frac{1}{1+F^\e_1 (K_0 + K_2) - \frac{2 F^{c(s)}_0) F^\e_1 K^2_1}{1 + F^\e_0 K_0}}, \nonumber                                                                                                                  \\
  \bar\Lambda^\e_0(s)                       & =- \Lambda^\e_1 \frac{
  \sqrt{2} F^\e_0 K_1}{1+F^\e_1 (K_0 + K_2) - \frac{2 F^\e_0 F^\e_1 K^2_1}{1 + F^\e_0 K_0}},
\end{align}
where $K_{0,1,2}$ are defined by Eq.~(\ref{new_2_2}). Substituting the last two equations into Eq.~(\ref{eq:susc-full-consistent}), we obtain
\begin{widetext}
\begin{equation}
\chi_1^{
\text{long}}(s) =
\nu_F
(\Lambda^\e_1)^2 \frac{K_0 + K_2-  \frac{2F^\e_0 K^2_1}{1 + F^\e_0 K_0}}{1+F^\e_1 (K_0 + K_2) - \frac{2 F^\e_0 F^\e_1 K^2_1}{1 + F^\e_0 K_0}}.
\end{equation}
\end{widetext}
Comparing the last result with  $\chi^{\text{long}}
_{\text{qp},1}$ in Eq. (\ref{extra_2}), we see that $\chi_1^{
\text{long}}(s)  = (\Lambda^\e_1)^2 \chi^
{\text{long}}
_{\text{qp},1}
(s)
$, exactly as in the static case. This result implies that the residue of the pole is proportional to $(\Lambda^\e_1)^2
 \propto (1 + F^\e_1)^2$ and thus vanishes at the Pomeranchuk instability also for the case of two non-zero Landau parameters, when the pole structure of the susceptibility becomes more involved.
 Still, like in the case when only $F^\e_1$ is non-zero,
 $(\Lambda^\e_1)^2$ is independent of $s = \omega/(v^*_F q)$, and it does not cancel the poles in  $\chi^{\text{long}}
_{\text{qp},1}$.
At  $1 + F^\e_1 <0$, one pole moves into the upper frequency half-plane,
signaling a Pomeranchuk instability.

 \subsection{Perturbation theory for the vertex in the dynamical case}
\label{sec:perturbation-theory} \def\ee{\epsilon}
\def\bee{\boldsymbol\ee} \def\w{\omega}
\def\q{{\bf q}} \def\p{{\bf p}}
\def\k{{\bf k}} \def\l{{\bf l}}

\begin{figure*}
    \includegraphics[width=\hsize]{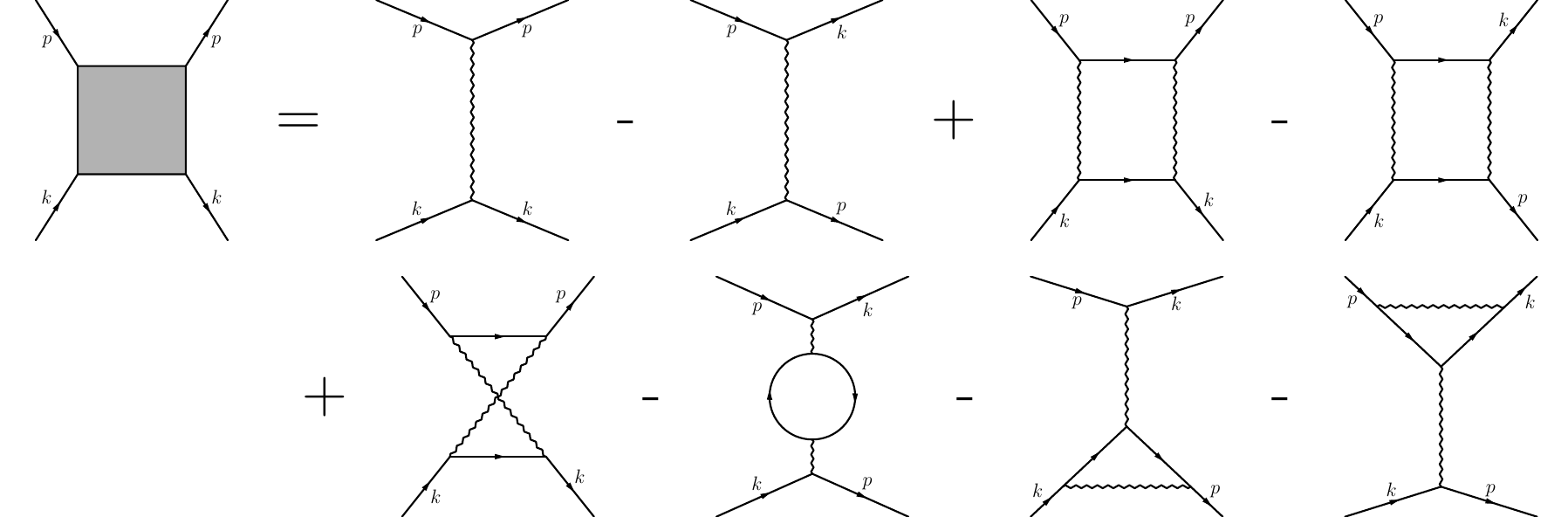}
      \caption{
    Diagrams for the four-fermion vertex $\Gamma^\omega_{\alpha\beta,\gamma\delta} $ to second order in the Hubbard-like interaction.
    One of external     $2+1$      momenta,     $k=(\k,\omega_k)$,     is chosen to be on the FS,
    i.e., $|\k|=k_F$ and $\omega_k=0$,     while the other, $     p$, is generically away from the FS.
    All internal momenta are away from the FS.     In this sense,     renormalization of $\Gamma^\omega   _{\alpha\beta,\gamma\delta}$ comes from high-energy fermions.
  }   \label{fig:gamma-w}
\end{figure*}

\begin{figure}
  \includegraphics[width=\hsize]{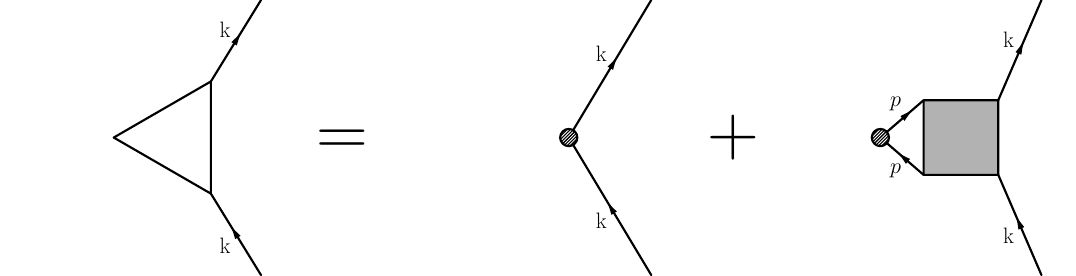}
  \caption{
    Diagrammatic representation of the high-energy triple vertex $\Lambda_1^s$.     The shaded box is $\Gamma^\omega_{\alpha\beta,\gamma\delta}$.
 \label{fig:lambda-w}}
\end{figure}

 We now return to the case of a single Landau parameter $F^\e_1$
 and verify by a perturbative calculation that  $(\Lambda^\e_1)^2$ does not cancel the dynamical poles in the full $l=1$ susceptibility.
 We perform the calculation to second order in the Hubbard (point-like) interaction $U$. For simplicity we limit our attention to the spin channel and also consider a Galilean-invariant system. We will show that the dynamical vertex $\bar\Lambda_1^{s} (s)$ has the pole structure of Eq. (\ref{new_9})
 with $s$-independent $\Lambda^\e_1$.
 To demonstrate this, it suffices to show that the vertex $\Lambda_1^{s}$, which acts as a source for the dynamical vertex $\bar\Lambda_1^s(s) $ in Eq.~(\ref{eq:lambda-lambda-relation}), does not vanish at $s$ which corresponds to the pole of the dynamical vertex.
Instead of calculating $\Lambda_1^s$ directly, we compute the product $\Lambda_1^s Z$ for reasons that will become clear later in the section. Using the Ward identity associated with the Galilean invariance,
we express quasiparticle $Z$
as \cite{Landau1980}
\begin{equation}
  \label{eq:Z-GI-FL}   \frac{1}{Z} = 1
  -  \frac{i}{2 k_F}\sum_{\alpha\beta}\int \frac{d^3p}{(2\pi)^3} \Gamma^\omega_{\alpha\beta,\alpha\beta}(k,p)(G_p^2)^\omega  (\hat k  \cdot\p),
\end{equation}
where the $2+1$-momentum
$p=(\p,\omega_p)$
 is not necessary close to the FS, and  $k=(k_F\hat k, 0) + \ee$
 is infinitesimally close to the FS, i.e.,
$\ee=(
\q,\omega)$
 with both $|\q|$ and $\omega$
 being infinitesimally small.
 The
direction
 of $\hat k$ in Eq. \eqref{eq:Z-GI-FL} is arbitrary.
 The  $\Gamma^\omega_{\alpha\beta,\alpha\beta}$  is the dressed four-fermion vertex and
$(G_p^2)^\omega  (\hat k  \cdot\p)$
is the regular part of the product $G_{p+\ee/2}G_{p-\ee/2}$ of two exact Green's functions, whose arguments differ by $\ee$.
 (Note that the $\q$ and $\omega$ in the definition of the $Z$-factor do not need to coincide with the corresponding variables describing the collective mode, but we choose them to be the same for simplicity.)
The product $G_{p+\ee/2}G_{p-\ee/2}$
 can be written as the sum of a regular part and a singular contribution from the FS\cite{Landau1980}:
\begin{align}
  G_{p+\ee/2}G_{p-\ee/2} &= (G_p^2)^\omega
   \nonumber \\
                         &
                           +\frac{2 \pi i Z^2}{v_F^*}\frac{\hat p\cdot \hat q}{s-\hat p\cdot \hat q
                        +i\delta\text{sgn}\omega_p }
 \delta(\omega_p)\delta(|\p|-k_F),
 \label{ll7}
\end{align}
where $\hat p=\p/|\p|$
 and, as before,
 $s= \omega/v_F^*|\q|$.

 To second order in $U$, $\Gamma^\omega_{\alpha\beta,\alpha\beta}$ is given by the diagrams shown in Fig.~\ref{fig:gamma-w}. Explicitly,
\begin{widetext}
\begin{align}
  \label{eq:gamma-w-u2}   \Gamma^\omega_{\alpha\beta,\gamma\delta}(k,p) &= \frac{1}{2}\delta_{\alpha\gamma}\delta_{\beta\delta}\left[U + i U^2 \int \frac{d^3k'}{(2\pi)^3}\left(2 G_{k'} G_{p-k + k'} + G_{k'} G_{p+ k-k'}\right)\right] \nonumber \\
  & -\frac{1}{2}  \boldsymbol{\sigma}_{\alpha\gamma}\cdot
    \boldsymbol{\sigma}_{\beta\delta}\left[U + i U^2 \int \frac{d^3k'}{(2\pi)^3} G_{k'} G_{p+ k-k'}\right]
    \equiv \delta_{\alpha\gamma}\delta_{\beta\delta}\Gamma^c(k,p)+ \boldsymbol{\sigma}_{\alpha\gamma}\cdot
  \boldsymbol{\sigma}_{\beta\delta}\Gamma^s(k,p),
   \end{align}
 \end{widetext}
 where at the last step we defined the charge and spin parts of the four-fermion vertex, $\Gamma^c(k,p)$  and $\Gamma^s(k,p)$, respectively.

  The renormalized spin-current vertex can be written as (see Fig. \ref{fig:lambda-w})
\begin{align}   \label{eq:delta-lambda-1s-FL}
  &\Lambda_{
  1}^s \sigma^z_{\beta\beta} = \sigma^z_{\beta\beta}
   \nonumber \\   &\qquad\qquad-\frac{i}{k_F}\sum_{\alpha}
  \int \frac{d^3p}{(2\pi)^3}   \Gamma^\omega_{\alpha\beta,\alpha\beta}(
  k,p)(G_p^2)^\omega
 (\hat k \cdot \p)\sigma^z_{\alpha\alpha}.
\end{align}
 where the internal momentum
 $p = ({\bf p}, \omega_p)$
is again not confined to the FS.
It is to be understood that $\Lambda_1^s$ is a function of the $2+1$ momentum $\ee$,
 and
 $\Lambda_1^s \neq 0$ even for $\ee = 0$.
Substituting the last formula in Eq.~(\ref{eq:gamma-w-u2}) into Eqs.~(\ref{eq:Z-GI-FL}) and (\ref{eq:delta-lambda-1s-FL}) and summing over spin indices, we obtain
\begin{equation}
  \label{eq:Z-GI-FL2}   \frac{1}{Z} = 1
  -  \frac{2i}{k_F}\int \frac{d^3p}{(2\pi)^3} \Gamma^c(k,p)(G_p^2)^\omega (\hat k \cdot\p)\end{equation}
and
\begin{align}   \label{eq:delta-lambda-1s-FL}
  &\Lambda_{1}^s = 1 -\frac{2i}{k_F}
  \int \frac{d^3p}{(2\pi)^3}   \Gamma^s(k,p)(G_p^2)^\omega (\hat k \cdot \p).
\end{align}
To second order in $U$, the product $\Lambda
^s_1 Z$ can then be written as
\begin{align}   \label{eq:delta-lambdaZ-u2}
  &\Lambda^s_1  Z =  1 \nonumber \\
  &\qquad-\frac{2 i}{k_F}\int \frac{d^3p}{(2\pi)^3}\left[\Gamma^c(k,p)-\Gamma^s(k,p)\right]\hat k \cdot \p (G_p^2)^\omega. \end{align}

We now show that the product $\Lambda_1^s Z$ does not depend on $s=\omega/v_F^*|\q|$. To see this, we add a term $k_F s$ to
$\hat k \cdot \p$ on the right-hand side of Eq.~(\ref{eq:delta-lambdaZ-u2}) and then subtract off the same term.
Equation (\ref{eq:delta-lambdaZ-u2})
then goes over to
\begin{equation}
  \label{eq:lambda-z-parts}
  \Lambda_1^s Z = 1 - Q_1 - Q_2,
   \end{equation}
where
\begin{align}
  \label{eq:Q1-new}
  Q_1 = 2 i s
  \int \frac{d^3p}{(2\pi)^3}\left[\Gamma^c(
  k,
  p)-\Gamma^s(
  k,
  p)\right]
   (G_p^2)^\omega
\end{align}
and
\begin{equation}
  \label{eq:Q2-new}
  Q_2 = 2 i\int \frac{d^3p}{(2\pi)^3}\left[\Gamma^c(
  k,
  p)-\Gamma^s(
  k,
  p)\right](\hat k \cdot \frac{\bf p}{k_F} -s) (G_p^2)^\omega.
\end{equation}
We now use the fact that conservation of charge and spin allows one to derive two independent relations for $Z$ (Refs.~\onlinecite{Landau1980,Kondratenko1964,Kondratenko1965,Chubukov2018}):
\begin{align}
  \label{eq:Z-rels}
  \frac{1}{Z} &= 1 -  \frac{i}{2 k_F}\sum_{\alpha\beta}\int \frac{d^3p}{(2\pi)^3} \Gamma^c(
  k,
  p)(G_p^2)^\omega ~\mbox{(charge)}; \nonumber\\
  \frac{1}{Z}&= 1 -  \frac{i}{2 k_F}\sum_{\alpha\beta}\int \frac{d^3p}{(2\pi)^3} \Gamma^s(
                                   k,
                                   p)(G_p^2)^\omega ~\mbox{(spin)}.
\end{align}
Combining the two, we find that
$Q_1=0$, i.e., $\Lambda_1^s Z = 1 - Q_2$.

We now analyze $Q_2$.
To first order in $ U$,
 the vertex remains static. Then
$\mathcal{O}(U)$ terms in $\Gamma^{c(s)}$ in Eq.~(\ref{eq:Q2-new}) vanish because of the double pole of $(G_p^2)^\omega$. Let us focus on the $\mathcal{O}(U^2)$ terms.
Substituting Eq. (\ref{ll7}) into Eq.~(\ref{eq:Q2-new}) for $Q_2$
and choosing the direction of
 ${\bf k}$
 to be along
 ${\bf q}$,
we obtain
 \begin{widetext}
  \begin{align}   \label{eq:Q-2}
  Q_2 &=  2i \int \frac{d^3p}{(2\pi)^3}\left(\Gamma^c(k,p)-\Gamma^s(k,p)\right)(\hat k
   \cdot \frac{\bf p}{k_F} - s)\left( G_{p+\ee/2}G_{p-\ee/2}
    - \frac{2\pi i Z^2}{v_F^*}\frac{\hat k \cdot \hat p}{s - \hat k\cdot \hat p + i\delta}\delta(\omega_p)\delta(
  |\p| - p_F)\right)\\
     &= 2i \int \frac{d^3p}{(2\pi)^3}\left(\Gamma^c(k,p)-\Gamma^s(k,p)\right)\left[(v_F^* |\q|)^{-1}(G_{p-\ee/2}-G_{p+\ee/2})
     + \frac{2\pi i Z^2}{v_F^*}\hat k \cdot \hat p\delta(\omega_p)\delta(
    |\p| - p_F)\right]\label{eq:Q-2-3}.
\end{align}
\end{widetext}
 One can now verify that the first term in the r.h.s. of Eq.~(\ref{eq:Q-2-3}) is zero, i.e.,
\begin{equation}
  \label{eq:Q2-zero-term}
  \int \frac{d^3p}{(2\pi)^3}\left(\Gamma^c(k,p)-\Gamma^s(k,p)\right)(G_{p-\ee/2}-G_{p+\ee/2}) = 0.
\end{equation}
This can be done by substituting the explicit expressions for $\Gamma^{c(s)}$ from Eq. (\ref{eq:gamma-w-u2}) and changing integration variables\cite{Chubukov2010}.
  We are then left with a contribution coming solely from the FS,
\begin{align}
  \label{eq:Q2-FS}
  Q_2 &= \frac{Z^2}{2\pi v_F^*}\int \frac{d\theta'}{\pi}\cos\theta' \left(\Gamma^c(\theta,\theta') - \Gamma^s(\theta,\theta')\right) \nonumber \\
  &= F_1^c - F_1^s,
\end{align}
where $\theta$ and $\theta'$ are the azimuthal angles of $\hat k$ and $\hat p$, respectively.
This term is independent of $s$ and only contributes to the static vertex~\cite{Wu2018}.
 Going back to Eq.~(\ref{eq:lambda-z-parts}), we obtain
\begin{equation}
  \label{eq:lambda-z-final}
  \Lambda_1^s Z = 1 - F_1^c + F_1^s = \frac{m}{m^*}(1+F_1^s),
\end{equation}
where we used the relation
$m^*/m = 1+F_1^c$
valid for a Galilean-invariant system.
This result agrees with the analysis in the previous Section.

 \subsection{
 Charge/spin current order parameter:
 Ginzburg-Landau
  functional and
time evolution}

We now analyze the structure of the Landau functional that describes the $l=1$ Pomeranchuk transition. Our purpose is to reconcile
 the apparent contradiction that on one hand
the FL ground state becomes unstable for $F_1^\e < -1$, while on the other hand the static $l=1$ susceptibility remains finite at $F_1^\e = -1$. The Ginzburg-Landau functional can be derived from the Hamiltonian of interacting fermions, coupled to an infinitesimal external perturbation $h_1^\e$, via a Hubbard-Stratonovich (HS) transformation with an auxiliary field $\Delta_1^\e$. For a generic
$l=1$
order parameter, the vertex remains finite at the Pomeranchuk transition. In this case, it is sufficient to consider only the quasiparticle part of the Hamiltonian and neglect the
contributions from high-energy fermions. Then the coupling to the external field is given by a bilinear term
$h_1^\e \Delta_1^\e $, and the total
susceptibility is identical to the quasiparticle susceptibility. For
the charge/spin current order, the coupling is still proportional to  $h_1^\e\Delta_1^\e$ term, but the
contributions from high-energy fermions to the proportionality coefficient cannot be neglected, as
with these contributions the fully dressed coupling vanishes at the transition. To see this, we explicitly separate the four-fermion interaction into the
 components coming from the states near and away from the FS.

Our point of departure is the effective, antisymmetrized interaction between fermions, expressed via the vertex function $\Gamma_{\alpha\beta;\gamma\delta} ({\mathbf k}, {\mathbf k'};{\mathbf q})$, where $
  \q$ is a small momentum transfer:
\begin{align}
  \label{eq:interaction-vertex}
& {\cal H}
_{\text{int}} =  \\
    & \sum_{
{\mathbf k},{\mathbf k}',{\mathbf q},
 \alpha,\beta,\gamma,\delta}
\Gamma_{\alpha\beta;\gamma \delta} ({\mathbf k}, {\mathbf k'};{\mathbf q})
a^\dagger_{\mathbf k+\frac{\mathbf q}{2},
\alpha} a^{\phantom{dagger}}
   _{\mathbf k
   -\frac{\mathbf q}{2},
   \gamma}
a^\dagger_{\mathbf k'-\frac{\mathbf q}{2},\beta}
 a^{\phantom{dagger}}
 _{\mathbf k'+\frac{\mathbf q}{2},
      \delta }
  \nonumber
\end{align}
The  generic form of the
$l=1$ component of $\Gamma_{\alpha\beta;\gamma\delta} ({\mathbf k}, {\mathbf k'};{\mathbf q})$ is
\begin{align}
  &\Gamma^{l=1}_{\alpha\beta;\gamma\delta} ({\mathbf k}, {\mathbf k'};{\mathbf q}) =
    -\tilde{\mathbf{k}}\cdot\tilde{\mathbf{k}}'
  \\
 & \times \left(U^c_1 f^c(|\tilde\k|,|\tilde\k'|) \delta_{\alpha
 \gamma}
  \delta_{
  \beta
   \delta} + U^s_1 f^s (|\tilde\k|,|\tilde\k'|)
  \boldsymbol{\sigma}
 _{\alpha\gamma } \cdot
 \boldsymbol{\sigma}_{\beta \delta}\right),\nonumber
\end{align}
where $\tilde{\k}=\k/k_F$ and $\tilde{\k}'=\k'/k_F$,
and $k_F$ should be treated here as just a normalization constant, which we choose for convenience to match the Fermi wavenumber of the quasiparticles.
For charge/spin current orders, we replace the formfactors $f^{c,s}(|\tilde\k|,|\tilde\k'|)$ by constants and incorporate them into $U^\e_1$. A instability in the $l=1$ channel occurs if $U^\e_1>0$. Below we approximate the full vertex function $\Gamma_{\alpha\beta;\gamma\delta}$ by its $l=1$ component. Other components are not necessarily small, but we assume they are irrelevant for the low-energy theory near the
$l=1$ Pomeranchuk instability. In this approximation, the effective interaction is separable into
two parts that depend on
$\tilde\k$ and $
\tilde\k'$, and can be written
 as the sum of the charge and spin components: ${\cal H}_{\text{int}} = {\cal H}^c_{\text{int}} + {\cal H}^s_{\text{int}}$, wehere
\begin{align}
  \label{eq:interaction_sep}
{\cal H}^\e_{\text{int}} &=
- U^\e_1 \sum_{{\mathbf q}}\left(  \sum_{\mathbf k, \alpha,
\gamma}
\tilde\k
a^\dagger_{\mathbf k+\frac{\mathbf q}{2},
\alpha} t^\e_{\alpha\gamma} a^{\phantom{dagger}}
_{\mathbf k-\frac{\mathbf q}{2},
\gamma}\right)\nn\\
& \boldsymbol{\cdot}\left(
  \sum_{{\mathbf k}',
  \beta,\delta}
\tilde\k'
a^\dagger_{\mathbf k'-\frac{\mathbf q}{2},
\beta} t^\e_{\beta\delta} a^{\phantom{dagger}}
_{\mathbf k'+\frac{\mathbf q}{2},\delta}\right),
\end{align}
and, as before, $t^c_{
\mu\nu}=\delta_{\mu\nu}$
and
$t^s_{
\mu\nu}=\sigma^z_{\mu\nu}$.
We next rewrite the sums over the fermionic momenta
 as
\begin{align}
  \label{eq:separation}
& \sum_{\mathbf k}a^\dagger_{\mathbf k+\frac{\mathbf q}{2},
 \alpha} t_{\alpha\gamma}^\e
a^{\phantom{dagger}}
_{\mathbf k-\frac{\mathbf q}{2},
\gamma
}
 \\
&=
 \sum_{\mathbf k}a^\dagger_{\mathbf k+\frac{\mathbf q}{2},
 \alpha}
t^\e_{\alpha\gamma}
a^{\phantom{\dagger}}
_{\mathbf k-\frac{\mathbf q}{2},
\gamma} \delta^\epsilon_{|\mathbf k+\frac{\mathbf q}{2}|,k_F}\delta^\epsilon_{|\mathbf k-\frac{\mathbf q}{2}|,k_F} \nonumber \\
& + \sum_{\mathbf k}a^\dagger_{\mathbf k+\frac{\mathbf q}{2},
  \alpha  }
  t^\e_{\alpha\gamma}
  a^{\phantom{\dagger}}
  _{\mathbf k-\frac{\mathbf q}{2},
  \gamma} (1-\delta^\epsilon_{|\mathbf k+\frac{\mathbf q}{2}|,k_F}\delta^\epsilon_{|\mathbf k-\frac{\mathbf q}{2}|,k_F})
  \nn
\end{align}
(and the same for the sum over $\k'$). Here,
$\delta^\epsilon_{a,b}$ is nonzero only for $|\mathbf a- \mathbf b| < \epsilon$, and
$\epsilon$ is small compared to $k_F$ and will be taken to zero at the end of the calculation. The purpose of the projectors
$\delta^\epsilon_{a,b}$ is to split the fermions into those near the FS, which form the FL of quasiparticles, and those away from the FS, whose role is to renormalize the interaction between quasiparticles and their coupling to an external perturbation. Below, we denote fermions near the FS as $\psi^\dagger
(\psi)$, and fermions away from the FS as $\tilde\psi^\dagger
(\tilde\psi)$. Using (\ref{eq:separation}), we rewrite (\ref{eq:interaction_sep}) as
\begin{widetext}
\begin{flalign}
  \label{eq:sep-interaction}
{\cal H}_{\text{int}} =
    - U^\e_1 \sum_{{\mathbf q}}
    \left[\sum_{{\mathbf k}, \alpha,
\gamma}
\tilde{\k}
 \left(\psi^\dagger_{\mathbf k+\frac{\mathbf q}{2},
 \alpha }
 t^\e_{\alpha \gamma} \psi^{\phantom{\dagger}}
 _{\mathbf k-\frac{\mathbf q}{2},
 \gamma} + \tilde\psi^\dagger_{\mathbf k+\frac{\mathbf q}{2},
 \alpha}
 t^\e_{\alpha\gamma} \tilde\psi^{\phantom{\dagger}}
 _{\mathbf k-\frac{\mathbf q}{2},
\gamma} \right)\right]
\qquad\qquad \nonumber \\
    \boldsymbol{\cdot}\left[\sum_{{\mathbf k'},
 \beta,
 \delta}
\tilde{\k}'
 \left(\psi^\dagger_{\mathbf k'-\frac{\mathbf q}{2},
 \beta }
 t^\e_{
 \beta\delta} \psi^{\phantom{\dagger}}
 _{\mathbf k'+\frac{\mathbf q}{2},
 \delta} + \tilde\psi^\dagger_{\mathbf k'-\frac{\mathbf q}{2},\beta}
 t^\e_{\beta\delta} \tilde\psi^{\phantom{\dagger}}
 _{\mathbf k'+\frac{\mathbf q}{2},\delta}\right) \right].
\end{flalign}
\end{widetext}
The coupling of the
charge/spin current order parameter to a
weak external field
$\mathbf{h}_{1}^\e (\mathbf q,t)$ (which may  be time-dependent)
can be split into the low- and high-energy parts in the same way:
\begin{align}
  \label{eq:sep-ext}
{\cal H}_h &=
             \sum_{\q}\mathbf{h}_{1}^\e (\mathbf q,
t)\nn\\
  &\quad\boldsymbol{\cdot}
\sum_{\k,\alpha,\gamma}\tilde{\k}
\left(\psi^\dagger_{\mathbf k+\frac{\mathbf q}{2},\alpha}t^\e_{\alpha\gamma}\psi^{\phantom{\dagger}}
_{\mathbf k-\frac{\mathbf q}{2},\gamma}+\tilde\psi^\dagger_{\mathbf k+\frac{\mathbf q}{2},\alpha}t^\e_{\alpha\gamma}\tilde\psi^{\phantom{\dagger}}
_{\mathbf k-\frac{\mathbf q}{2},\gamma}\right). \nonumber
\end{align}
Note that the field couples to both
$\psi$ and
 $\tilde\psi$.

The interaction in
 Eq.~(\ref{eq:sep-interaction}) contains one term involving four fermions near the FS, one term involving four fermions away from the FS, and two mixed terms involving two fermions at the FS and two away from the FS. We decouple the quartic term with fermions away from the FS by
 going from a Hamiltonian representation to a Lagrangian one, upon which
 $\psi^\dagger_n$ and $\psi_n$ with $n=k,p,q$ become Grassman fields, which depend on the $2+1$ momentum $n=({\bf n},\omega_n)$. Next, we introduce
 a momentum- and time-dependent HS vector field
 $\tilde{\boldsymbol{\Delta}}_1^\e(\q,t)$ to decouple only the term in $H_{\text{int}}$ that involves fermions away from the FS, leaving the term involving fermions near the FS and the two mixed terms untouched.
 We then integrate out the
high-energy fermions and obtain  the effective action for $
\dd_{1}$. The action is given by
\begin{widetext}
\begin{equation}
  \label{eq:S-Delta-h}
  \mathcal{S}[{\tilde\Delta}_1^\e] =
                                    \sum_{
q} \frac{(\dd_{1}(q))^2}{U^\e_1}
                                   + \sum_{k,\alpha}
\ln M^\e_{\alpha,\beta}(k,k')
\vert_{\beta=\alpha,k=k'}
\end{equation}
with
\begin{equation}
  \label{eq:M-def}
  M^\e_{\alpha\beta}(k,k') =  G^{-1}_0(k)\delta_{k,k'}
\delta_{\alpha\beta}
 - t^\e_{\alpha\beta} \left.
\frac{\tilde{\k}+\tilde{\k}'}{2} \cdot\left[\mathbf{h}_{1}^\e(
q
)-
2 \left(\dd_{1}(
q)
 + U^\e_1 \sum_{p,\gamma,\delta}
\tilde{\p}
\psi^\dagger_{ p - \frac{ q}{2},\gamma}t^\e_{\gamma\delta} \psi_{ p + \frac{ q}{2},\delta} \right)\right]\right|_{q = k' - k},
\end{equation}
\end{widetext}
where
$G_0(k)$ is the free-fermion propagator.

Corrections to the low-energy theory from high-energy fermions are obtained by expanding $\mathcal{S}[\tilde\Delta_1^\e]$ up to first order in $\mathbf{h}_{1}^\e(
q)$ and up to second order in $\psi^\dagger \psi$. For definiteness, we consider the longitudinal $l=1$ channel and restrict the external field to a longitudinal component, i.e., we
set $\mathbf{h}_{1}^\e(
q)
 = \hat q h_{1}^\e(q)$.
  The new terms generated by integration over
  $\dd_{1}(q)$
   affect the action for low-energy fermions in three ways:
(i) the propagator of
 low-energy fermions acquires a
$Z$-factor and $v_F$ gets renormalized into $v^*_F$;
(ii) the coupling to the external field
acquires a factor of $\Lambda^\e_1$;
and (iii) the coupling
 constant of the interaction between low-energy fermions is renormalized from $-U^\e_1$ to
 $F_1/\nu_F$. With these modifications, the action for properly normalized low-energy fermions ($\psi^\dagger_k$ and $\psi_k$)
 becomes
\begin{widetext}
\begin{align}
  \label{eq:low-S}
  & \mathcal{S}[\psi]
    = -\sum_k \psi^\dagger_k Z^{-1}(\omega_k - v_F^*(|\mathbf{k}|-k_F))\psi_k
    + \Lambda^\e_1 \sum_qh_{1}^\e
(q)\sum_{k,\alpha,\beta}
\cos\theta_\k\psi^\dagger_{k
+\frac{q}{2},
\alpha}t^\e_{\alpha\beta}\psi_{k
-\frac{q}{2},\beta}\nonumber \\
  &
    +
    \frac{1}{\nu_F} \sum_{k,k',
    q,
    \alpha,\beta,
    \gamma,\delta
    }F_1^\e\cos\theta_\k\cos\theta_{\k'}\psi^\dagger_{k+\frac{q}{2},
    \gamma} t^\e_{\alpha
    \gamma} \psi_{k-\frac{q}{2},
    \alpha}
    \psi^\dagger_{k'-\frac{q}{2},
    \delta} t^\e_{
    \beta\delta}\psi_{k'+\frac{q}{2},
    \beta}.
\end{align}
\end{widetext}
where the summation over
$k$ is confined to the vicinity of the FS. We see that the factor $\Lambda_1^\e$ only changes the \emph{response} function to an external perturbation, but does not affect the \emph{thermodynamic} stability of the FL state. If we compute the
response function
by differentiating the partition function twice with respect to
 $h_1^\e (q)$, we find the same expression as
in Eq.\eqref{new_1}:
\begin{align}
  \label{eq:response-tilde-Lambda}
 & \chi_{1}^\e (q) =  \frac{\partial^2\mathcal{Z}}{\left(\partial h_{1}^\e (q)\right)^2}\nonumber \\
 & = \left\langle \left(\sum_{k,\alpha} \Lambda^\e_1\psi^\dagger_{ k+q/2,\alpha}\psi_{ k-q/2,\alpha}\cos\theta_\k\right)^2\right\rangle_{\mathcal S[\psi]} +
                \chi_{\text{inc},1}^\e
                \nonumber \\
 &= (\Lambda_1^\e)^2 \chi_{\text{qp},1}^\e (q) + \chi_{\text{inc},1}^\e,
\end{align} where $\mathcal Z$ is the partition function,
$\langle\dots\rangle_{\mathcal{S}[\psi]}$ denotes averaging with action ${\mathcal{S}[\psi]}$, and
$\chi_{\text{inc},1}$ is obtained by
differentiating
$\mathcal{{\tilde Z}} = \int e^{-\mathcal{S}[\tilde\Delta_1^\e]}$ twice with respect to
$h_1^\e$ without taking into account the contribution from low-energy fermions
(the $\psi^\dagger \psi$ term in \eqref{eq:M-def}).
We recall that the static susceptibility does not diverge at $F^\e_1 =-1$ because
 $
\Lambda_1^\e = (m/m^*Z) (1 + F^\e_1)$ vanishes at $F^\e_1 =-1$.

 We now introduce a
 low-energy HS field $\Delta_1^\e(q)$
   to decouple the quartic term in
$\mathcal{S}[\psi]$. Integrating out
 low-energy fermions, we obtain the effective action for $\Delta_1^\e (q)$ in the form
\begin{align}
  \label{eq:low-func}
  S[\Delta_1^\e] &= \sum_
  q\left(  a
  |\Delta_1^\e(
   q)|^2 +
  b
   |\Delta_1^\e(
   q)|^4 \right.\\
   &
   \left.
                 +
    \Lambda_1^\e h_1^\e(
                 q)\Delta_1^\e(
                 q)
                   \chi_{1,\text{free}}^\e (q)
                   +\text{c.c.}\right)\nn
\end{align}
plus higher order terms.

In Eq.~\eqref{eq:low-func},
$ a \propto 1+F_1^\e$ changes sign at the critical point, i.e., fluctuations of the order parameter $\Delta_1^\e$ diverge at the critical point, like for any other order parameter. In this sense, Pomeranchuk order with the structure of spin/charge current does develop when $1+ F^\e_1$ becomes negative. What makes the case of spin/charge current special is that the response to an external field gets critically reduced because of destructive interference from high-energy fermions.

The presence of $\Lambda^\e_1 \propto
1 + F^\e_1$ in the response function changes the time evolution of
$ \Delta^\e_1 (t^*)$ after an instant perturbation
 $h_1^\e (t^*) = h^\e_1 \delta(t^*)$. For $1+F_1^\e > 0$, we have
\begin{flalign}
  \Delta^\e_1 (t^*) \propto h_1^\e(1+F_1^\e)^2 t^*  e^{-(1+F^\e_1) t^*/4} \nonumber \\
 \times
  \frac{\sin{\sqrt{\frac{1+ F^\e_1}{2}} t^*}}{\sqrt{\frac{1+ F^\e_1}{2}} t^*},
  \label{eq:stable-time}
\end{flalign}
The functional form of  $\Delta^\e_1 (t^*)$ is the same as for a generic $l=1$ order parameter, when
high-energy renormalizations can be neglected, just the amplitude is smaller. For $1+F_1^\e < 0$, a
deviation from the normal state grows as
\begin{equation}
  \Delta^\e_1
  (t^*)  \propto h_1^\e(1+F_1^\e)^2  t^*   \frac{\sinh{\sqrt{\frac{|1+ F^\e_1|}{2}} t^*}}{\sqrt{\frac{|1+ F^\e_1|}{2}} t^*}.
  \label{eq:unstable-time}
\end{equation}
The functional form is again the same as for a generic $l=1$ order parameter. The presence of the overall small factor $(1+F_1^\e)^2$ just implies that it takes a longer time for a deviation
 to develop. In particular, the ratio of $
\Delta^\e_1
(t^*)$ in \eqref{eq:unstable-time}
and the initial perturbation $h_1^\e$ becomes $O(1)$ only after
$
\Delta^\e_1
(t^*)$ begins to grow exponentially.

\section{Conclusions}
\label{sec:conclusions}
In this paper we analyzed zero-sound collective bosonic excitations in different angular momentum channels in a metal with an isotropic, but not necessary parabolic dispersion $\epsilon_k$. We explicitly computed the longitudinal and transverse dynamical susceptibility $\chi^\e_l (q, \omega)$ in charge and spin channels for $l=0, l=1$, and $l=2$, and extracted zero-sound modes at $\omega = s v_F^* q$ from the poles of $\chi^\e_l (q, \omega)$. We also presented the generic structure of zero-sound excitations for arbitrary frequency.   Our key goal was to identify, in each case, the mode, whose frequency moves from the lower to the upper half-plane as the system undergoes a Pomeranchuk instability,
  when the corresponding Landau parameter $F^\e_l =-1$.
Right at the transition, the mode is located at $\omega=0$, i.e. the static susceptibility diverges. At $F^\e_l <-1$, the mode moves to the upper frequency half-plane, and a perturbation around a state with no Pomeranchuk order grows exponentially with time, i.e., the system becomes unstable towards spontaneous development of a uniform order parameter, bilinear in fermions.

 We also discussed the evolution of the poles with $F^\e_l >-1$ both for infinitesimally small and for finite fermionic damping rate.  For infinitesimally small damping, we found that in some channels,
  the poles are located very close to a real frequency axis and outside particle-hole continuum already for negative (attractive) $F^\e_l$. This result is at a first glance an unexpected one as naively one would expect the poles to be located inside the continuum. We found that these poles are located below the branch cut and cannot be gradually moved to real axis without simultaneously moving from the physical Riemann sheet to an unphysical one. As the consequence, these poles are silent in the sense that, although they do exist infinitesimally close to the real axis, they are not visible in Im $\chi^\e_l (\omega)$ for real $\omega$.  Besides, we found that for $l >0$,  zero-sound poles for positive $F^\e_l$ exist only if $F^\e_l$ is below a certain value. For larger $F^\e_1$, the poles move from the physical Riemann sheet to an unphysical one.  This does not eliminate the zero-sound peak in Im $\chi^\e_l (\omega)$ for real $\omega$, but the width of the peak becomes larger than fermionic damping $\gamma$.  We argued that in this situation the behavior of time-dependent susceptibility $\chi^\e_l (t)$  at large $t$ is determined by the end point of the branch cut ($\omega = \pm v_F q$) rather than by the zero-sound peak.

We next showed that the situation is somewhat different for $l=1$ order parameters with the same form-factors as that of spin or charge currents. In these two cases, the bosonic response has a zero-sound pole that crosses to the upper half-plane at $F_1^{c(s)} < -1$, but its residue vanishes precisely at $F_1^{c(s)} = -1$. We argued that in this situation static uniform susceptibility does not diverge at  $F_1^{c(s)} = -1$, yet at $1 + F^\e_1 <0$ the system still develops long-range Pomeranchuk order, and the shape of the FS gets modified. It just takes more time for the system to reach the steady ordered state.

 \begin{acknowledgments}
   We thank L. Levitov, J. Schmalian, P. Woelfle, and Y-M. Wu for stimulating discussions. This work was supported by the NSF DMR-1523036 (A.K. and A.V.C.) and NSF-DMR-1720816 (D.L.M.) \end{acknowledgments}

\bibliography{FLtheory}
\end{document}